\documentclass[preprint, 10pt, a4paper, fleqn]{elsarticle}
\usepackage{custom_style}

\usepackage[utf8]{inputenc}
\usepackage[T1]{fontenc}
\usepackage{amsfonts}
\usepackage{amsmath,amssymb,amsthm,setspace,fancybox,mathtools}
\usepackage{graphicx}
\usepackage{import}
\usepackage{subcaption}
\usepackage{bm}
\usepackage[english]{babel}
\usepackage{geometry}
\usepackage[dvipsnames, table]{xcolor}
\usepackage{todonotes}
\usepackage{float}
\usepackage{epstopdf}
\usepackage[hypertexnames=false,colorlinks=true]{hyperref}
\hypersetup{pdfauthor={KL}}
\usepackage{placeins}
\usepackage{booktabs}
\usepackage{nth}
\usepackage{natbib}
\usepackage{bbm}
\usepackage[most]{tcolorbox}
\usepackage{multirow}
\usepackage{tikz}
\usepackage{siunitx}
\usepackage{colortbl}
\usepackage{hhline}
\usepackage{xfrac}
\usepackage{tabularx}
\usepackage{xcolor}

\definecolor{Bordeaux}{RGB}{161,16,53}
\definecolor{Petrol}{RGB}{0,97,101}

\theoremstyle{definition}

\usepackage[ruled,lined,linesnumbered,resetcount]{algorithm2e} 
\usepackage{setspace}
\SetKwComment{Comment}{$\triangleright$\ }{}
\SetCommentSty{}
\SetKwInOut{Given}{Given}
 
\colorlet{author}{magenta}
\colorlet{reviewer1}{blue}
\colorlet{reviewer2}{ForestGreen}
\colorlet{reviewer3}{red}

\definecolor{traincolor}{RGB}{210,230,255}  
\definecolor{valcolor}{RGB}{255,230,200}    

\usepackage{siunitx}
\graphicspath{{src/}} 

\newcommand{\tns}[1]{\bm{\mathrm{#1}}}

\newcommand{\partialder}[2]{\frac{\partial #1}{\partial #2}}
\newcommand{\norm}[1]{\left\lVert#1\right\rVert}

\newcommand{\gb}{\mathrm{gb}}

\newcommand{\bu}{\boldsymbol{u}}
\newcommand{\bv}{\boldsymbol{v}}
\newcommand{\bb}{\boldsymbol{b}}
\newcommand{\bn}{\boldsymbol{n}}
\newcommand{\bP}{\boldsymbol{P}}
\newcommand{\bF}{\boldsymbol{F}}

\newcommand{\bG}{\boldsymbol{G}}
\newcommand{\bR}{\boldsymbol{R}}
\newcommand{\bU}{\boldsymbol{U}}
\newcommand{\bUe}{\boldsymbol{U}^{e}}

\newcommand{\bK}{\boldsymbol{K}}

\newcommand{\bt}{\boldsymbol{t}}
\newcommand{\bX}{\boldsymbol{X}}

\newcommand{\wb}{\mathrm{wb}}
\newcommand{\Cmat}{\boldsymbol{C}}
\newcommand{\Smat}{\boldsymbol{S}}
\newcommand{\Imat}{\boldsymbol{I}}

\newcommand{\Grad}{\nabla}
\newcommand{\Div}{\nabla \cdot}
\newcommand{\tr}{\operatorname{tr}}

\newcommand{\U}{\mathcal{U}}
\newcommand{\V}{\mathcal{V}}
\newcommand{\Fweak}{\mathcal{F}}
\newcommand{\DFweak}{\mathrm{D}\mathcal{F}}

\newcommand{\Ctens}{\mathbb{C}}
\newcommand{\Ke}{\boldsymbol{K}^{e}}

\newcommand{\pd}[2]{\frac{\partial #1}{\partial #2}}

\newcommand{\dV}{\,\mathrm{d}V}
\newcommand{\dA}{\,\mathrm{d}A}

\newcommand{\deltabu}{\delta \bu}

\newcommand{\btheta}{\boldsymbol{\theta}}
\newcommand{\blambda}{\boldsymbol{\lambda}}
\newcommand{\btau}{\boldsymbol{\tau}}

\newcommand{\buobs}{\boldsymbol{u}^{\mathrm{obs}}}

\newcommand{\bFFE}{\boldsymbol{F}^{\mathrm{FE}}}

\newcommand{\bkFE}{\boldsymbol{\kappa}^{\mathrm{FE}}}
\newcommand{\bkobs}{\boldsymbol{\kappa}^{\mathrm{obs}}}

\newcommand{\bQu}{\boldsymbol{Q}_u}

\DeclareMathOperator{\cof}{cof}

\newtcolorbox[
  use counter=myboxnum,
  number freestyle={\noexpand\arabic{\tcbcounter}}
  ]{mybox}[3][]{%
  ams gather,
  enhanced,
  colframe=black,
  boxrule=1pt,
  fonttitle=\bfseries,
  boxed title style={
    boxrule=0pt,
    colback=white,
    colframe=white,
    },
  title=Box~\thetcbcounter: #2,
  #1
}

\makeatletter
\def\ps@pprintTitle{%
    \let\@oddhead\@empty
    \let\@evenhead\@empty
    \def\@oddfoot{\footnotesize\itshape
         {Submitted preprint} \hfill}%
    \let\@evenfoot\@oddfoot
    }
\makeatother

\begin{document}
	
	\begin{frontmatter}

        \title{Automatic-Differentiation-Enabled Finite Element Model Updating: Learning Constitutive Neural Networks from Structural Deformation Data}
        
        \author[1]{Matthias Knipper}

        \author[1]{Chenyi Ji}
  
        \author[1,2]{Malte Brand}

        \author[1]{Kevin Linka \corref{cor1}}

        \address[1]{Computational Mechanics in Medicine, Applied Medical Engineering, RWTH Aachen University\\
        Pauwelsstraße 20, 52074 Aachen, Germany}

        \address[2]{Institute for Continuum and Material Mechanics, Hamburg University of Technology,\\ Ei\ss endorfer Stra\ss e 42, 21073 Hamburg, Germany}

        \cortext[cor1]{Corresponding author: linka@ame.rwth-aachen.de}
		
\begin{abstract}
The identification of constitutive neural network models from heterogeneous full-field deformation data provides a robust alternative to traditional calibration methods based on homogeneous stress–strain experiments, particularly given the high dimensionality of trainable parameters. Existing approaches must balance generality, robustness, and computational efficiency: Conventional finite element model updating is broadly applicable but computationally demanding; weak-form methods offer efficiency but are sensitive to noise and data scarcity; neural operator models are highly expressive but require extensive training datasets. This work presents ADE-FEMU (Automatic-Differentiation-Enabled Finite Element Model Updating), an end-to-end differentiable framework that integrates a constitutive neural network model within a JAX-FEM nonlinear solver and identifies its parameters through gradient-based minimization of a measurement-mismatch loss, allowing for training on structural data with limited kinetic information. Thus, it can be considered as an extension of finite element model updating (FEMU) for hyperelastic materials, implemented within a GPU-accelerated, fully automatically differentiable finite element framework  Newton tangent stiffness and loss gradients are computed automatically using forward- and reverse-mode automatic differentiation throughout the entire pipeline, thereby removing the need for analytic adjoints or offline surrogate models. ADE-FEMU is demonstrated for two architectures: a grey-box Constitutive Artificial Neural Network (CANN), a polyconvex, fully connected model with high flexibility, and a white-box CANN, an expert-system network with phenomenologically interpretable strain-energy terms. Focusing on incompressible isotropic hyperelasticity, ADE-FEMU is evaluated on three open experimental datasets: (1) full digital image correlation (DIC) of a perforated tensile specimen, (2) a reduced-data scenario with a one-dimensional stretch profile and global force–displacement curve, and (3) a heterogeneous matrix–inclusion system in which both phases’ constitutive laws are identified and generalized to twenty-two previously unseen samples. The grey-box CANN demonstrates superior performance with comprehensive, full-field, or heterogeneous data, whereas the interpretable white-box CANN is more reliable for model discovery from sparse datasets. The source code will be released upon publication.
\end{abstract}
		
		
	\end{frontmatter}

    
	\section{Introduction}\label{sec:intro}

Accurate constitutive modeling is central to predictive computational mechanics. In particular, for soft materials and biological tissues, the identification of hyperelastic constitutive laws remains a key prerequisite for reliable finite element simulations under large deformation. Traditionally, constitutive characterization has relied on homogeneous experiments on simple specimen geometries and subsequent fitting of a preselected strain-energy density function to stress--strain data, with comprehensive comparative studies of phenomenological models reported in the rubber-like and soft-tissue literature \cite{Dal2021}. While this paradigm has been successful for many engineering materials, it becomes increasingly restrictive when the underlying material behavior is complex, when the constitutive model class is not known \emph{a priori}, or when the available measurements arise from heterogeneous deformation states rather than from idealized homogeneous tests \cite{Tan2026,Fuhg2024Review}. As recently emphasized, the simultaneous measurement of all tensorial components of strain and stress is in fact practically infeasible, so that unsupervised identification strategies operating on indirect observables such as displacement fields and reaction forces are the only realistic route in many situations of practical interest \cite{Regazzoni2026,Roux2020}.

\paragraph{Full-field identification and data-driven constitutive modeling}\label{sec:intro:sota}

For this reason, modern material characterization has increasingly shifted toward \emph{full-field} experimental approaches, in which heterogeneous displacement data and global reaction forces are measured on complex specimen geometries \cite{Avril2008,PierronGrediac2020,RomerEtAl2025}. Such experiments are particularly attractive because they probe a richer set of deformation modes within a single test and thereby provide a substantially more informative basis for constitutive identification. In this context, a broad class of so-called unsupervised identification strategies has emerged, in which material parameters are inferred not from direct stress--strain pairs, but from indirect observables such as local displacements and boundary forces. Among these methods, finite element model updating (FEMU) remains one of the most widely used approaches for constitutive calibration \cite{CollinsEtAl1974,WiesheierEtAl2024,WiesheierEtAl2026}. FEMU identifies material parameters by repeatedly solving the forward boundary value problem and minimizing the mismatch between measured and simulated responses. Its main strengths are flexibility and robustness, since, in principle, it can be applied to virtually any constitutive model and experimental configuration that can be represented in a forward finite element simulation. However, this generality comes at the price of high computational cost, and the resulting optimization problems may be difficult to solve robustly in the presence of nonconvex objective landscapes and multiple local minima, especially when the parameter space is large and the forward model expensive \cite{RomerEtAl2025,Tan2026}. Recent extensions to FEMU leverage automatic differentiation for efficient gradient computation \cite{Hassanieh2025,Ferreira2026}, also considering path-dependent behavior. However, integrating these approaches into existing large-deformation finite element codes remains nontrivial due to the need for differentiable implementations of the solver. 

In contrast to conventional iterative FEMU approaches, Physics-Informed Neural Networks (PINNs) enable the direct calibration of constitutive material models from full-field displacement data by embedding the governing mechanical equations into the learning process \cite{Anton2022pinns}. In particular, PINN-based approaches provide a natural framework for uncertainty-aware material identification, allowing noisy measurements and probabilistic parameter representations to be incorporated directly into the training procedure \cite{Anton2025deter}.

An alternative family of methods also avoids repeated forward solves by enforcing the governing equations directly in weak form. Prominent examples include the Virtual Fields Method (VFM) \cite{Grediac1989,GrediacVautrin1990,GrediacEtAl2006}, the Equilibrium Gap Method (EGM) \cite{ClaireEtAl2004}, and more recent extensions such as EUCLID for constitutive law identification and discovery \cite{FlaschelKumarDeLorenzis2021,FlaschelKumarDeLorenzis2022,FlaschelKumarDeLorenzis2023}. The EUCLID family has subsequently been extended in several directions, including deep-learning hyperelasticity without stress data \cite{Thakolkaran2022}, Bayesian discovery with uncertainty quantification \cite{Joshi2022}, automated identification of linear viscoelastic laws \cite{Marino2023}, non-associated pressure-sensitive plasticity \cite{Xu2025}, optimal experimental design through topology optimization of test specimens \cite{Ghouli2025}, and most recently the direct validation on experimental data for hyperelastic rubber \cite{Abbasi2026}. These approaches can be highly efficient, since they operate directly on experimentally measured displacement fields and reaction forces. At the same time, they typically require differentiation of the measured displacement field to obtain strain-like quantities, which makes them sensitive to noise and potentially less robust when the data are incomplete, spatially sparse, or experimentally imperfect \cite{Avril2008,PierronGrediac2020,RomerEtAl2025,Wessels2026b}. As a further consequence, weak-form approaches in their current form rely on the availability of the complete displacement field across the specimen, which limits their applicability in settings where only limited kinematic data is accessible, such as boundary-only or surface-only observations of three-dimensional specimens \cite{Regazzoni2026}. Hence, current full-field identification strategies often face a trade-off between robustness and computational efficiency: methods based on repeated finite element simulations are general but expensive, whereas weak-form-based approaches are fast but can be fragile with respect to measurement quality and observability.

In parallel with these developments, data-driven constitutive modeling has gained enormous momentum, with a recent comprehensive review summarizing the rapidly growing landscape \cite{Fuhg2024Review}. Neural networks and related machine learning models have shown considerable promise in representing complex constitutive responses beyond the limitations of hand-crafted phenomenological laws \cite{GhaboussiEtAl1991,Linka2021,AsadFarhat2022,KleinEtAl2022,LindenEtAl2023,KalinaEtAl2025,Huang2020,tepole2022node}. Related approaches dispense with an explicit constitutive model altogether and instead perform simulations directly guided by experimental data \cite{KirchdoerferOrtiz2016,IbanezEtAl2017}. At the same time, a complementary research direction seeks to retain interpretability by discovering constitutive relations automatically through symbolic or sparse regression \cite{SchoenauerEtAl1996,AbdusalamovEtAl2023,LinkaKuhl2023,LinkaKuhl2024,Kissas2024}. Recent contributions have begun to merge these two perspectives by combining physics-augmented neural networks with sparsification, transfer learning, and multi-modal data sources, thereby enabling rapid model discovery that is both flexible and interpretable across material classes \cite{Tan2026}. A related and equally important line of research has focused on architectural design, ensuring that the learned strain-energy functions satisfy frame indifference, material symmetry, polyconvexity, coercivity, and a stress-free reference configuration by construction \cite{Regazzoni2026,LindenEtAl2023,Klein2023,Chen2022,Geuken2025,Flaschel2025JMPS}, so that well-posedness of the underlying boundary value problem is preserved throughout the learning process. A complementary line is concerned with model discovery that bypasses the optimization problem altogether through database lookup, exemplified by the recent Material Fingerprinting framework \cite{Flaschel2026MF}.

Yet, even with these advances, most current workflows remain fragmented at the implementation level: simulation data are generated in one finite element environment, constitutive models are trained in a separate machine learning framework, and the resulting model must then be re-implemented in a finite element code for deployment \cite{XueEtAl2023}. Beyond this practical inconvenience, learning constitutive models from full-field data introduces an additional computational challenge. If the learning objective depends implicitly on the solution of a nonlinear boundary value problem, then efficient training requires sensitivities of the full-field response with respect to constitutive model parameters. For nonlinear hyperelasticity, the derivation and implementation of these sensitivities are often tedious, problem-specific, and error-prone \cite{VanKeulenEtAl2005,ReesEtAl2010}.

Recent progress in differentiable simulation offers a compelling route to overcome these limitations. In particular, JAX-FEM has demonstrated that finite element workflows can be reformulated in a differentiable programming setting by building directly on JAX \cite{BradburyEtAl2018,XueEtAl2023}. Within this framework, the weak form is implemented explicitly, while the linearization required by Newton's method and the sensitivities required for inverse problems are obtained automatically via automatic differentiation \cite{GriewankWalther2008,XueEtAl2023}. This avoids the manual derivation of tangent operators and adjoint sensitivities, which is especially attractive for complex constitutive relations. Moreover, JAX-FEM combines differentiable simulation, GPU acceleration, and seamless integration with machine learning within a unified computational ecosystem, and has been shown to deliver order-of-magnitude speed-ups over conventional finite element codes for representative inverse-design problems in nonlinear solid mechanics \cite{XueEtAl2023}. These properties make it a promising foundation not only for classical inverse design problems, but also for constitutive learning problems in which neural material models must be trained through the solution of finite element boundary value problems. Building on this idea, recent works have begun to embed differentiable finite element solvers directly into the training loop of neural constitutive models, thereby coupling the discovery of material laws to the elastostatic equilibrium problem associated with each candidate model \cite{Regazzoni2026,Tan2026,Bleyer2025}.

\paragraph{Finite-element-based framework for learning constitutive neural network models}\label{sec:intro:gap}
Despite this rapid progress, several limitations persist that motivate the present contribution. First, classical FEMU and most differentiable finite-element-based identification strategies have so far been formulated for constitutive families with relatively few parameters and a fixed functional form, while genuine \emph{model discovery} from full-field data with high-capacity neural constitutive representations remains comparatively underexplored \cite{Tan2026}. Second, weak-form discovery frameworks such as EUCLID achieve impressive efficiency, yet rely on differentiation of measured displacement fields and on the availability of dense full-field kinematic information, which restricts their applicability to settings in which the entire displacement field is observable and only mildly affected by noise \cite{FlaschelKumarDeLorenzis2021,Regazzoni2026}. Third, although architectural strategies that enforce physical admissibility \emph{by construction} have matured significantly \cite{LindenEtAl2023,Regazzoni2026}, their integration with simulation-based identification still typically requires either a dedicated differentiable finite element implementation tailored to a specific constitutive ansatz, or fragile coupling between separate simulation and machine learning environments. As a result, a unified workflow that combines (i) flexible neural constitutive representations with strong physical inductive biases, (ii) end-to-end differentiable finite element analysis at structural scale, and (iii) identification from realistic, possibly partial full-field measurements is still missing.

In response to these challenges, the present work proposes a finite-element-based framework for learning hyperelastic constitutive neural network models directly from full-field deformation data using automatic differentiation. Rather than relying on homogeneous stress--strain pairs, we exploit heterogeneous full-field displacement measurements generated by a structural experiment and embed the constitutive neural network directly into the finite element model. The parameters of the material network are then identified by minimizing the mismatch between measured and simulated structural responses, while gradients with respect to all trainable parameters are obtained automatically through the differentiable finite element solver. In this way, the proposed approach preserves the robustness and physical consistency of simulation-based full-field identification while substantially simplifying the sensitivity analysis that usually constitutes a major implementation bottleneck.

The proposed framework is methodologically closest to classical FEMU, in the sense that constitutive parameters are identified through simulation-based calibration against structural full-field measurements. From this perspective, the present approach may be interpreted as a differentiable FEMU formulation for constitutive neural network models, in which gradients with respect to material parameters are obtained automatically through end-to-end differentiation of the finite element model rather than through manually derived sensitivities or external sensitivity analyses cf. \cite{Tan2026}. In this way, automatic differentiation substantially reduces a traditional implementation bottleneck while naturally enabling the use of high-capacity neural constitutive representations within a unified computational framework implemented in JAX. In contrast to weak-form identification approaches, the method does not require differentiation of measured displacement fields and instead differentiates the computational model itself, which improves robustness with respect to noisy or incomplete observations. Likewise, compared with two-stage workflows that first generate simulation data in a conventional finite element environment and subsequently train constitutive surrogates offline, the present formulation removes the separation between simulation and learning by embedding constitutive neural networks directly into a single differentiable finite element workflow \cite{Regazzoni2026,XueEtAl2023}.

\paragraph{Objective and outline}\label{sec:intro:outline}
Motivated by the above observations, this work develops a fast and general framework for constitutive learning from full-field data based on the combination of nonlinear finite elements, automatic differentiation, and neural constitutive modeling. Focusing on hyperelasticity, we demonstrate how differentiable finite element analysis can be used to infer complex material behavior from heterogeneous deformation fields in a structurally consistent manner, and how this framework integrates the strengths of simulation-based identification, weak-form discovery, and physics-augmented neural network design within a single end-to-end differentiable pipeline. More broadly, this work contributes to FEMU methods by learning data-driven material models directly at the governing boundary-value-problem level within a GPU-accelerated, fully differentiable nonlinear finite element framework, rather than through decoupled stress–strain fitting. This enables material identification from both heterogeneous full-field and sparse data. Importantly, the information content of full-field measurements depends not only on spatial resolution or data volume, but on the diversity of deformation states and their sensitivity to the constitutive parameters. Thus, heterogeneous experiments do not inherently provide better parameter identifiability than multiple well-designed homogeneous tests. Their advantage lies in exploiting informative deformation fields when specimen designs activate complementary material responses and sufficiently constrain the constitutive model.
The open datasets used here concern incompressible hyperelastic materials, for which the volume-preserving constraint allows through-thickness deformation to be inferred from planar full-field measurements. This choice reflects the available datasets rather than a limitation of ADE-FEMU. The framework also supports compressible and nearly incompressible materials, extending its applicability to more general constitutive models.

The remainder of this manuscript is structured as follows. Section~\ref{sec:theory} summarizes the kinematic and continuum-mechanical foundations and introduces the two families of constitutive artificial neural networks employed in this work, namely a grey-box and a white-box CANN. Section~\ref{sec:framework} formulates the forward nonlinear finite element problem in the JAX-FEM \cite{XueEtAl2023} setting and the inverse PDE-constrained identification problem, derives the associated reduced-gradient and adjoint expressions, and describes the numerical implementation of the full-field training procedure. Section~\ref{sec:results} demonstrates the performance of the proposed framework on three representative cases based on openly available experimental data: a full-field DIC dataset of perforated soft-tissue specimens (Case \hyperref[sec:results:case1]{1}), a reduced-data setting in which only the maximum stretch along a single path is available (Case \hyperref[sec:results:case2]{2}), and a heterogeneous material system with twenty-three specimens of varying matrix--inclusion topology (Case \hyperref[sec:results:case3]{3}). Section~\ref{sec:discussion} discusses the results in the context of existing FEMU \cite{CollinsEtAl1974,WiesheierEtAl2024,WiesheierEtAl2026,Roux2020,Tan2026}, VFM \cite{Grediac1989,GrediacVautrin1990,GrediacEtAl2006}, EUCLID \cite{FlaschelKumarDeLorenzis2021,FlaschelKumarDeLorenzis2022,FlaschelKumarDeLorenzis2023,Thakolkaran2022,Joshi2022,Marino2023,Abbasi2026}, and Neural-DFEM-type approaches \cite{Regazzoni2026}, summarizes the current limitations, and outlines directions for future work.

\section{Theoretical background}\label{sec:theory}
        
    Starting with the deformation gradient $\boldsymbol{F}$, with $J=\det\boldsymbol{F}>0$, we define the right Cauchy--Green deformation tensor $\Cmat=\boldsymbol{F}^\mathrm{T}\boldsymbol{F}$. The principal invariants of $\Cmat$ are defined by,
    \begin{equation}\label{eq:invars}
        I_1 = \tr (\Cmat) , \qquad I_2 = \frac{1}{2} \left[ (\tr\Cmat)^2 - \tr(\Cmat^2) \right], \qquad I_3 = \det \Cmat = J^2.
    \end{equation}
    In this work, we consider perfectly incompressible materials such that $I_1, I_2 \geq 3$, with equality in the undeformed reference configuration ($\Cmat=\tns{I}$), and $I_3=1$. Incompressible hyperelasticity postulates the existence of a free energy function
    \begin{equation}
        \Psi = \hat{\Psi}(\Cmat) - p(J-1),
    \end{equation}
    where $p$ is an indeterminate Lagrange multiplier enforcing the incompressibility constraint $J=1$. Following the second law of thermodynamics \cite{Truesdell1984}, the second Piola--Kirchhoff stress $\Smat$ is given by,   
    \begin{equation}
        \Smat = 2\partialder{\Psi}{\Cmat} = - p\Cmat^{-1} + 2\partialder{\hat{\Psi}}{\Cmat},
    \end{equation}
    The dependency of $\Psi$ on $\Cmat$ a priori ensures material objectivity. Specifically, for isotropic materials, the free energy must depend on the first two principal invariants of $\Cmat$, i.e., $\Psi=\Psi(I_1, I_2)$. Using the differentiation relations
    \begin{equation}
     \partialder{I_1}{\Cmat}=\tns{I}, \qquad \partialder{I_2}{\Cmat}=I_1\tns{I} - \Cmat, \qquad \partialder{J}{\Cmat} = \frac{J}{2}\Cmat^{-1},
    \end{equation}
    this results in the following expression for the stress,
    \begin{equation}
        \Smat = -p\Cmat^{-1} + 2 \partialder{\Psi}{I_1} + 2 \partialder{\Psi}{I_2} \left(I_1\tns{I} - \Cmat \right).
    \end{equation}

    Typically, several physically reasonable and mathematically convenient requirements are typically imposed on the free energy. For example, the normalization condition requires that both energy and stress vanish in the undeformed reference configuration:
	\begin{equation}\label{eq:normalization_condition}
		\Psi(\Cmat=\tns{I})=0, \qquad \Smat(\Cmat=\tns{I}) = \tns{0},
	\end{equation}
	Another desirable property of the free energy is polyconvexity, meaning $\Psi$ is a convex function of $\boldsymbol{F}$, $\cof\boldsymbol{F}$, and $\det \boldsymbol{F}$ \cite{Ball1976}. Although a mathematical concept and not a constitutive requirement, polyconvexity is essential to ensure material stability. The invariants $I_1$ and $I_2$ are convex in $\boldsymbol{F}$ and $\cof \boldsymbol{F}$, respectively, rendering $\Psi$ polyconvex.
 
    
    It remains to design an appropriate free energy function to accurately describe the mechanical behavior of a specific material. However, instead of relying on human expertise to model a free energy function, we automate this process and use constitutive artificial neural networks.        
    
	
    \subsection{Constitutive artificial neural networks}

    Constitutive artificial neural networks (CANNs) can be divided into two main categories. The first category uses general fully connected feedforward neural networks (FFNNs) to represent the free energy \cite{Linka2021}. These FFNNs, with their flexible width and depth, offer powerful approximation capabilities. By carefully selecting activation functions, imposing weight constraints, and adding corrections, they inherently satisfy all constitutive constraints from Section~\ref{sec:theory}. However, the interpretability of free energy may be challenging, particularly with large FFNNs, due to their \emph{grey-box} nature.

    The second category of CANNs functions as an expert system, generalizing widely accepted free energies and inherently satisfying essential constitutive constraints \cite{Linka2022a}. While these \emph{white-box} models lack the flexibility of grey-box CANNs, their constrained architecture allows for a more interpretable free energy with physically meaningful parameters \cite{tepole2024canns}. 

    \begin{figure}[htbp]
        \centering
        \begin{subfigure}{0.47\linewidth}
            \includegraphics[height=5.1cm]{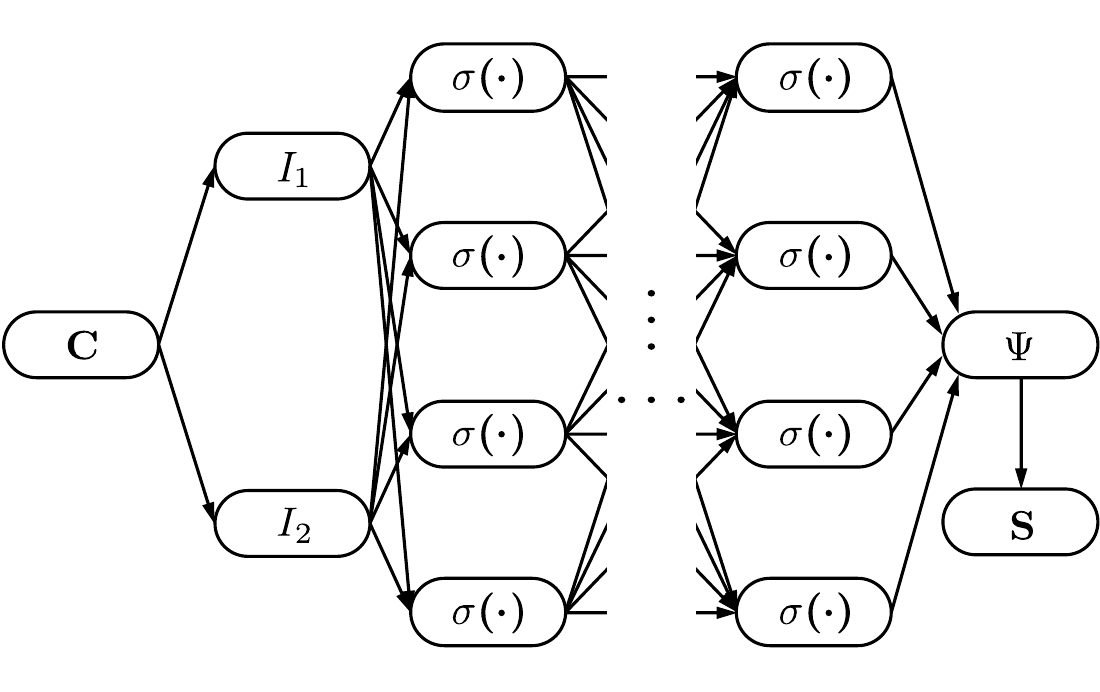}
            \caption{\textbf{Grey-box CANN.} The free-energy function $\Psi$ is represented by a general fully connected feedforward neural network taking the principal invariants $I_1$ and $I_2$ of $\Cmat$ as input. The number of layers and neurons per layer is arbitrary. Polyconvexity is enforced by construction: the activation functions $\sigma(\cdot)$ are chosen convex and non-decreasing, and the weights are constrained to be positive.}
            \label{fig:results:vhb_a}
        \end{subfigure}
        \hfill
        \begin{subfigure}{0.47\linewidth}
          \centering
            \def\svgwidth{0.95\linewidth}
            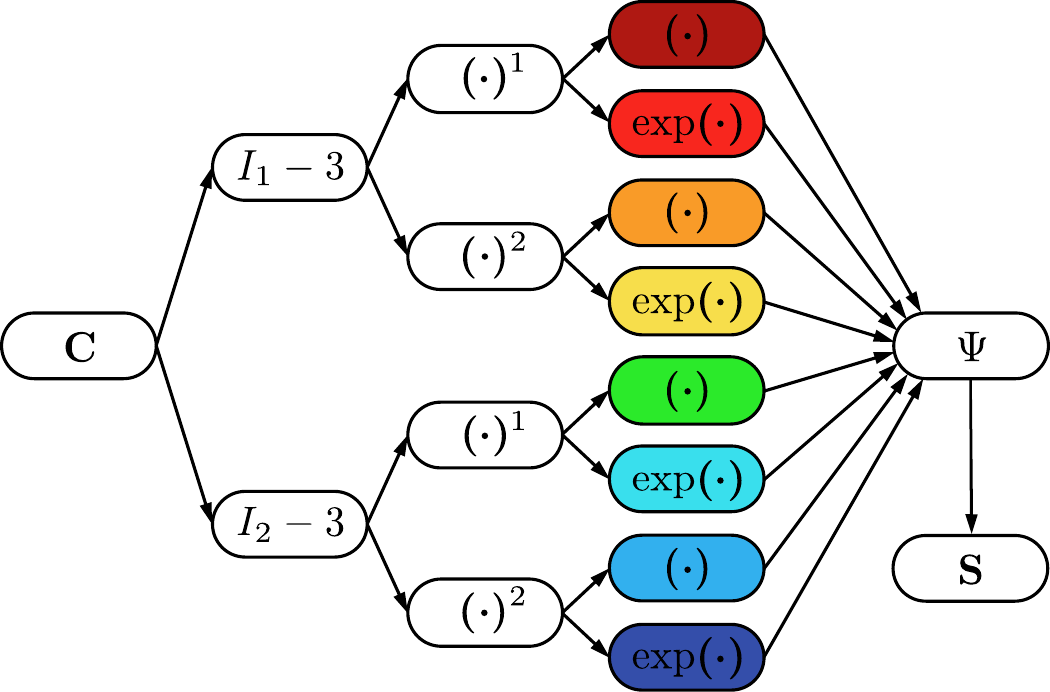
            \caption{\textbf{White-box CANN.} The network has two partially connected hidden layers and represents the free-energy function $\Psi$ as a sum of eight interpretable terms built from the principal invariants $I_1$ and $I_2$ of $\Cmat$. The first layer generates the powers $(\cdot)^1$ and $(\cdot)^2$ of the network input. The second layer applies the identity and the exponential function $\exp(\cdot)$ to these powers, and the network output is the sum of the resulting eight contributions.}
            \label{fig:results:vhb_b}
        \end{subfigure}
        \caption{\textbf{Constitutive artificial neural network architectures employed in this work.} (a) Grey-box CANN: a fully connected feedforward network with arbitrary depth and width that yields a polyconvex free-energy function by construction. (b) White-box CANN: a partially connected, two-layer expert system whose eight terms recover classical phenomenological constitutive blocks (neo-Hookean, Demiray, Gent type) and therefore provide a directly interpretable free-energy density.}
        \label{fig:results:vhb}
    \end{figure}

   \subsection{Constitutive artificial neural networks as a grey box model}
    \label{ss:grey-box}

    Using the grey-box approach, we represent a scalar-valued free energy function $\Psi^\mathcal{N}$ using an FFNN $\mathcal{N}_\Psi$, with the principal invariants $I_1$ and $I_2$ as inputs:
    \begin{equation}
        \Psi^\mathcal{N}(I_1, I_2) = \mathcal{N}_\Psi(I_1, I_2).
    \end{equation}
    By using the principal invariants as inputs, the principle of material objectivity is automatically satisfied, and material symmetry, here isotropy, is preserved. The stress is calculated as the gradient of the free energy function, $\Smat^\mathcal{N} = 2 \, \partial\Psi^\mathcal{N} / \partial \Cmat$, which ensures thermodynamic consistency and symmetry of the second Piola–Kirchhoff stress tensor. However, the invariant-based FFNN $\mathcal{N}_\Psi$ does not inherently satisfy the normalization condition. To address this, we introduce correction terms $\Psi^\sigma$ and $\Psi^\varepsilon$ to enforce a stress- and energy-free reference configuration. Following \cite{Abdolazizi2023}, the stress correction is 
    \begin{equation}
        \Psi^\sigma( I_1, I_2 ) \coloneqq - 2\left( \partialder{\Psi^\mathcal{N}}{I_1} + 2 \partialder{\Psi^\mathcal{N}}{I_2} \right)\bigg|_{\Cmat=\tns{I}} (J-1),
    \end{equation}
    and the energy correction is simply the constant
    \begin{equation}
         \Psi^\varepsilon \coloneqq - \Psi^\mathcal{N}( I_1, I_2 )|_{\Cmat=\tns{I}} 
    \end{equation}

    Thus, the total free energy of the grey-box model reads
    \begin{equation}\label{eq:SEF_grey}
        \Psi^{\gb}( I_1, I_2 ) \coloneqq \Psi^\mathcal{N} ( I_1, I_2 ) + \Psi^\sigma ( I_1, I_2) + \Psi^\varepsilon,
    \end{equation}
    and the associated stress is
    \begin{equation}
        \Smat^{\gb} \coloneqq 2\partialder{\Psi^{\gb}}{\Cmat} = -\left[ p + \left( \partialder{\Psi^\mathcal{N}}{I_1} + 2 \partialder{\Psi^\mathcal{N}}{I_2} \right)\bigg|_{\Cmat=\tns{I}} \, \right] \Cmat^{-1} +  2 \left[ \left(\partialder{\Psi^\mathcal{N}}{I_1} + I_1\partialder{\Psi^\mathcal{N}}{I_2} \right) \tns{I} - \partialder{\Psi^\mathcal{N}}{I_2}\Cmat \right].
    \end{equation}
    
    Since $\Psi^{\gb}$ in \eqref{eq:SEF_grey} is a positively weighted sum of free energy contributions, it is polyconvex if each contribution is polyconvex \cite{Boyd2004}. Given that $\Psi^\sigma$ is polyconvex by construction, and $\Psi^\varepsilon$ is a constant, we need only ensure that $\Psi^\mathcal{N}$ is polyconvex.
    
    For this, we recall a fundamental property of convex functions: the composition of a convex function and a convex, non-decreasing function is also convex \cite[6.]{Wilber2002}. Since the principal invariants $I_1$ and $I_2$ are already (poly)convex, it remains to ensure that the FFNN $\mathcal{N}_\Psi$ is convex and non-decreasing. To achieve this, we impose constraints on $\mathcal{N}_\Psi$. Specifically, we require that all weights of $\mathcal{N}_\Psi$ are non-negative, and we use convex, non-decreasing activation functions in each layer \cite{Amos2017}. In the hidden layers, we use the softplus activation function, $\sigma(x)=\ln[1+\exp(x)]$, while in the final layer, we apply a linear activation function, $\sigma(x)=x$, both of which are convex and non-decreasing. As a result, the composition of the (poly)convex generalized invariants and the convex, non-decreasing FFNN $\mathcal{N}_\Psi$ ensures the polyconvexity of the equilibrium free energy $\Psi$.
    In summary, we satisfy by construction all of the important requirements for the free energy function.

\subsection{Constitutive artificial neural networks as a white box model}
\label{ss:white-box}
As an interpretable alternative to general feedforward free-energy networks, we consider a selectively connected constitutive artificial neural network (CANN) in the spirit of \cite{Linka2023}, in which the strain-energy density is represented through a prescribed library of admissible constitutive building blocks. To guarantee a stress-free and energy-free reference configuration, the network is expressed in terms of the shifted invariants
\begin{equation}
\hat{I}_1 = I_1 - 3,
\qquad
\hat{I}_2 = I_2 - 3.
\end{equation}
The white-box CANN combines powers of these shifted invariants with a predefined set of nonlinear activation functions. Specifically, the first hidden layer generates the linear and quadratic terms
\begin{equation}
\hat{I}_1, \quad \hat{I}_1^2, \quad \hat{I}_2, \quad \hat{I}_2^2,
\end{equation}
while the second hidden layer applies the identity and exponential mappings to each of these quantities. The resulting strain-energy density reads
\begin{align}
\Psi^{\wb}(I_1,I_2)
&=
w_{2,1}  \hat{I}_1
+ w_{2,2} \left[\exp\!\left(w_{1,2}\hat{I}_1\right)-1\right]
\nonumber\\
&\quad
+ w_{2,3} \hat{I}_1^{2}
+ w_{2,4} \left[\exp\!\left(w_{1,4}\hat{I}_1^{2}\right)-1\right]
\nonumber\\
&\quad
+ w_{2,5} \hat{I}_2
+ w_{2,6} \left[\exp\!\left(w_{1,6}\hat{I}_2\right)-1\right]
\nonumber\\
&\quad
+ w_{2,7}  \hat{I}_2^{2}
+ w_{2,8} \left[\exp\!\left(w_{1,8}\hat{I}_2^{2}\right)-1\right]
\label{eq:SEF_white}
\end{align}

In order to ensure non-negativity, numerical stability and stable differentiability, each weight is activated by a softplus activation function: $w_{i,j} = \ln[1+\exp(\hat w_{i,j})]$. This architecture yields an interpretable constitutive representation, since each active term corresponds to a familiar constitutive mechanism: linear invariant terms recover neo-Hookean- or Mooney--Rivlin-type contributions and exponential terms recover Demiray- or Holzapfel-type behavior. Owing to this explicit structure, the model can be interpreted as a neural constitutive expert system that interpolates between classical phenomenological energy functions.

The associated stress response follows from thermodynamic consistency by differentiating the strain-energy density with respect to the chosen strain measure. In the invariant-based formulation used here, the second Piola--Kirchhoff stress is obtained as
\begin{equation}
\Smat^{\wb}
=
2 \frac{\partial \Psi^{\wb}}{\partial \Cmat}
=
2\frac{\partial \Psi^{\wb}}{\partial I_1}\frac{\partial I_1}{\partial \Cmat}
+
2\frac{\partial \Psi^{\wb}}{\partial I_2}\frac{\partial I_2}{\partial \Cmat},
\end{equation}
with
\begin{equation}
\frac{\partial I_1}{\partial \Cmat} = \Imat,
\qquad
\frac{\partial I_2}{\partial \Cmat} = I_1 \Imat - \Cmat.
\end{equation}
The invariant derivatives of the strain-energy density are given by
\begin{align}
\frac{\partial \Psi^{\wb}}{\partial I_1}
&=
w_{2,1}
+ w_{2,2}w_{1,2}\exp(w_{1,2}\hat{I}_1)
\nonumber\\
&\quad
+ 2\hat{I}_1\!\left[
w_{2,3}
+ w_{2,4}w_{1,4}\exp(w_{1,4}\hat{I}_1^2)
\right],
\\
\frac{\partial \Psi^{\wb}}{\partial I_2}
&=
w_{2,5}
+ w_{2,6}w_{1,6}\exp(w_{1,6}\hat{I}_2)
\nonumber\\
&\quad
+ 2\hat{I}_2\!\left[
w_{2,7}
+ w_{2,8}w_{1,8}\exp(w_{1,8}\hat{I}_2^2)
\right].
\end{align}

Accordingly, the final stress representation becomes
\begin{equation}
\Smat^{\wb}
=
2\left[
\left(\frac{\partial \Psi^{\wb}}{\partial I_1}
+ I_1 \frac{\partial \Psi^{\wb}}{\partial I_2}\right)\Imat
-
\frac{\partial \Psi^{\wb}}{\partial I_2}\Cmat
\right].
\end{equation}

By construction, this white-box architecture preserves material objectivity, isotropy, and thermodynamic consistency. At the same time, its constrained and interpretable functional form allows the identified parameters to retain a direct constitutive meaning, in contrast to the more flexible but less transparent grey-box CANNs.

\section{Computational model framework}\label{sec:framework}

\subsection{Forward problem: nonlinear finite element formulation for hyperelasticity}

Let $\Omega \subset \mathbb{R}^3$ denote the reference configuration of a hyperelastic body with boundary $\partial \Omega = \Gamma_D \cup \Gamma_N$, where $\Gamma_D \cap \Gamma_N = \emptyset$. The forward problem consists in finding the displacement field $\bu : \Omega \to \mathbb{R}^3$ such that
\begin{align}
-\Div \bP &= \bb \qquad \text{in } \Omega, \\
\bu &= \bar{\bu} \qquad \text{on } \Gamma_D, \\
\bP \bn &= \bar{\bt} \qquad \text{on } \Gamma_N,
\end{align}
where $\bb$ denotes the body force per unit reference volume, $\bar{\bu}$ the prescribed displacement, $\bar{\bt}$ the prescribed traction, $\bn$ the outward unit normal on $\Gamma_N$, and $\bP$ the first Piola--Kirchhoff stress tensor. In the JAX-FEM framework, hyperelasticity is treated as a specific instance of a general nonlinear PDE posed in weak form, with the constitutive flux identified as the stress response\cite{XueEtAl2023}. The corresponding weak form reads: find $\bu \in \U$ such that
\begin{equation}
\Fweak(\bu;\bv) = 0
\qquad \forall \bv \in \V,
\end{equation}
with
\begin{equation}
\Fweak(\bu;\bv)
=
\int_{\Omega} \bP(\bF) : \Grad \bv \, \dV
-
\int_{\Gamma_N} \bar{\bt} \cdot \bv \, \dA
-
\int_{\Omega} \bb \cdot \bv \, \dV,
\end{equation}
where the trial and test spaces are defined as
\begin{align}
\U &= \left\{ \bu \in H^1(\Omega;\mathbb{R}^3) \,:\, \bu = \bar{\bu} \text{ on } \Gamma_D \right\}, \\
\V &= \left\{ \bv \in H^1(\Omega;\mathbb{R}^3) \,:\, \bv = \mathbf{0} \text{ on } \Gamma_D \right\}.
\end{align}
Since $\bP$ depends nonlinearly on $\bu$ through $\bF$, the residual equation is solved by Newton's method \cite{XueEtAl2023}. 

Given a current iterate $\bu$, the Newton increment $\deltabu \in \V$ is determined from the linearized problem
\begin{equation}
\DFweak(\bu)[\deltabu;\bv] = -\Fweak(\bu;\bv)
\qquad \forall \bv \in \V,
\end{equation}
where $\DFweak(\bu)[\deltabu;\bv]$ denotes the G\^ateaux derivative of the weak form with respect to $\bu$ in the direction $\deltabu$. In a classical nonlinear finite element setting, this linearization involves the consistent material tangent
\begin{equation}
\Ctens = \pd{\bP}{\bF}.
\end{equation}

For the finite element discretization, the displacement field is approximated as
\begin{equation}
\bu^h(\bX) = \sum_{a=1}^{n_{\mathrm{node}}} N_a(\bX)\,\bU_a,
\end{equation}
where $N_a$ are the shape functions and $\bU_a$ are the nodal displacement unknowns. This leads to the discrete nonlinear residual problem
\begin{equation}
\boldsymbol{G}(\bU) = \mathbf{0},
\end{equation}
where $\bU$ collects all global degrees of freedom (DOFs). At the element level, JAX-FEM defines the residual vector as
\begin{equation}
G_i^e
=
\int_{\Omega_e} \bP(\bF^h) : \Grad \phi_i \, \dV,
\end{equation}
where $\phi_i$ denotes the local basis function associated with the $i$th element degree of freedom. Rather than deriving the consistent tangent analytically, JAX-FEM computes the element stiffness matrix as the Jacobian of the element residual,
\begin{equation}
\Ke = \pd{\boldsymbol{g}_e}{\bUe},
\end{equation}
using automatic differentiation. This bypasses the explicit and manual derivation and implementation of fourth-order tangent operators and is particularly advantageous for complex constitutive laws \cite{XueEtAl2023}. Consequently, for hyperelastic materials, the user only needs to implement the strain energy density function. Automatic differentiation is then used to obtain both the stress measures required for the residual evaluation and the consistent tangent required for the Newton iterations. It should be noted that, while AD provides a convenient and robust approach for computing consistent tangent operators, standard AD does not automatically exploit the sparsity structure of the assembled finite element matrices. Therefore, for large-scale three-dimensional problems, the computational cost and memory requirements may become significant, motivating the development of sparsity-aware differentiation and matrix-free approaches. After global assembly, the Newton step reads
\begin{equation}
\bK\!\left(\bU^{(l)}\right)\,\Delta \bU^{(l)} = -\bG\!\left(\bU^{(l)}\right),
\end{equation}
followed by the update
\begin{equation}
\bU^{(l+1)} = \bU^{(l)} + \Delta \bU^{(l)}.
\end{equation}
This defines the nonlinear forward solve for hyperelasticity within the differentiable finite element framework adopted in this work \cite{XueEtAl2023}.

For incompressible or nearly incompressible hyperelasticity, the formulation is extended using a mixed displacement--pressure approach based on the Herrmann formulation \cite{Herrmann1965}. In this setting, the pressure field is introduced as an additional unknown acting as a Lagrange multiplier enforcing the volumetric constraint
\begin{equation}
J - 1 = 0,
\qquad
J = \det \bF.
\end{equation}
The first Piola--Kirchhoff stress tensor is correspondingly decomposed into deviatoric and volumetric contributions according to
\begin{equation}
\bP = \bP_{\mathrm{dev}} - p\,J\,\bF^{-T},
\label{equ:Psplit}
\end{equation}
so that the equilibrium equation depends on both displacement and pressure. The implementation follows the hybrid element concept commonly used in Abaqus incompressible solid elements \cite{SimoTaylor1991,AbaqusTheoryGuide}. 
As established in Eq. \eqref{equ:Psplit}, only the deviatoric part of the stress tensor must be evaluated. Consequently, the CANNs are employed exclusively to predict the deviatoric stress response. To ensure a proper volumetric–isochoric split within the hybrid formulation, the standard invariants are replaced with the isochoric invariants
    \begin{equation}\label{eq:invars_incomp}
        \bar I_1 = I_1 J^{-2/3} , \qquad \bar I_2 = I_2J^{-4/3}, \qquad J = \sqrt{\det\Cmat}.
    \end{equation}

While these invariants coincide with the standard invariants in the exact incompressible case $(J=1)$, the isochoric formulation remains necessary in the discrete setting, since $J$ must still be treated as a kinematic variable whose incompressibility constraint is enforced only weakly through the pressure field. Therefore, finite element approximations may exhibit small local deviations from $J=1$. 
The displacement field is discretized using linear HEX8 interpolation, while the pressure field is represented by a single scalar pressure degree of freedom per element, corresponding to a piecewise-constant $P_0$ interpolation. 
The resulting nonlinear system is therefore written in terms of the coupled unknown vector $[\bU,\bP]^T$, where $\bU$ collects nodal displacement degrees of freedom and $\bP$ the elemental pressure unknowns. After linearization, the Newton system takes the block form
\begin{equation}
\begin{bmatrix}
\bK_{uu} & \bK_{up} \\
\bK_{pu} & \bK_{pp}
\end{bmatrix}
\begin{bmatrix}
\Delta \bU \\
\Delta \bP
\end{bmatrix}
=
-
\begin{bmatrix}
\bG_u \\
\bG_p
\end{bmatrix},
\end{equation}
where the off-diagonal blocks represent the coupling between the equilibrium and incompressibility equations. Within JAX-FEM, the coupled element Jacobian is obtained automatically through differentiation of the residual vector using automatic differentiation.
\color{black}

\subsection{Inverse problem: PDE-constrained identification of hyperelastic material parameters}

Parameter identification in continuum mechanics can be formulated as a PDE-constrained optimization problem, the standard mathematical framework for inverse problems governed by partial differential equations \cite{Lions_1971,Vogel_2002,Aster_2018,Hinze2008_1}. Its extensive theoretical and computational development includes well-posedness, regularization, optimality conditions, and efficient large-scale algorithms \cite{Vogel_2002,Hinze2008_1,Herzog2010}. Adjoint-based methods are particularly important because they provide gradients at a cost that is essentially independent of the number of optimization variables.
PDE-constrained optimization is commonly formulated either in full space, where state and control variables are optimized simultaneously subject to the PDE constraints, or in reduced space, where the state is treated as an implicit function of the control variables through the forward solution \cite{Hinze2008_1,Vogel_2002,Herzog2010}. Despite requiring repeated forward solves, the reduced-space formulation is well suited to large-scale parameter identification because adjoint methods enable efficient gradient evaluation.
A further distinction concerns the order of discretization and optimization \cite{Hinze2008_2}. The discretize-then-optimize approach first discretizes the governing PDE and subsequently derives the finite-dimensional optimization problem and its optimality conditions. In contrast, optimize-then-discretize derives the continuous optimality system before discretization. While the two approaches are often equivalent for consistent discretizations, differences can arise from discretization errors, particularly for nonlinear problems. We adopt discretize-then-optimize because it integrates naturally with the finite element formulation developed above and enables efficient implementation within JAX-FEM \cite{XueEtAl2023}.

Let $\btheta \in \mathbb{R}^{M}$ denote the vector of unknown material parameters of the hyperelastic model, for example the parameters of a phenomenological strain-energy density function or the trainable weights of a constitutive neural network. For a given parameter vector $\btheta$, the nonlinear finite element problem introduced above yields the discrete equilibrium solution $\bU \in \mathbb{R}^{N}$ through the constraint
\begin{equation}
\bG(\bU,\btheta) = \mathbf{0},
\end{equation}
where $\bG : \mathbb{R}^{N} \times \mathbb{R}^{M} \to \mathbb{R}^{N}$ denotes the assembled residual vector of the discretized weak form after imposing Dirichlet boundary conditions. In agreement with JAX-FEM, we adopt the discretize-then-optimize viewpoint and write the inverse problem in the form \cite{XueEtAl2023}
\begin{equation}
\min_{\bU \in \mathbb{R}^{N},\,\btheta \in \mathbb{R}^{M}} \mathcal{L}(\bU,\btheta)
\qquad
\text{s.t. } \bG(\bU,\btheta)=\mathbf{0}.
\end{equation}

For constitutive identification from experiments, the objective functional $\mathcal{L}$ measures the mismatch between observed and simulated structural responses. A general choice consistent with full-field identification, but also suited for sparse data scenarios, is
\begin{equation}
\mathcal{L}(\bU,\btheta)
=
\frac{\omega_{\kappa}}{2}\norm{ \bkFE(\bU,\btheta)  - \bkobs}^{2}
+
\frac{\omega_F}{2}\norm{ \bFFE(\bU,\btheta) - \boldsymbol{F}^{\mathrm{obs}}}^{2}
\label{eq:generalized_Loss}
\end{equation}
where $\bkobs$ denotes the observed kinematic information, $\boldsymbol{F}^{\mathrm{obs}}$ the measured reaction forces, $\bkFE$ the simulated kinematic information, $\bFFE$ the simulated reaction forces, $\omega_{\kappa}$ and $\omega_F$ are weighting factors. Depending on the available data, the objective may be based on kinematic fields only, reaction forces only, or a combination of both. For most applications displacement fields are the primary kinematic information, although there might be cases where only limited kinematic data is at hand, such as localized measurements with fiber optic sensors or strain gauges. In all cases, the optimization is constrained by the discrete nonlinear equilibrium equations.

\begin{figure}[htbp]
        \centering
        \hfill
          \centering
            \def\svgwidth{\linewidth}
            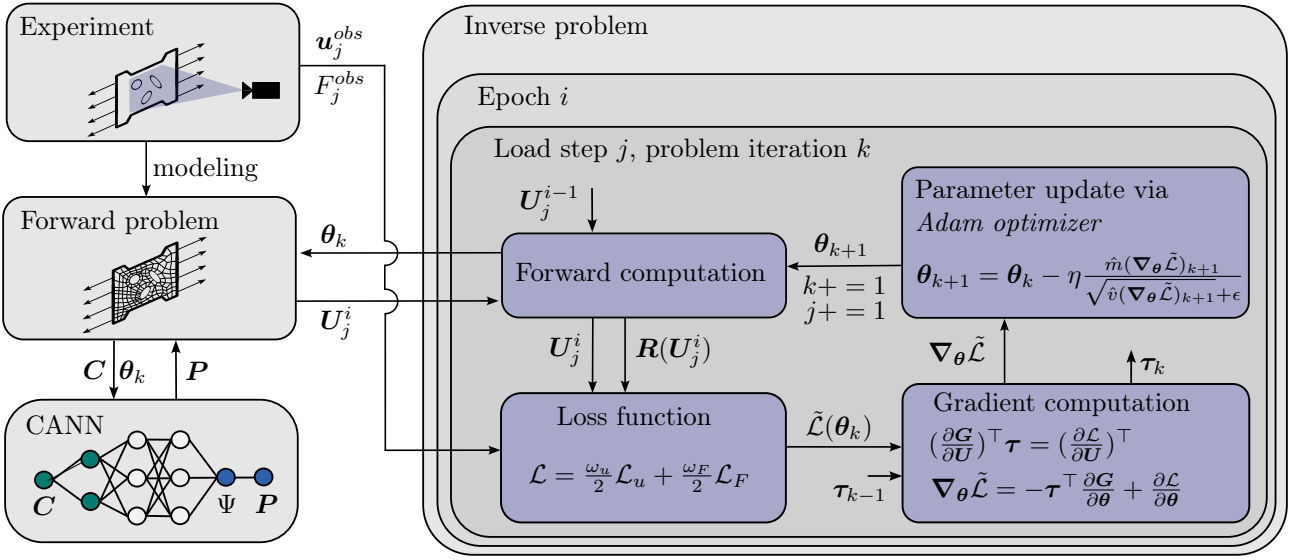
        \caption{\textbf{Schematic of the proposed end-to-end ADE-FEMU training loop.} Full-field displacement measurements $\bu_j^{\mathrm{ref}}$ obtained, e.g., from digital image correlation, together with the global reaction-force measurements $F_j^{\mathrm{obs}}$, define the experimental observables. For each candidate parameter vector $\btheta$ of the constitutive neural network, the JAX-FEM solver assembles the nonlinear residual, performs a Newton solve with automatically differentiated tangent stiffness, and returns the simulated displacement field $\bU$ and reaction forces $\boldsymbol{R}$. Observation operators project the simulated quantities onto the measurement modality and feed them into the data-mismatch loss $\mathcal{L}$ of Eq.~\eqref{eq:loss_total}. The gradient with respect to $\btheta$ is propagated automatically through the entire pipeline by reverse-mode automatic differentiation, and a first-order optimizer updates the parameters. The cycle is iterated until convergence, after which the trained constitutive network can be redeployed in standard finite element analyses.}
        \label{fig:FEMAD_workflow}
\end{figure}

As in JAX-FEM, a reduced-space formulation \cite{Vogel_2002,Herzog2010}  is obtained by eliminating the state variable $\bU$ through the implicit solution map $\bU=\bU(\btheta)$ defined by the forward problem \cite{XueEtAl2023}. The inverse problem can then be written as
\begin{equation}
\min_{\btheta \in \mathbb{R}^{M}} \tilde{\mathcal{L}}(\btheta),
\qquad
\tilde{\mathcal{L}}(\btheta) := \mathcal{L}(\bU(\btheta),\btheta).
\end{equation}
Efficient solution of this optimization problem requires the gradient of the reduced objective with respect to the material parameters. By the chain rule,
\begin{equation}
\boldsymbol{\nabla}_{\btheta} \tilde{\mathcal{L}}
=
\frac{\mathrm{d}\tilde{\mathcal{L}}}{\mathrm{d}\btheta}
=
\frac{\partial \mathcal{L}}{\partial \bU}\frac{\mathrm{d}\bU}{\mathrm{d}\btheta}
+
\frac{\partial \mathcal{L}}{\partial \btheta}.
\end{equation}
Differentiating the equilibrium constraint $\bG(\bU,\btheta)=\mathbf{0}$ with respect to $\btheta$ gives
\begin{equation}
\frac{\partial \bG}{\partial \bU}\frac{\mathrm{d}\bU}{\mathrm{d}\btheta}
+
\frac{\partial \bG}{\partial \btheta}
=
\mathbf{0},
\end{equation}
and therefore
\begin{equation}
\frac{\mathrm{d}\bU}{\mathrm{d}\btheta}
=
-
\left(\frac{\partial \bG}{\partial \bU}\right)^{-1}
\frac{\partial \bG}{\partial \btheta}.
\end{equation}
Substitution yields
\begin{equation}
\boldsymbol{\nabla}_{\btheta} \tilde{\mathcal{L}}
=
-
\frac{\partial \mathcal{L}}{\partial \bU}
\left(\frac{\partial \bG}{\partial \bU}\right)^{-1}
\frac{\partial \bG}{\partial \btheta}
+
\frac{\partial \mathcal{L}}{\partial \btheta},
\end{equation}
which is the reduced gradient formula underlying PDE-constrained optimization in JAX-FEM. Since the number of material parameters is typically much larger than the dimension of the scalar objective, JAX-FEM advocates the adjoint method for efficiency \cite{XueEtAl2023}. Introducing the adjoint vector $\btau\in \mathbb{R}^{N}$ as the solution of
\begin{equation}
\left(\frac{\partial \bG}{\partial \bU}\right)^{\top}\btau
=
\left(\frac{\partial \mathcal{L}}{\partial \bU}\right)^{\top},
\end{equation}
the reduced gradient becomes
\begin{equation}
\boldsymbol{\nabla}_{\btheta} \tilde{\mathcal{L}}
=
-
\btau^{\top}\frac{\partial \bG}{\partial \btheta}
+
\frac{\partial \mathcal{L}}{\partial \btheta}.
\end{equation}
The main computational expense arises from the nonlinear forward solve for $\bU$, while the adjoint problem remains linear once the forward state is determined \cite{farrell2013automated}. JAX-FEM evaluates derivatives in the adjoint formulation through automatic differentiation. The Jacobian $\partial \bG / \partial \bU$ is derived from the forward residual, while vector–Jacobian products such as $\btau^{\top} \partial \bG / \partial \btheta$ are computed using reverse-mode differentiation (for example, \texttt{jax.vjp}) \cite{BradburyEtAl2018,XueEtAl2023}. This approach is particularly advantageous for hyperelastic constitutive identification, where the residual's dependence on material parameters is often highly nonlinear and may originate from either closed-form or neural constitutive models. Each optimization iteration consists of: (1) solving the nonlinear forward problem for the current $\btheta$; (2) evaluating the objective and solving the adjoint problem; and (3) computing the reduced gradient to update $\btheta$ using a gradient-based optimizer, as depicted in Figure \ref{fig:FEMAD_workflow}. This procedure aligns with the general PDE-constrained optimization algorithm implemented in JAX-FEM \cite{XueEtAl2023}. In this context, the unknown parameter vector $\btheta$ represents a hyperelastic material law. The inverse problem aims to identify material parameters that minimize the discrepancy between measured and simulated full-field responses, while ensuring mechanical admissibility through nonlinear finite-element equilibrium. This framework facilitates simulation-based constitutive identification using heterogeneous deformation data.

For incompressible hyperelasticity, the state vector $\bU$ in $\bG(\bU,\btheta)=\mathbf{0}$ comprises both displacement and pressure, as defined by the mixed formulation. In this setting, $\bG$ denotes the residual of equilibrium and incompressibility ($J-1=0$), with pressure serving as a Lagrange multiplier. Consequently, the sensitivity system $\partial \bG/\partial \bU$ constitutes the full block Jacobian, and the inverse problem is subject to both equilibrium and incompressibility constraints \cite{sussman1987finite}.

\subsection{Numerical implementation of the full-field model training}\label{sec:framework:impl}

The PDE-constrained inverse problem introduced in Section~\ref{sec:framework} is solved entirely within a single differentiable programming environment. The forward nonlinear finite element solver, the constitutive neural network, and the optimization loop for the material parameters are all implemented in JAX \cite{BradburyEtAl2018,XueEtAl2023}. This integrated setting is essential, because it allows the gradients required for parameter updates to be propagated automatically through the nonlinear forward solve and the constitutive ansatz, removing the need for manually derived sensitivities and ad hoc coupling between independent simulation and machine learning packages \cite{XueEtAl2023,Tan2026}.

\paragraph{Discretization and forward solve} The reference configuration $\Omega \subset \mathbb{R}^3$ is discretized using standard isoparametric finite elements. For each load step, the discrete nonlinear residual $\bG(\bU,\btheta) = \mathbf{0}$ is solved by Newton's method, where the global tangent stiffness $\bK = \partial \bG / \partial \bU$ is assembled element-wise and is obtained automatically through forward-mode differentiation of the element residual with respect to the local degrees of freedom \cite{XueEtAl2023}.

\paragraph{Constitutive artificial neural networks} The two CANNs are implemented in JAX using the Flax Linen API, which enables the flexible realization of a broad range of neural network architectures. The grey-box CANN is implemented according to the formulation presented in Section \ref{ss:grey-box}, using an architecture with two hidden layers comprising 12 neurons each. Softplus activation functions are employed in all hidden layers. The white-box CANN is implemented as described in Section \ref{ss:white-box}. 

\paragraph{Loss function and observation operators} For each training specimen $s = 1, \ldots, N_{\mathrm{exp}}$ and each load step $j = 1, \ldots, N_j^{(s)}$, we evaluate the discrete objective function introduced in Section~\ref{sec:framework},
\begin{equation}\label{eq:loss_total}
\mathcal{L}(\btheta)
=
\frac{\omega_{\kappa}}{2}\,\left\| \bU^{(s,j)}(\btheta) - \bQu\, \boldsymbol{u}^{\mathrm{obs},(s,j)} \right\|^{2}
+
\frac{\omega_F}{2}\,\left\|\boldsymbol{Q}_F\, \boldsymbol{R}^{\mathrm{ext},(s,j)}(\btheta) - F^{\mathrm{obs},(s,j)} \right\|^{2}
+
\mathcal{R}(\btheta).
\end{equation}
In contrast to the more generalized form of the loss function  \eqref{eq:generalized_Loss}, the implemented loss function \eqref{eq:loss_total} is specifically suited to incorporate data obtained by measurements with digital image correlation (DIC). Since only displacement fields are considered here, we replaced the arbitrary kinematic variable $\boldsymbol{\kappa}$ with a displacement variable $\boldsymbol{u}$.
The observation operator $\bQu$ maps the experimentally observable quantity to the finite element space, that is, displacements at material points covered by the DIC measurement. $\boldsymbol{Q}_F$ is an operator that maps the calculated reaction force to the measured force quantities. Their use is essential because experimental data are typically available on a subset of the boundary or on the visible specimen surface only, and the framework must remain valid even in such partial-observability scenarios \cite{Regazzoni2026}. Both contributions in \eqref{eq:loss_total} can be normalized by the empirical magnitude of the corresponding measurements, $\|\buobs\|_{2}^{2}$ and $\|F^{\mathrm{obs}}\|_{2}^{2}$, before applying the user-defined weights $\omega_{\kappa}$ and $\omega_F$. This normalization ensures that the displacement and reaction-force terms enter the objective at a comparable scale, as recently advocated for multi-modal full-field identification \cite{Tan2026}.

\paragraph{Regularization} The regularization term $\mathcal{R}(\btheta)$ in \eqref{eq:loss_total} serves two purposes. First, for the white-box CANN of Section~\ref{sec:theory}, an $L_1$ penalty on the outer-layer weights $w_{2,k}$ promotes sparsity and thereby drives the network toward an interpretable model with only a few active constitutive terms, in the spirit of automated model discovery \cite{LinkaKuhl2023,LinkaKuhl2024,Tan2026}. Second, for the grey-box CANN, a small $L_2$ penalty on the trainable weights stabilizes the training when measurements are limited or noisy. Both contributions are scaled by problem-dependent factors $\lambda_1$ and $\lambda_2$ that are calibrated once and then kept fixed across the different experimental settings.

\paragraph{Gradient computation} The gradient $\boldsymbol{\nabla}_{\btheta} \tilde{\mathcal{L}}$ of the reduced objective is computed by reverse-mode automatic differentiation through the entire forward pipeline. In agreement with the discretize-then-optimize formulation derived in Section~\ref{sec:framework}, the implicit dependence $\bU = \bU(\btheta)$ is differentiated by JAX-FEM via an implicit-function rule that internally solves the linear adjoint system in \eqref{eq:loss_total}-related notation, see \cite{XueEtAl2023}. From a user perspective, this is invisible: a single call to \texttt{jax.grad} returns the gradient of the loss with respect to all trainable network parameters, including those entering the strain-energy density in a strongly nonlinear way. This avoids the manual derivation of tangent and adjoint operators, which has historically constituted a major implementation bottleneck for FEMU and full-field identification \cite{VanKeulenEtAl2005,ReesEtAl2010,Tan2026}.

\paragraph{Optimization strategy} In the related neural constitutive identification literature, the loss is frequently minimized in two stages: a first stage based on an adaptive first-order method such as Adam, which is robust against poor initial conditions and helps escape regions of vanishing gradient, followed by a second stage based on a quasi-Newton optimizer such as L-BFGS that exploits curvature information close to the optimum \cite{Regazzoni2026}. In the present work, this two-stage scheme is conceptually applicable, but we found that the first-order Adam optimizer alone, combined with the cyclic learning-rate schedule described below, was sufficient to reach a satisfactory plateau of the loss for all three cases considered in Section~\ref{sec:results}. We therefore use Adam throughout, both for its simplicity and for the resulting parameter-update behavior, which is well suited to the noisy gradients arising from realistic full-field measurements. Convergence is monitored via the relative change of the loss between successive epochs, and training is terminated once this change drops below a prescribed threshold or once a maximum number of epochs is reached.

\paragraph{Learning-rate schedule} To improve robustness against shallow local minima that are characteristic of full-field identification with high-capacity constitutive networks, we employ a cyclic learning-rate schedule in which the learning rate oscillates between a base value $\eta_{\min}$ and a peak value $\eta_{\max}$ over a fixed step size of typically a few tens of epochs. The cyclic behavior periodically increases the gradient norm and helps the optimizer leave flat regions of the loss landscape, while still allowing fine-grained refinement during the descending half-cycles.

\paragraph{Initialization and well-posedness during training} A central practical challenge of embedding a nonlinear FE solver into the training loop is that every intermediate candidate constitutive model must yield a solvable equilibrium problem; failure of the forward solve at any iteration would otherwise terminate the training prematurely \cite{Regazzoni2026}. To safeguard this, the network weights are initialized such that the initial strain-energy density is close to a neo-Hookean reference response and satisfies the normalization, polyconvexity, and stress-free conditions of Section~\ref{sec:theory} by construction. This combination of physics-augmented architecture and careful initialization ensures that the forward problem remains well-posed throughout the optimization \cite{Regazzoni2026,LindenEtAl2023}. In addition, for each forward solve, the displacement solution $\boldsymbol{U}^{(i-1,j)}$ corresponding to the same load step $j$ from the previous epoch $i-1$ is provided to the Newton solver as the initial guess, substantially improving both computational efficiency and numerical stability.

\paragraph{Solver and hardware} All computations are carried out within a single Python environment built on JAX \cite{BradburyEtAl2018} and JAX-FEM \cite{XueEtAl2023}. The entire pipeline encompassing FE assembly, Newton solves, neural network evaluation, loss computation, and gradient propagation---is just-in-time compiled with \texttt{jax.jit} and executed on a single GPU. This integrated execution model removes the data round-trips between simulator and machine learning framework that typically dominate the runtime of two-stage workflows \cite{XueEtAl2023}, and is one of the key enablers of the proposed end-to-end training procedure. The complete training procedure is summarized in Algorithm~\ref{algo:training}.

In summary, the proposed implementation realizes a fully differentiable training loop in which (i) the forward nonlinear FE problem is solved by a Newton scheme with automatically differentiated tangent stiffness, (ii) the loss is evaluated through observation operators that mirror the actual experimental measurement modality, and (iii) the gradient with respect to all trainable constitutive parameters is obtained automatically through the entire computational pipeline. This setting unifies the generality of simulation-based identification with the efficiency of differentiable programming and provides the algorithmic basis for the numerical results presented in Section~\ref{sec:results}.

\begin{algorithm}[H]
    \setstretch{1.2}
    \caption{\textbf{End-to-end training of the constitutive neural network from full-field data.}}
    \label{algo:training}
    \DontPrintSemicolon
    \KwIn{Training data $\{\boldsymbol{u}^{\mathrm{obs},(s,j)}, F^{\mathrm{obs},(s,j)}\}_{s,j}$, FE meshes and BCs $\{\Omega^{(s)}, \Gamma_D^{(s)}, \Gamma_N^{(s)}\}$, initial parameters $\btheta^{(0)}$, weights $\omega_{\kappa}, \omega_F$, regularization term $\mathcal{R}(\btheta)$.}
    \KwOut{Trained constitutive network parameters $\btheta^{*}$.}
    \BlankLine
    $k \leftarrow 0$\;
    \For{epoch $i = 0, 1, 2, \ldots$ \textbf{until convergence}}{
        \For{specimen $s = 1, \ldots, N_{\mathrm{exp}}$}{
            \For{load step $j = 1, \ldots, N_j^{(s)}$}{
                 $\mathcal{L}^{(k)} \leftarrow 0$\;
                Solve nonlinear FE residual $\bG(\bU^{(s,j)},\btheta^{(k)})=\mathbf{0}$ via Newton's method, with $\bK = \partial \bG/\partial \bU$ obtained by AD\;
                Evaluate kinematic loss term: $\mathcal{L}_\kappa =\| \bU^{(s,j)}(\btheta) - \bQu\, \boldsymbol{u}^{\mathrm{obs},(s,j)} \|_{2}^{2}$\;
                Evaluate reaction force loss term: $\mathcal{L}_F =\norm{\boldsymbol{Q}_F \bR^{\mathrm{ext},(s,j)}(\bU,\btheta) - F^{\mathrm{obs},(s,j)}}_{2}^{2}$\;
                Add regularization $\mathcal{R}(\btheta^{(k)})$ to $\mathcal{L}^{(k)}$\;
                Compute $\boldsymbol{\nabla}_{\btheta} \mathcal{L}^{(k)}$ via reverse-mode AD through the entire forward pipeline\;
                Update parameters: $\btheta^{(k+1)} \leftarrow \mathrm{Optimizer}(\btheta^{(k)}, \boldsymbol{\nabla}_{\btheta}\mathcal{L}^{(k)})$ \Comment*[r]{Adam optimizer}
                $k \gets k + 1$
            }
        }

    }
    \Return $\boldsymbol{\theta}^{*} = \boldsymbol{\theta}^{(k)}$\;
\end{algorithm}

\paragraph{Offline and online stages} In practice, the training procedure separates naturally into an offline preparation stage and an online optimization stage. The offline stage builds the FE counterpart of the experiment: the specimen geometry is meshed, the Dirichlet boundary conditions are reconstructed from the DIC by fitting smooth curves to the visible boundary, the displacements are projected from the DIC discretization onto the FE discretization, and a subset of load steps along the force--displacement curve is selected to expose the material nonlinearity. The loss function in Eq.~\eqref{eq:loss_total} is constructed from these components, and an initial forward solve over all load steps yields displacement solutions for the initial parameter set $\btheta^{(0)}$. These precomputed solutions are subsequently used as initial guesses for the Newton solver for the initial problem iteration $k = 0$. In the online stage, the inverse problem is solved by iterating the loop of Algorithm~\ref{algo:training}: for each load step $j$ in each epoch $i$, the forward problem is solved (using the previous-epoch solution as initial guess, which improves stability and reduces the number of Newton iterations), the loss is evaluated, the reverse-mode AD pass returns the gradient with respect to $\btheta^{(k)}$, and a single Adam step produces $\btheta^{(k+1)}$. The adjoint linear system that arises internally during the gradient computation is also seeded with the solution of the previous-iteration adjoint problem $\btau^{(k-1)}$ as initial guess to accelerate convergence.

    
\section{Results}\label{sec:results}

We illustrate the capabilities of the proposed ADE-FEMU framework on three representative cases, all of which are based on openly available experimental data. \emph{Case \hyperref[sec:results:case1]{1}} considers a homogeneous, perforated polymer membrane subjected to tension and combines full-field digital image correlation (DIC) data of two complex tensile specimens with two homogeneous reference tests (uniaxial tension and pure shear), from the data set provided in \cite{Abbasi2026}. \emph{Case \hyperref[sec:results:case2]{2}} addresses a more challenging setting in which a second polymer is tested on a five-hole specimen, but the only available kinematic information consists of the maximum principal stretch measured along a single one-dimensional path through the specimen, together with the resulting load--displacement curve. The data for \emph{Case \hyperref[sec:results:case2]{2}} is available in \cite{meunier2008mechanical}. \emph{Case \hyperref[sec:results:case3]{3}} finally turns to heterogeneous material systems and identifies the constitutive parameters of two distinct phases simultaneously from a dataset of twenty-three thin plate specimens with varying matrix--inclusion topology, provided in \cite{MechanicalMNISTChallenge}. For each case, both the grey-box and the white-box CANN of Section~\ref{sec:theory} are trained, and we systematically compare the influence of the training data, the architecture, and the experimental observability on the identification result. All cases considered here are simulated in three dimensions; however, the small out-of-plane thickness and predominantly planar loading conditions render their mechanical response effectively two-dimensional.

\subsection{Case 1: Full-field displacement data}\label{sec:results:case1}

\begin{figure}[htbp]
        \centering
        \hfill
          \centering
            \def\svgwidth{\linewidth}
            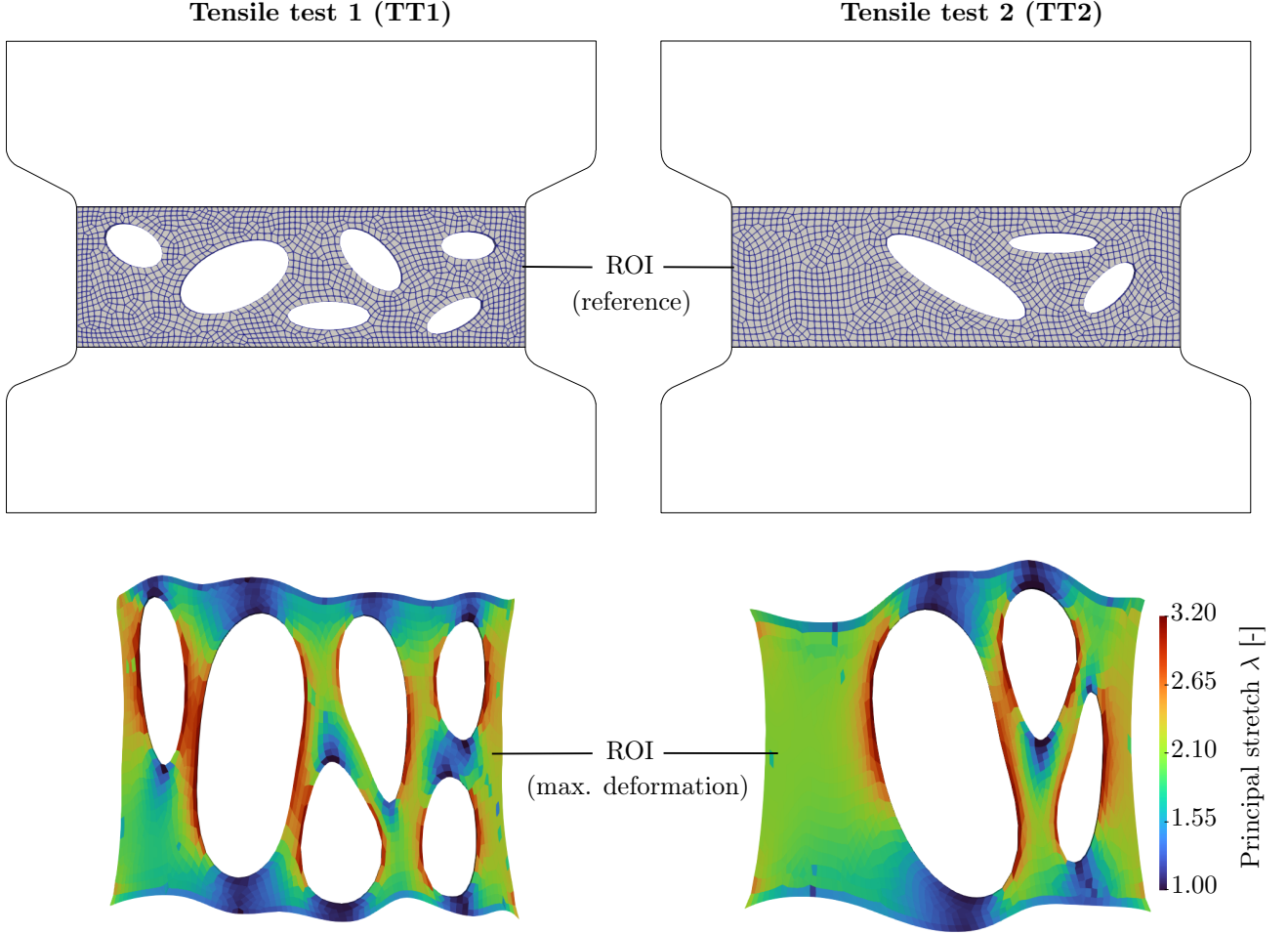
        \caption{\textbf{Case \hyperref[sec:results:case1]{1}, perforated tensile specimens TT1 and TT2.} Undeformed and deformed wide dogbone-shaped specimens with the region of interest (ROI) used for the FE model highlighted. Only the ROI is meshed with finite elements. The deformed ROIs are color-coded by the maximum principal stretch $\lambda_1$ obtained from the FE simulation. Pronounced stretch concentrations are visible at the hole boundaries, in particular at the narrow ligaments between adjacent holes, with peak values of approximately $\lambda_1 \approx 3.2$. Data are obtained from \cite{Abbasi2026}.}
        \label{fig:LS_Overview}
\end{figure}

\paragraph{Data and experimental setting} The first case is based on tensile experiments on a perforated polymer membrane reported in the open-access dataset of \cite{Abbasi2026}. The data set consists of two homogeneous reference tests---uniaxial tension (UT) and pure shear (PS)---and two perforated tensile specimens (TT1 and TT2) of identical material but with different hole configurations. For each experiment, both the global force--displacement curve and the full DIC displacement field are available. We use the homogeneous tests UT and PS only as one-dimensional stress--stretch data, with assumed deformation gradients $\boldsymbol{F}_{\mathrm{UT}} = \mathrm{diag(\lambda,1/\sqrt\lambda,1/\sqrt\lambda)}$ and $\boldsymbol{F}_{\mathrm{PS}} = \mathrm{diag(\lambda,1,1/\lambda)}$   .TT1 and TT2 are used as full-field, heterogeneous benchmarks. Figure~\ref{fig:LS_Overview} shows the two perforated specimens together with the meshed region of interest (ROI) and the maximum principal stretch at peak loading. The ROI of TT1 is discretized using $1{,}518$ elements and $3{,}402$ nodes, resulting in $10{,}206$ displacement DOFs and $1{,}518$ pressure DOFs, for a total of $11{,}724$ DOFs. To solve the optimization problem, the 2D DIC displacement field is projected onto $3{,}174$ nodes, yielding $6{,}348$ displacement DOFs over which the optimization is performed. TT2 is discretized using $1{,}343$ elements and $2{,}952$ nodes, resulting in $8{,}856$ displacement DOFs and $1{,}343$ pressure DOFs, for a total of $10{,}199$ DOFs. The Dirichlet boundary of the ROI is reconstructed from the DIC by fitting B-spline curves to the upper and lower edges, which yields a smooth representation of the experimentally observed boundary displacement and is then prescribed in the forward FE solve. Stretch concentrations are clearly visible at the hole boundaries, in particular at the narrow ligaments between adjacent holes, with peak principal stretches of approximately~$\lambda_1^{max} \approx 3.2$.

\paragraph{Training and testing protocol} To probe the role of the training data, we train both the grey-box and the white-box CANN on three distinct training sets: (i)~UT only, (ii)~the combination UT~$\&$~PS, and (iii)~the heterogeneous tensile test TT1. In each case, the resulting models are evaluated on all four datasets (UT, PS, TT1, TT2). The configuration that trains on TT2 is deliberately omitted from the matrix and reserved as an entirely unseen benchmark, which is discussed separately in Section~\ref{sec:results:case1:unseen}. This protocol allows us to disentangle the influence of the constitutive architecture from that of the experimental loading conditions captured by the training data. For training on TT1, the loss function defined in Eq.~\eqref{eq:loss_total} is employed with $N_j = 11$ load steps and $N_s = 1$ structure. The loss scaling parameters are selected as $\omega_{\kappa} = 0.02$ and $\omega_F = 2$. No normalization or regularization techniques are applied. The networks are trained using the Adam optimizer with a cyclic learning-rate schedule oscillating between a base value of $0.0001$ and a peak value of $0.05$ over a step size of $15$ epochs.

\begin{figure}[htbp]
        \centering
        \includegraphics[width = \linewidth]{ 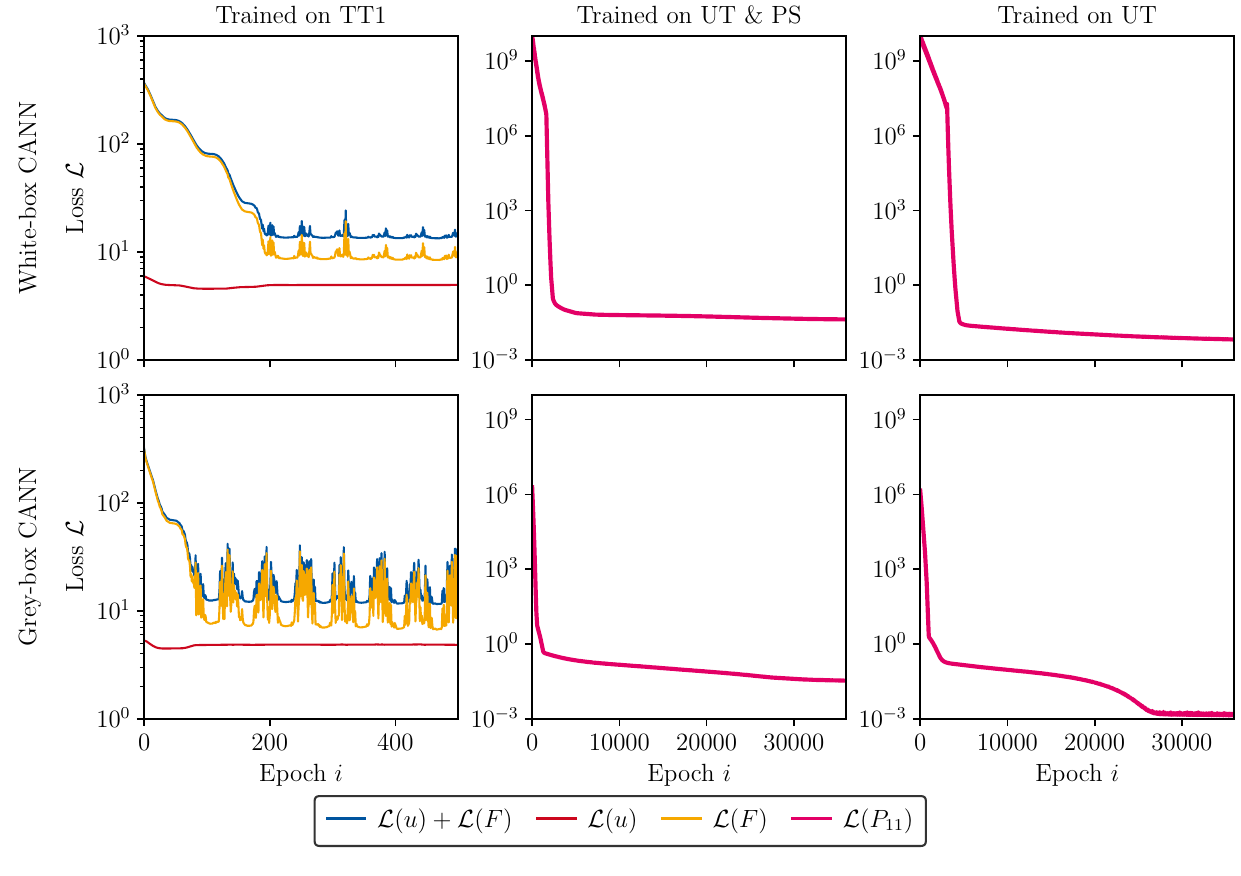}   
        \caption{\textbf{Case \hyperref[sec:results:case1]{1}, loss evolution during training.} Evolution of the kinematic loss $\mathcal{L}(u)$ (red) and the reaction-force loss $\mathcal{L}(F)$ (orange) as a function of the training epoch for the three training configurations (columns: TT1, UT~$\&$~PS, UT) and the two architectures (rows: grey-box CANN, white-box CANN). The total loss minimized at each parameter update is the sum of both contributions. The high-frequency oscillations reflect the cyclic learning-rate schedule of Section~\ref{sec:framework:impl}. Models trained on the heterogeneous TT1 data converge within about $200$ epochs, whereas models trained on homogeneous data require approximately $1\,000$ epochs to reach a comparable plateau.}
        \label{fig:LS_loss_evolutions}
\end{figure}

\paragraph{Loss evolution} Figure~\ref{fig:LS_loss_evolutions} reports the evolution of the kinematic loss $\mathcal{L}(u)$ and the reaction-force loss $\mathcal{L}(F)$ during training. The total loss minimized at each parameter update is the sum of both contributions, evaluated at every load step. Across all training sets, the loss decreases smoothly within the first few hundred epochs and reaches a stable plateau. Models trained on the heterogeneous TT1 data converge particularly fast---typically within about $200$ epochs---whereas models trained on homogeneous data require approximately $1000$ epochs to reach a comparable plateau, which is consistent with the smaller information content of the homogeneous tests. The grey-box CANN exhibits more pronounced fluctuations in the force-loss component than the white-box CANN, reflecting its higher capacity and therefore higher sensitivity to local parameter changes. The high-frequency oscillations visible in all curves originate from the cyclic learning-rate schedule introduced in Section~\ref{sec:framework:impl} and are not a sign of poor convergence. We note in passing that, for some of the configurations trained on heterogeneous data, the kinematic loss $\mathcal{L}(u)$ reaches a temporary minimum before slightly increasing again as the reaction-force loss continues to decrease---a hallmark of multi-modal training in which the two loss terms are not perfectly aligned.

\begin{figure}[htbp]
        \centering
        \includegraphics[width = \linewidth]{ 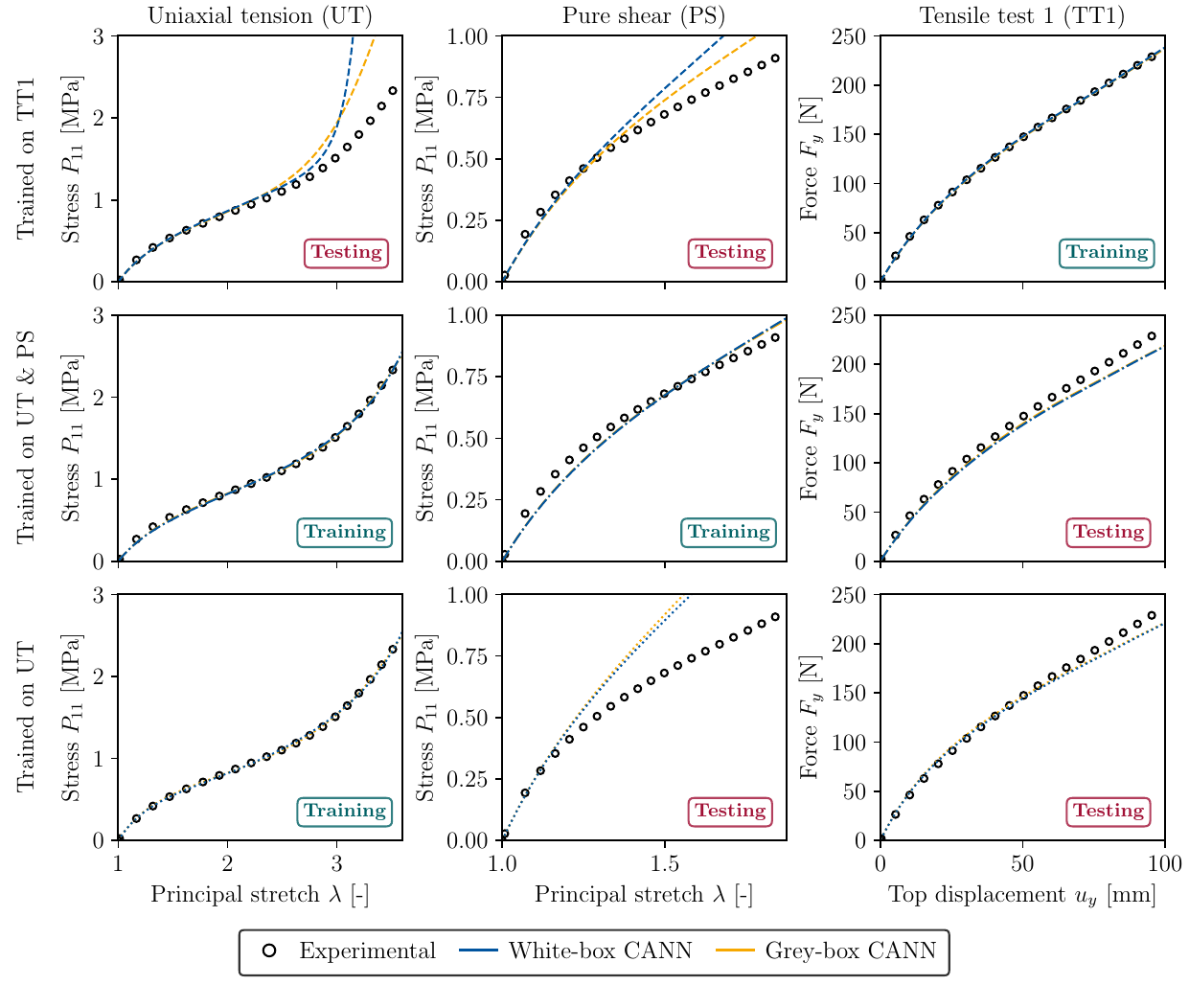}   
        \caption{\textbf{Case \hyperref[sec:results:case1]{1}, training--testing matrix of stress--stretch curves.} Stress--stretch responses of all six trained models (grey-box and white-box CANN, three training configurations) on the homogeneous tests UT and PS and on the reaction-force curve of the heterogeneous specimen TT1. Each row corresponds to a training configuration; the training data for each row is explicitly labeled. The grey-box and the white-box CANN trained on the same data deliver essentially indistinguishable curves, while the choice of training data has a much stronger influence than the constitutive architecture.}
        \label{fig:LS_training_testing_matrix}
\end{figure}

\paragraph{Training--testing matrix} Figure~\ref{fig:LS_training_testing_matrix} compares all six trained models across the four datasets. A first, robust observation is that the training data dominate the performance much more than the constitutive architecture: for each row of the matrix, the grey-box and the white-box CANN deliver almost indistinguishable predictions. Models trained on TT1 reproduce the heterogeneous reaction force essentially exactly and capture UT well up to a stretch of about~$2.5$, beyond which both architectures progressively deviate from the experiment, with the white-box CANN diverging more strongly. PS is reproduced reasonably up to $\lambda \approx 1.25$ and then deteriorates. The models trained on UT~$\&$~PS achieve the most balanced performance on the homogeneous tests but slightly underestimate the reaction force of TT1. Finally, models trained on UT alone provide the best fit on UT but show the weakest extrapolation to PS and a small but systematic underestimation of the heterogeneous reaction force, consistent with the limited variety of deformation modes contained in a single uniaxial test.

Overall, this matrix establishes two recurring themes that we will encounter again in the remaining cases. First, the training data dictate the predictive performance more than the architectural choice; this is a direct consequence of the strong physical inductive biases shared by both CANNs. Second, the configuration trained on UT~$\&$~PS produces the most versatile model because it covers the largest portion of the invariant plane, while still missing the extreme stretches accessible in the heterogeneous TT1 experiment.

\begin{figure}[htbp]
        \centering
        \includegraphics[width=\linewidth]{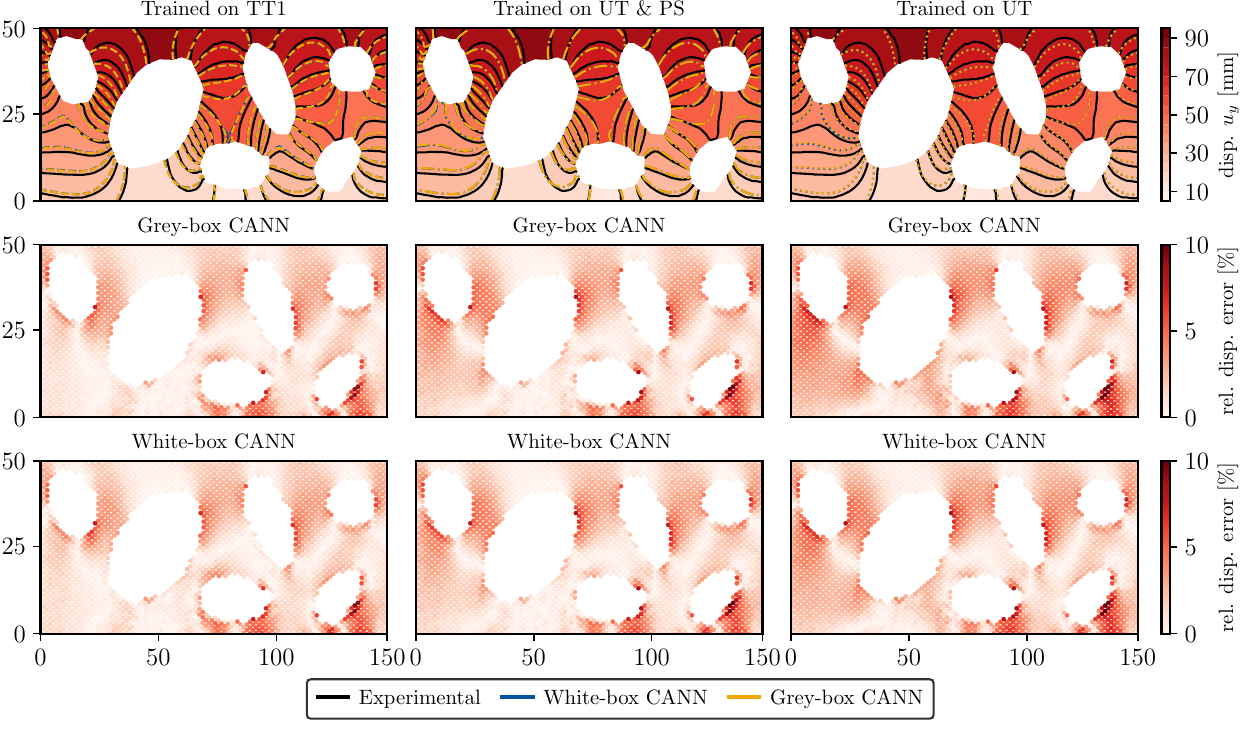}
        \caption{\textbf{Case \hyperref[sec:results:case1]{1}, displacement field on TT1.} For each training configuration (columns: UT, UT~$\&$~PS, TT1), the figure compares the DIC-measured vertical displacement field $u_y$ (top row, black contour lines on filled colormap) with the FE prediction (overlaid blue contours for the white-box CANN and yellow contours for the grey-box CANN), and reports the pointwise relative error of the grey-box prediction (middle row) and of the white-box prediction (bottom row) with respect to the DIC reference. The predicted contours track the DIC reference closely, and the differences between the two architectures within a column are essentially imperceptible at the contour level.}
        \label{fig:all_models_on_TT1_disp}
\end{figure}

\begin{figure}[htbp]
        \centering
        \includegraphics[width=\linewidth]{ 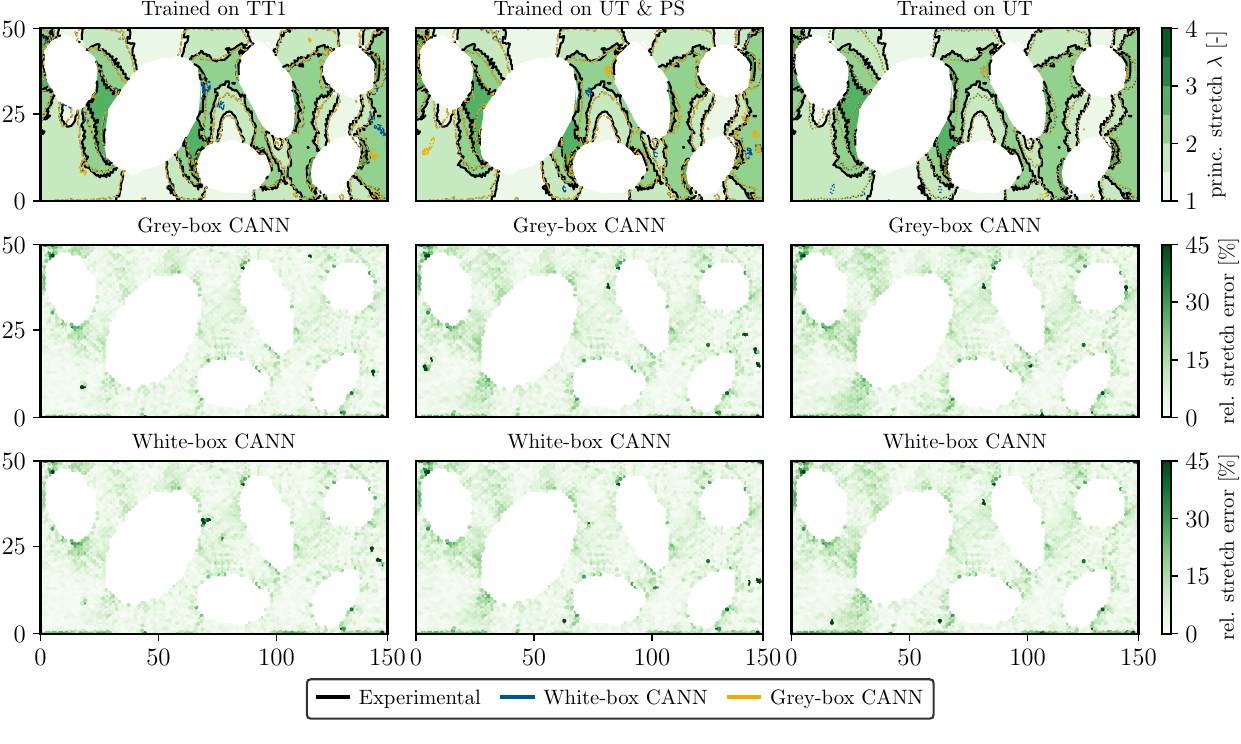}
        \caption{\textbf{Case \hyperref[sec:results:case1]{1}, maximum principal stretch field on TT1.} For each training configuration (columns: UT, UT~$\&$~PS, TT1), the figure compares the DIC-derived maximum principal stretch $\lambda_1$ (top row, black contour lines on filled colormap) with the FE prediction (overlaid blue contours for the white-box CANN and yellow contours for the grey-box CANN), and reports the pointwise relative error of the grey-box prediction (middle row) and of the white-box prediction (bottom row). Stretches are evaluated at the centroids of a triangulation of the DIC points and projected from the FE quadrature points via cubic interpolation. Errors are systematically larger than for the displacement field by roughly a factor of $1.5$--$2$, reflecting the higher derivative order of the stretch measure.}
        \label{fig:all_models_on_TT1_stretch}
\end{figure}

\paragraph{Full-field reproduction on TT1} Figures~\ref{fig:all_models_on_TT1_disp} and~\ref{fig:all_models_on_TT1_stretch} compare the measured and predicted displacement and stretch fields on TT1 at peak loading. The contour lines of the predicted vertical displacement (blue for the white-box CANN, yellow for the grey-box CANN) closely follow the DIC contours (black) for every training configuration, and the differences between the two architectures trained on the same data are essentially invisible at the contour level. The corresponding pointwise relative-error maps show smooth, low-amplitude error fields with localized peaks at the hole boundaries; these are the regions where the DIC measurement itself is typically the noisiest and where the geometric idealization of the FE model is most sensitive to small inaccuracies in the specimen drawings used to generate the mesh. The stretch fields in Figure~\ref{fig:all_models_on_TT1_stretch} show systematically higher relative errors than the displacement fields by roughly a factor of $1.5$--$2$, both because the stretch is a derived quantity that amplifies measurement noise and because the projection between the FE quadrature points and the DIC triangulation introduces additional interpolation error. In addition, the inter-architecture variability is slightly larger than in the displacement fields, and isolated high-error points appear at apparently random locations, which we attribute to local interpolation artifacts in the DIC-to-FE projection. The fact that the stretch field varies more steeply over the specimen and exhibits richer spatial features than the displacement field also suggests that the stretch could in principle be a more informative training target than the displacement, an observation we exploit explicitly in Case \hyperref[sec:results:case2]{2}.

\begin{figure}[htbp]
        \centering
        \includegraphics[width=1\linewidth]{ 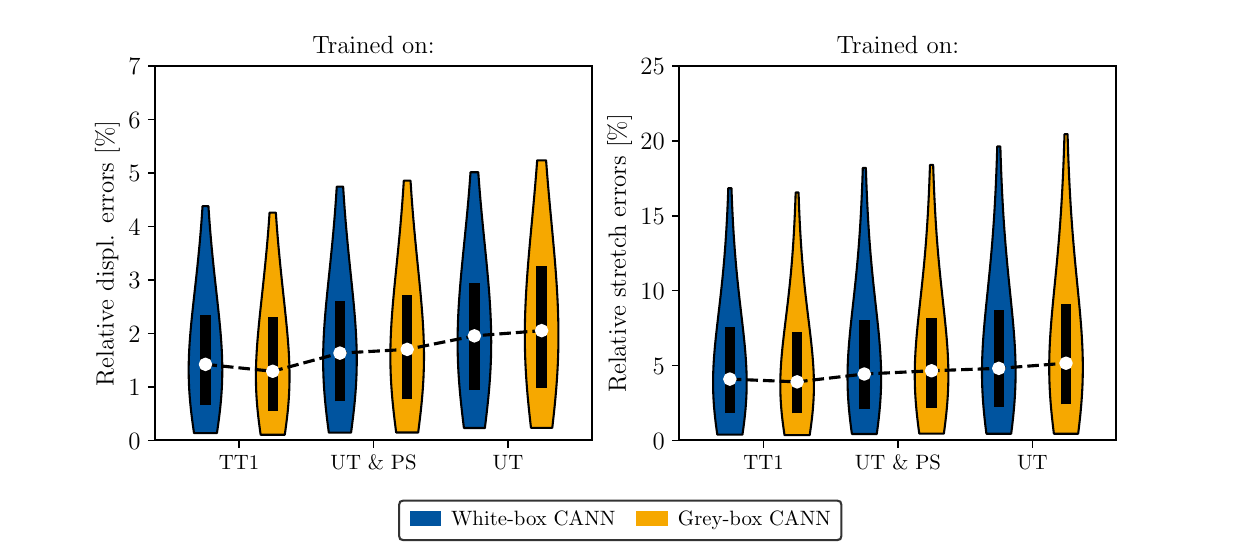}
        \caption{\textbf{Case \hyperref[sec:results:case1]{1}, distributions of the pointwise relative error on TT1.} Empirical distributions of the relative error of the predicted displacement field (left) and of the maximum principal stretch (right) on the training specimen TT1, derived from Figures~\ref{fig:all_models_on_TT1_disp} and~\ref{fig:all_models_on_TT1_stretch}. Each violin shows the central $90\,\%$ of the data (clipped between the $5\,\%$ and $95\,\%$ percentile); the black bar indicates the central $50\,\%$ (clipped between the $25\,\%$ and $75\,\%$ percentile); the white dot marks the median. The grey-box CANN trained on TT1 achieves the best overall fit, while the same grey-box architecture trained on UT alone is the worst, in line with the data-hungry character of the higher-capacity model.}
        \label{fig:TT1_violins}
\end{figure}

\paragraph{Error distributions on TT1} The violin plots in Figure~\ref{fig:TT1_violins} summarize the pointwise relative errors on TT1 as empirical distributions, restricted to the central~$90\,\%$ of the data. The displacement-error distributions are narrower than the stretch-error distributions, but their relative ordering across training configurations is identical: models trained on TT1 deliver the smallest median error, followed by those trained on UT~$\&$~PS, and finally by those trained on UT alone. The grey-box CANN trained on TT1 achieves the overall best fit, in line with the ``data-hungry'' character of higher-capacity networks. Conversely, the same grey-box architecture trained on UT alone is the worst performer in this comparison, illustrating that the additional flexibility of the grey-box model is beneficial only when the training data are sufficiently informative.

\subsubsection{Generalization to a completely unseen boundary-value problem}\label{sec:results:case1:unseen}

Because all configurations of the training--testing matrix use TT1 either as training or as evaluation data, the previous comparison only assesses interpolation within a single heterogeneous experiment. To probe genuine generalization, we now apply all six trained models to the second perforated specimen TT2, which differs from TT1 in its hole topology and was held out from training in all configurations.

\begin{figure}[htbp]
        \centering
        \includegraphics[width=\linewidth]{ 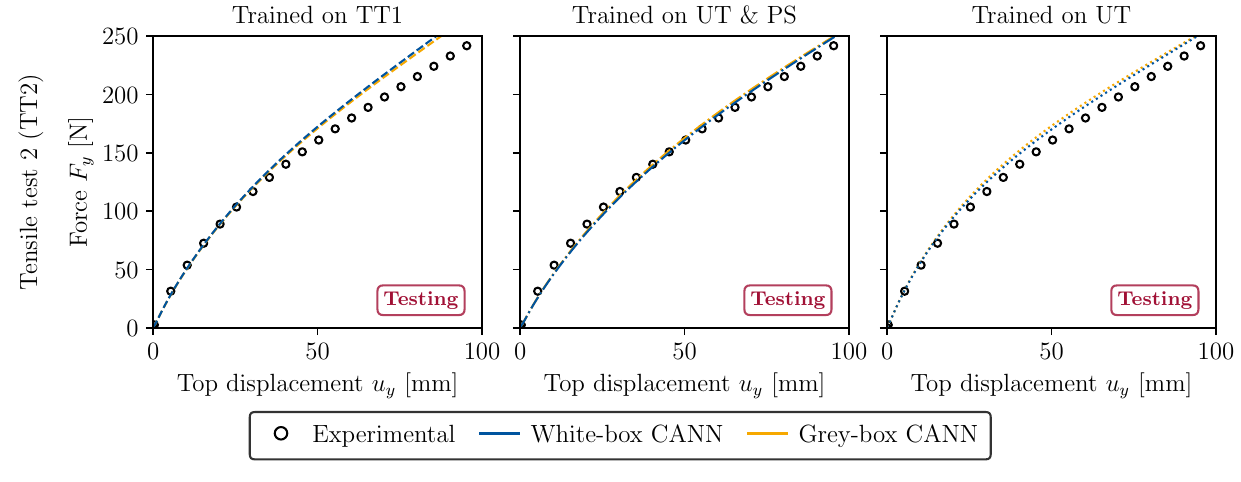}
        \caption{\textbf{Case \hyperref[sec:results:case1]{1}, reaction force on the unseen specimen TT2.} Comparison of the measured force--displacement curve of TT2 (black) with the responses simulated by all six trained models (grey-box and white-box CANN, three training configurations). The models trained on TT1 overestimate the reaction force by approximately $9\,\%$, whereas the models trained on UT~$\&$~PS and on UT alone match the measurement to within a few percent. TT2 is held out from all training configurations and therefore probes genuine generalization.}
        \label{fig:LS_all_models_TT2}
\end{figure}

\paragraph{Reaction force on unseen TT2} The reaction-force responses on TT2 are summarized in Figure~\ref{fig:LS_all_models_TT2}. Somewhat counter-intuitively, the models trained on TT1 \emph{overestimate} the reaction force of TT2 by approximately~$9\,\%$, although TT1 was reproduced essentially exactly during training. The models trained on UT~$\&$~PS reproduce TT2 to within a few percent, and the models trained on UT alone perform almost equally well. This indicates that the constitutive laws identified from a single heterogeneous experiment are not necessarily more transferable than those obtained from a well-chosen combination of homogeneous tests, in spite of the much richer kinematic content of the heterogeneous specimen.

\begin{figure}[htbp]
        \centering
        \includegraphics[width=\linewidth]{ 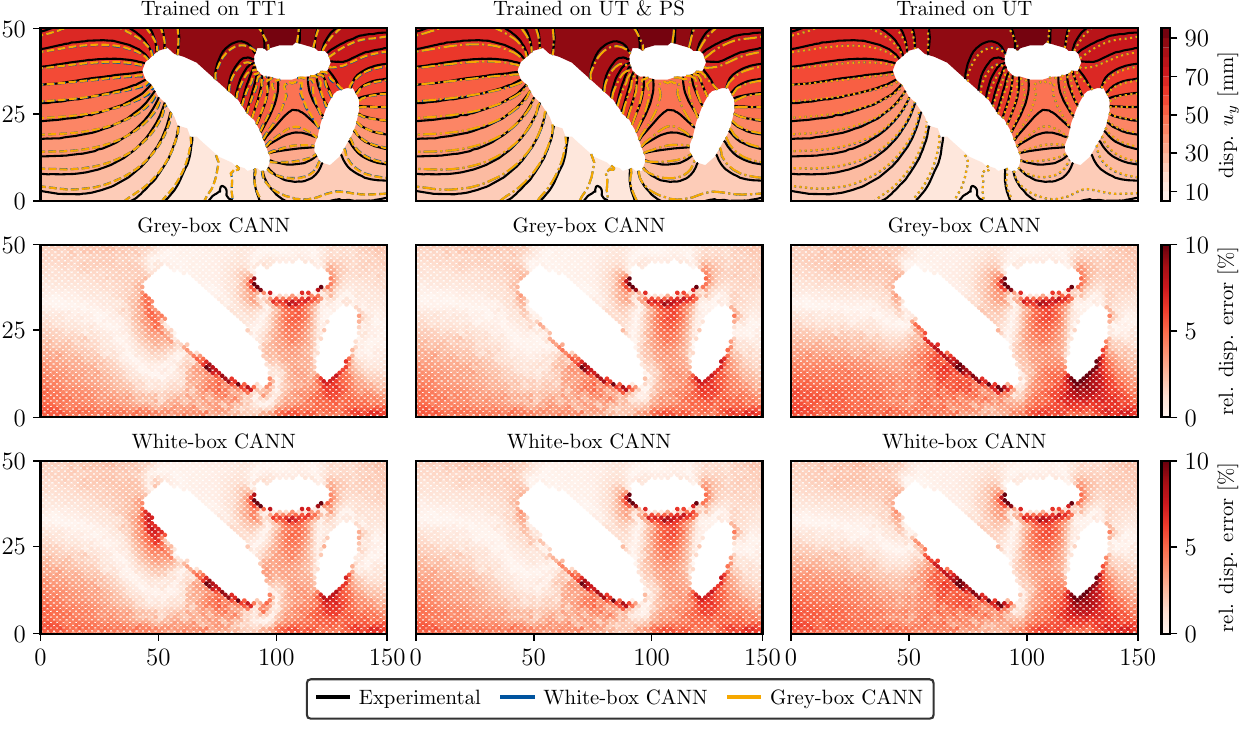}
        \caption{\textbf{Case \hyperref[sec:results:case1]{1}, displacement field on the unseen specimen TT2.} For each training configuration (columns: UT, UT~$\&$~PS, TT1) and each architecture (grey-box and white-box CANN), the figure compares the DIC-measured vertical displacement field $u_y$ (top row, black contour lines on filled colormap) with the FE prediction (overlaid blue contours for the white-box CANN and yellow contours for the grey-box CANN), and reports the pointwise relative error of the grey-box prediction (middle row) and of the white-box prediction (bottom row). The pointwise errors are systematically higher in width and amplitude than on TT1 (see Figure~\ref{fig:all_models_on_TT1_disp}), reflecting the unseen-BVP setting.}
        \label{fig:all_models_on_TT2_disp}
\end{figure}

\paragraph{Full-field reproduction on unseen TT2} The displacement field on TT2 (Figure~\ref{fig:all_models_on_TT2_disp}) and the corresponding stretch field (Figure~\ref{fig:all_models_on_TT2_stretch} in Appendix~\ref{app:case1}) confirm the picture: all six trained models reproduce the global field structure correctly, with localized error concentrations at the hole boundaries that follow a remarkably uniform pattern across architectures. The pointwise errors are however systematically higher in width and amplitude than on TT1, which is in part inherent to the unseen-BVP setting and in part attributable to the geometric idealization of the FE mesh, since the specimen drawings used to generate the mesh may not exactly match the as-manufactured geometry. Within each training configuration, the white-box and the grey-box CANN remain very close to each other and consistently produce the strongest error peaks at the same locations along the hole boundaries. The stretch fields show a slightly less smooth DIC reference than on TT1, but the predicted contours still track the measurement closely. The dominant stretch-error concentration appears near the upper-right hole boundary and is shared by all six models, which suggests that this localized discrepancy is dominated by the modeling assumptions (mesh geometry, boundary-condition reconstruction, incompressibility) rather than by the constitutive law itself. Additional stretch-error stripes along the upper and lower edges of the specimen are consistent with the imposed B-spline Dirichlet boundary and the kinematic constraint of incompressibility.

\paragraph{Error distributions on unseen TT2} The violin plots in Figure~\ref{fig:TT2_violins} (Appendix~\ref{app:case1}) quantify these observations. On the displacement field, the configurations trained on TT1 and on UT~$\&$~PS lead to comparable, broad error distributions, whereas the configuration trained on UT alone is shifted toward larger errors. On the stretch field, all six distributions collapse to a strikingly uniform shape with a low median of about $4\,\%$ and pronounced outliers. The grey-box CANN trained on TT1 remains the best performer in median terms, while the grey-box CANN trained on UT is again the worst, confirming the data-hungry character of the higher-capacity model.

\begin{figure}[htbp]
        \centering
        \includegraphics[width=\linewidth]{ 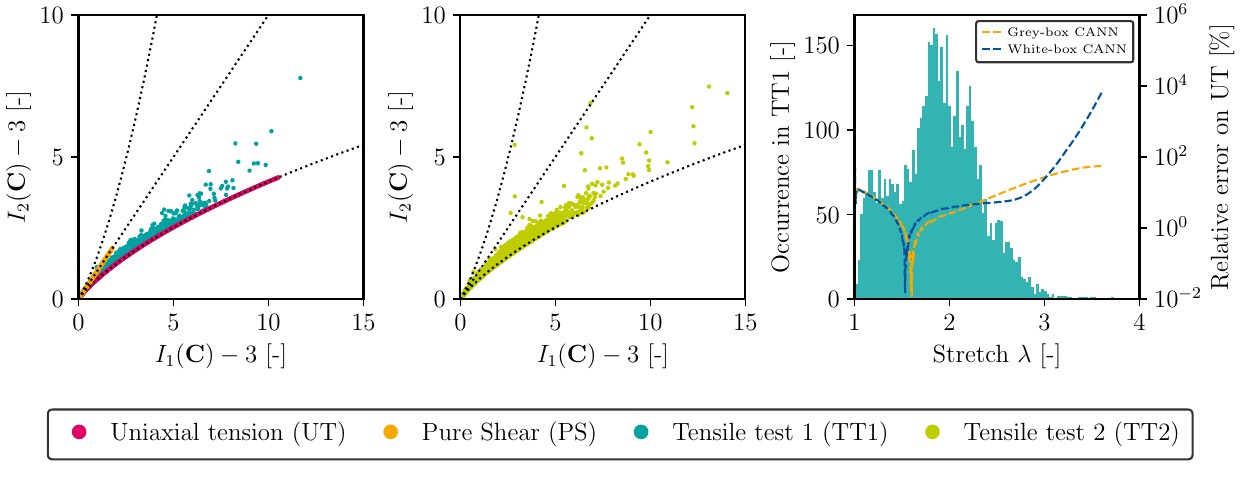}
        \caption{\textbf{Case \hyperref[sec:results:case1]{1}, kinematic content of the training and testing data.} Left: projection of the training (and testing) sets UT, PS, and TT1 into the $(I_1,I_2)$ invariant plane. Center: projection of the unseen testing set TT2 into the same plane. Right: histogram of the maximum principal stretches sampled in the TT1 training set, overlaid with the relative error on UT of the models trained on TT1. Stretches above $\lambda \approx 3$ are essentially absent from the TT1 training set, and the relative error of the uniaxial response climbs steeply beyond this threshold, with the white-box CANN diverging more aggressively than the grey-box CANN.}
        \label{fig:LS_data}
\end{figure}

\paragraph{Coverage of the invariant plane} The systematic differences between the training configurations are best understood from the kinematic content of the corresponding training data. Figure~\ref{fig:LS_data} projects all three training sets and the unseen TT2 test set into the $(I_1, I_2)$ invariant plane. The training data of TT1 and TT2 are dominated by uniaxial-like states with isolated excursions toward higher invariants; PS occupies the boundary closest to the biaxial line. Crucially, the stretch histogram on the right shows that principal stretches above $\lambda \approx 3$ are essentially absent from the TT1 training set. The relative error of the uniaxial response of TT1-trained models climbs steeply beyond this threshold, and the white-box CANN diverges much more aggressively than the grey-box CANN. This behavior is intrinsic to the exponential terms of the white-box ansatz, whose softplus-activated weights cause an exponential blow-up once the network is extrapolated beyond the stretch range covered during training. The grey-box CANN remains more graceful in its extrapolation but is not immune to the same effect. From a practical standpoint, this confirms that data-driven discovery of constitutive laws from a single heterogeneous test must be accompanied by an awareness of the regions of the invariant plane that are actually sampled.

\paragraph{Quantitative summary} The consolidated one- and two-dimensional error metrics for Case \hyperref[sec:results:case1]{1} are reported in Tables~\ref{tab:LS_1D_quantities} and~\ref{tab:LS_2D_quantities} of Appendix~\ref{app:case1}. Models trained on TT1 match the heterogeneous reaction force to within $\sim 0.5\,\%$ on TT1 itself but show pathological extrapolation errors on UT for the white-box CANN ($\sim 1.7 \cdot 10^{3}\,\%$), driven exclusively by the high-stretch regime described above. Models trained on UT~$\&$~PS yield the best overall balance, with stress errors below~$10\,\%$ on both homogeneous tests and reaction-force errors of about~$3$--$7\,\%$ on TT1 and TT2. The corresponding two-dimensional metrics, evaluated as the median pointwise relative error of the displacement and stretch field across all DIC points, lie between~$1.3$ and~$2.4\,\%$ for the displacement and between $3.4$ and~$5.2\,\%$ for the stretch field, with negligible inter-architecture variation. The full-field error is therefore an order of magnitude less sensitive to the choice of training data than the one-dimensional stress error---a finding we attribute to the global character of the displacement field, which is largely fixed by the imposed Dirichlet boundary conditions and the structural geometry, and which therefore constrains the constitutive model only weakly.

In summary, Case \hyperref[sec:results:case1]{1} establishes that the ADE-FEMU framework reliably identifies hyperelastic constitutive neural networks from full-field DIC data and that the resulting models generalize to an unseen, geometrically different boundary-value problem. The dominant factor controlling generalization is the diversity of deformation modes contained in the training data, with the combination of well-chosen homogeneous tests providing the most versatile constitutive description and a single heterogeneous test acting essentially as a regularization device that constrains the model only along the dominant deformation direction of that experiment.


\subsection{Case 2: Reduced-data setting based on a one-dimensional stretch profile}\label{sec:results:case2}

\begin{figure}[htbp]
        \centering
        \hfill
          \centering
            \def\svgwidth{\linewidth}
            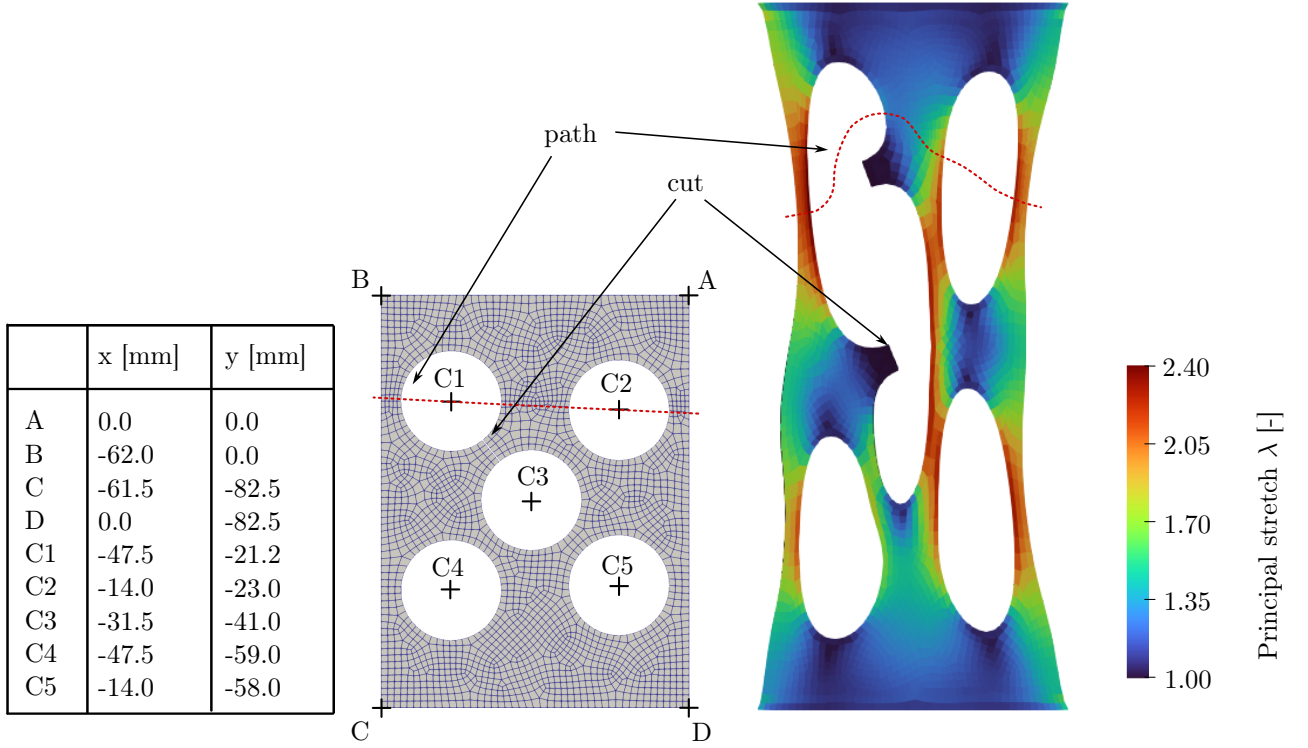
        \caption{\textbf{Case \hyperref[sec:results:case2]{2}, five-hole tensile specimen TT.} Specimen TT in the undeformed reference state (left) and at maximum deformation of $u_{y,top} =57.3\,\mathrm{mm}$ (right), color-coded by the maximum principal stretch $\lambda_1$. The narrow ligament between holes C1 and C3 is cut, resulting in a highly nonlinear deformation profile. The red curve indicates the path along which the maximum stretch is measured experimentally; in the undeformed state, this path is a straight line through the centers of holes C1 and C2. All data are extracted from \cite{meunier2008mechanical}. The relevant coordinates are given in Table \ref{tab:case2_coors} of Appendix \ref{app:case2}}
        \label{fig:Overview_FH}
\end{figure}

\begin{figure}[htbp]
        \centering
        \includegraphics[width=\linewidth]{ 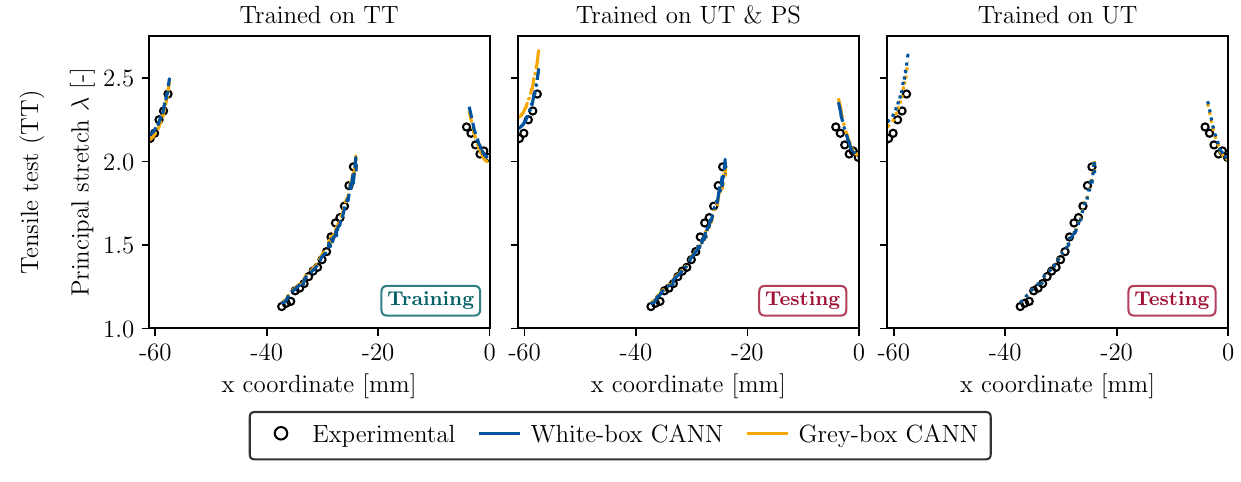}
        \caption{\textbf{Case \hyperref[sec:results:case2]{2}, maximum principal stretch along the measurement path of TT.} Comparison of the experimentally measured maximum principal stretch $\lambda_1$ along the path through holes C1--C2 (black markers) with the FE predictions of the grey-box CANN (yellow) and the white-box CANN (blue), for the three training configurations (TT, UT~$\&$~PS, UT). White-box models trained on TT and on UT~$\&$~PS recover the central peak well; models trained on UT~$\&$~PS exhibit the largest spread between architectures, especially in the outer regions of the path.}
        \label{fig:FH_stretch_along_path}
\end{figure}

\paragraph{Data and experimental setting} Case \hyperref[sec:results:case2]{2} examines the opposite end of the data-availability spectrum: instead of a complete DIC field, only a single one-dimensional stretch profile and the corresponding reaction force and machine displacement of a heterogeneous tensile test are available \cite{meunier2008mechanical}. Although this example may appear somewhat idealized, it illustrates an important limitation of equilibrium gap methods. In particular, training from such a limited dataset would generally not be feasible within an equilibrium gap framework. Moreover, recent advances in distributed sensing technologies, such as fiber optic sensors, enable the acquisition of spatially resolved strain fields that are qualitatively similar to the measurements considered in this example. Consequently, the reduced-data setting investigated here is not merely of academic interest but is representative of an increasingly relevant class of experimental measurements. The specimen TT is a five-hole tensile plate whose layout is shown in Figure~\ref{fig:Overview_FH}. The narrow ligament between holes C1 and C3 is intentionally cut, modeled with a separation thickness of $0.1\,\mathrm{mm}$, The narrow ligament between holes C1 and C3 is cut, resulting in a highly nonlinear deformation profile. The boundary value problem is discretized with $2{,}181$ elements and $4{,}826$ nodes. This results in $14{,}478$ displacement DOFs and $2{,}181$ pressure DOFs, in total $16{,}659$ DOFs. The experimentally available kinematic measurement is the maximum principal stretch along the red path that intersects the centers of holes C1 and C2 in the undeformed state. Figure \ref{fig:FH_stretch_along_path} illustrates the experimentally obtained principal stretch distribution. The maximum principal stretch is given for a maximum vertical displacement of $57.3\,\mathrm{mm}$. The path data are smoothed region-wise by a third-order polynomial fit, and the resulting stretch profile, together with the maximum reaction force of $20\,\mathrm{N}$ at maximum displacement, defines the training target. With only one available data point of the force---displacement curve, just one load step $N_j = 1$ per epoch $i$ is calculated during training. The networks are trained using the Adam optimizer with a cyclic learning-rate schedule oscillating between a base value of $0.0001$ and a peak value of $0.03$ over a step size of $15$ epochs.

\begin{figure}[htbp]
        \centering
        \includegraphics[width=\linewidth]{ 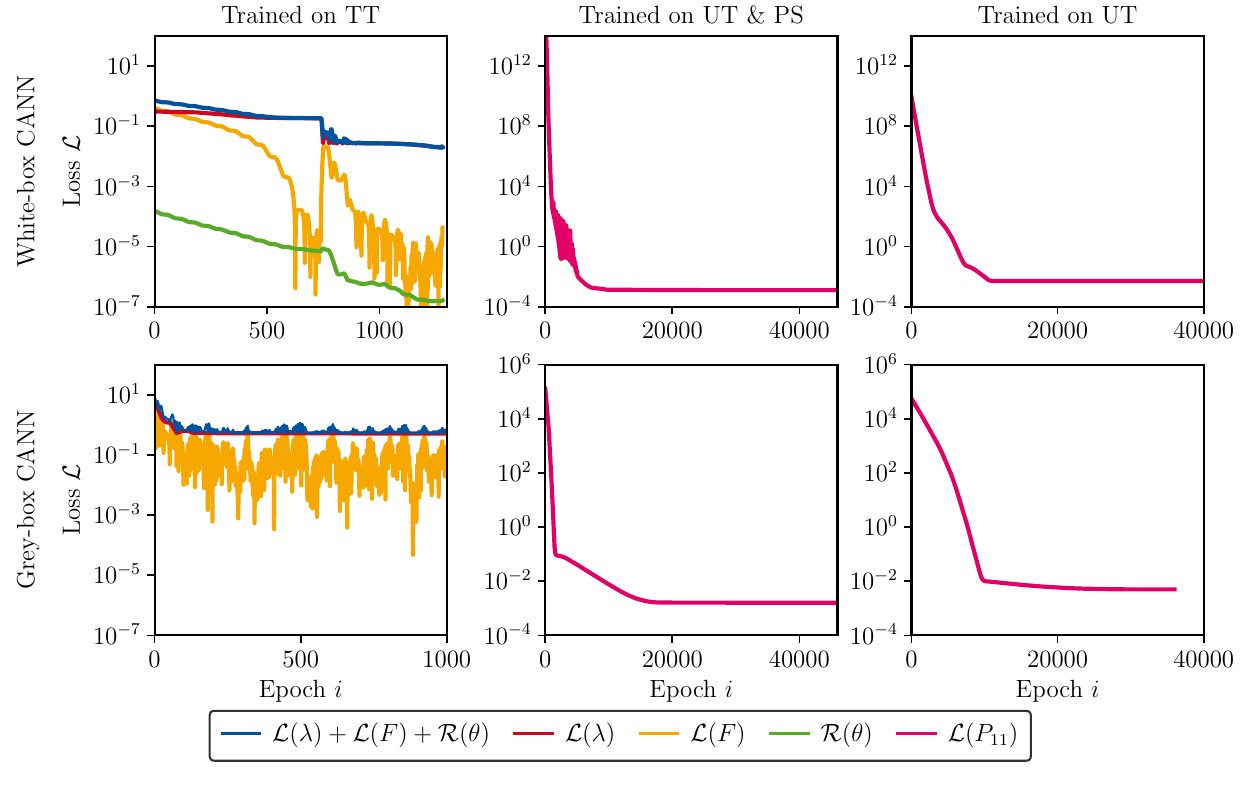}
        \caption{\textbf{Case \hyperref[sec:results:case2]{2}, loss evolution during training.} Evolution of the kinematic loss $\mathcal{L}(\lambda)$ (red), the reaction-force loss $\mathcal{L}_F$ (orange), the regularization term $\mathcal{R}(\btheta)$ and the stress loss $\mathcal{L}(P_{11})$ for the three training configurations (columns: TT, UT~$\&$~PS, UT) and the two architectures (rows: grey-box CANN, white-box CANN). In contrast to Case \hyperref[sec:results:case1]{1}, both loss contributions are now clearly visible during training because the stretch loss is markedly more sensitive to the constitutive parameters than the kinematic loss. The high-frequency oscillations originate from the cyclic learning-rate schedule.}
        \label{fig:FH_loss_evolution}
\end{figure}

\paragraph{Loss reformulation} Because no displacement field is available, the loss \eqref{eq:loss_total} introduced in Section~\ref{sec:framework:impl} can no longer be evaluated on $\bu^{\mathrm{obs}}$. We therefore replace the displacement-mismatch expression in the kinematic loss term by a stretch-mismatch expression \eqref{eq:loss_lambda} evaluated at the $200$ FE quadrature points closest to the measurement path. The maximum principal stretch at each such point is computed from the local deformation gradient, $\lambda_1 = \sqrt{\mathrm{eig}_{\max}(\Cmat)}$, and compared to the polynomial fit of the experimental measurement at the same arc-length coordinate. The reaction-force term remains unchanged. This case therefore demonstrates that the ADE-FEMU framework can be adapted to arbitrary observation operators that map the FE solution to whatever subset of kinematic quantities is experimentally accessible, in line with the partial-observability requirement emphasized in \cite{Regazzoni2026}. 

\begin{equation}
   {\mathcal{L}}(\btheta) =  
\frac{\omega_{\kappa}}{2} \left\| \boldsymbol{Q}_{\lambda}\blambda_1(\bU(\btheta)) - \blambda_1^{\mathrm{obs}} \right\|^{2}
+
\frac{\omega_F}{2}\left\| \boldsymbol{Q}_F\boldsymbol{R}^{\mathrm{ext}}(\btheta) - F^{\mathrm{obs}} \right\|^{2} + \mathcal{R}(\btheta)
\label{eq:loss_lambda}
\end{equation}

Additionally a regularization term $\mathcal{R}(\btheta)$ is added, which will be explained in more detail in the following paragraph. 
As in Case \hyperref[sec:results:case1]{1}, both architectures are trained on three different training sets---TT (stretch profile + reaction force), UT~$\&$~PS, and UT alone---and the resulting models are evaluated on a broader set of homogeneous deformation modes than were available during training. This setting deliberately probes the regime in which the data are spatially sparse, derived from the displacement field rather than measured directly, and limited to a single deformation mode.

\paragraph{Loss-scaling and L1-regularization} 
To investigate the influence of the loss-function scaling parameters, a small parameter study was conducted, see Figure \ref{fig:case_2_norm_and_reg_evolution} in Appendix \ref{app:case2}. The ratio between the mismatch term scaling parameters $\frac{\omega_F}{\omega_{\kappa}}$ is varied over eight orders of magnitude, ranging from ($10^{-4}\leq\frac{\omega_F}{\omega_{\kappa}} \leq 10^{4}$). The results show that increasing the weighting of the kinematic mismatch term substantially reduces the corresponding kinematic error, whereas the reduction in the force mismatch error is considerably less pronounced when the force term is weighted more strongly. Furthermore, larger weighting ratios lead to poorer predictions for the uniaxial tension (UT) experiment but significantly improve the predictions for the pure shear (PS) experiment. This behavior suggests that assigning a higher weight to the force mismatch term causes the calibration to treat the experiments more similarly to homogeneous deformation states, thereby favoring the PS response over the UT response. 
Additionally, an L1-regularization term, $\mathcal{R}(\boldsymbol{\theta}) = \lambda_{L1}\cdot \|w_{2,j}\|$, is incorporated into the loss function (\ref{eq:loss_lambda}) during the training of the white-box model to promote sparsity by reducing the number of active terms, as proposed in \cite{McCulloch2024}. In the previous example (Case~\hyperref[sec:results]{1}), the weights were activated using a softplus function, $w_{i,j} = \ln\left(1+\exp(\hat{w}_{i,j})\right)$, to enforce non-negativity. However, since the softplus function does not permit weights to become exactly zero, it was replaced in the present study by a quadratic activation, $w_{i,j} = \hat{w}_{i,j}^{2}$, which preserves non-negativity while allowing the optimizer to eliminate terms completely. The influence of the regularization parameter was investigated over the range $10^{-8} \leq \lambda_{L1} \leq 5 \cdot 10^{-6}$ using a fixed scaling parameter ratio of $\frac{\omega_F}{\omega_{\kappa}} = 10^{-3}$. The results, presented in Figure~\ref{fig:case_2_norm_and_reg_evolution} in Appendix \ref{app:case2}, show that for $\lambda_{L1} \geq 2 \cdot 10^{-6}$, all but one model term are eliminated, resulting in a substantial increase in the kinematic mismatch. At the same time, the introduction of the regularization consistently improves the prediction of the pure shear (PS) response across the investigated parameter range. An optimal trade-off between sparsity and predictive accuracy was obtained for $\lambda_{L1} = 10^{-6}$, where all considered error measures attain their minimum values. The corresponding results are presented in the following. Figures \ref{fig:stretch_reg} \& \ref{fig:UT_PF_F_reg} in Appendix \ref{app:case2} show data reproductions for different regularization parameters $\lambda_{L1}$.

\paragraph{Loss evolution} Figure~\ref{fig:FH_loss_evolution} shows the evolution of the stretch- and force-loss components during training. In contrast to Case \hyperref[sec:results:case1]{1}, both contributions now decrease in a clearly visible fashion for both architectures, since the stretch loss is markedly more sensitive to the constitutive parameters than the kinematic loss of the previous case. The grey-box model reaches a plateau within approximately $100$ epochs on TT. While the white-box model takes $> 1,000$ epochs for convergence, most likely due to the regularization term. The models converge more slowly, typically over several thousand epochs, when trained on the homogeneous data, where only a stress miss-match term is evaluated. The high-frequency oscillations again reflect the cyclic learning-rate schedule rather than instability of the training. On the homogeneous training sets, the convergence behavior of both architectures is more similar to each other than to their counterparts trained on TT, which once more confirms that the training data dominate the optimization landscape.
\paragraph{Training--testing matrix} The full training--testing matrix on UT, PS, and on the reaction force of TT, reported in Figure~\ref{fig:FH_training_testing_matrix} of Appendix~\ref{app:case2}, leads to a similar qualitative conclusions as in Case \hyperref[sec:results:case1]{1}, but in contrast to Case \hyperref[sec:results:case1]{1} the regularized white-box CANN produces much more stable results. The single training data point of the TT training configuration is reproduced to high accuracy by both architectures, and extrapolation to UT is also reasonable. Now the regularized white-box CANN reproduces PS to great accuracy, with the grey-box CANN being somewhat less stable than the white-box CANN. Training on UT~$\&$~PS again provides the most balanced fit on the homogeneous tests, with both models, but underestimates the reaction force on TT by approximately~$5\,\%$. Models trained on UT alone are excellent on uniaxial loading but produce the weakest reproduction of PS among all configurations. Across the matrix, the differences between the grey-box and the white-box CANN trained on the same data remain comparatively small.

\begin{figure}[htbp]
        \centering
        \includegraphics[width=\linewidth]{ 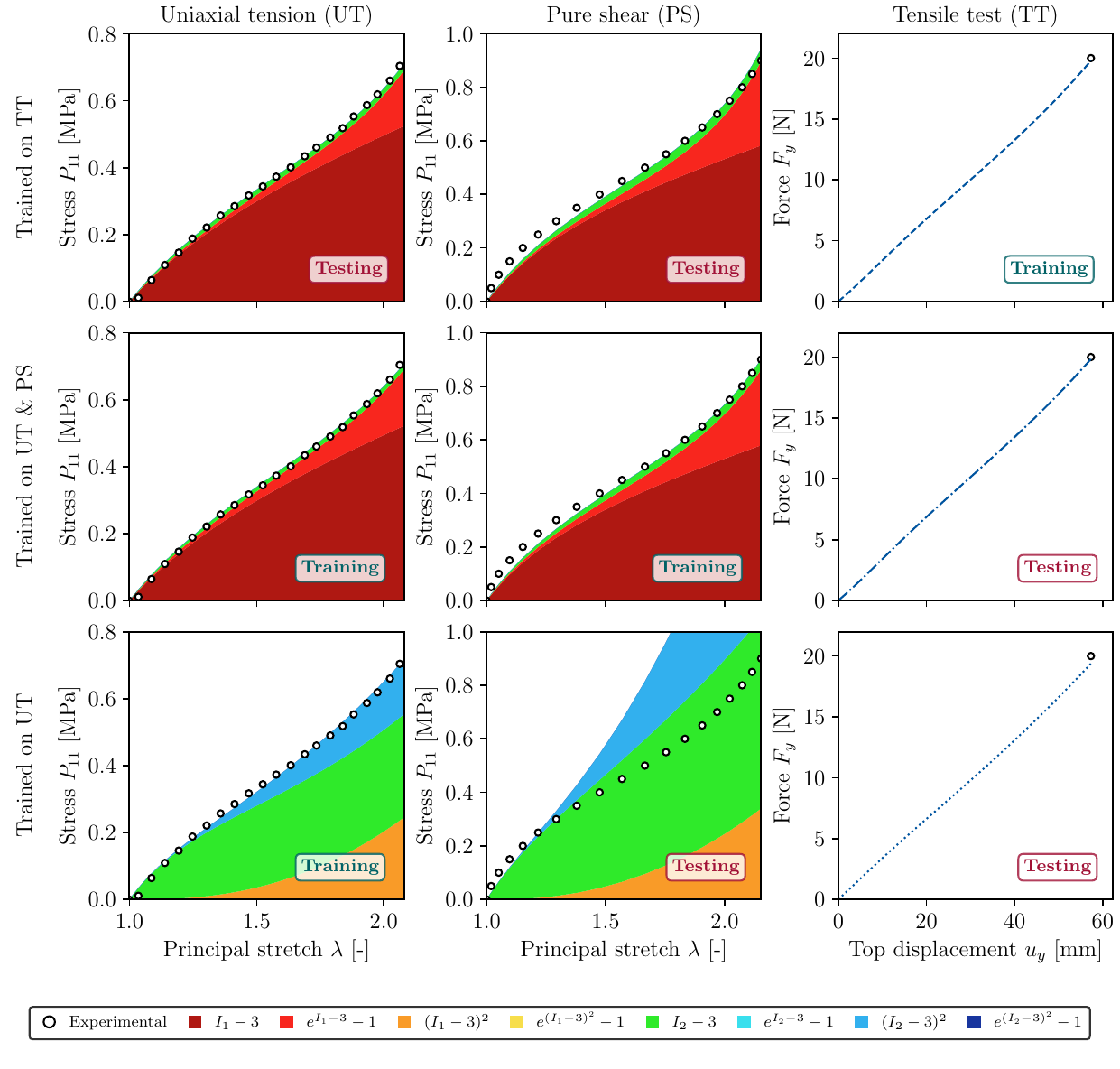}
        \caption{\textbf{Case \hyperref[sec:results:case2]{2}, model discovery with the white-box CANN.} Stress--stretch responses of the three white-box CANN models on the homogeneous tests UT and PS, and reaction-force response on the heterogeneous specimen TT. Each row corresponds to a training configuration (TT, UT~$\&$~PS, UT); the training data for each row is explicitly labeled. Stacked colored areas underneath the response curves decompose the predicted first Piola--Kirchhoff stress into the contributions of the individual interpretable terms of the white-box CANN, so that the number and identity of the activated terms can be read off directly. The term activation for the models trained on TT and on UT~$\&$~PS are very similar, while different terms are activated for the model trained on UT. In all training configurations three terms are activated.}
        \label{fig:FH_training_testing_matrix_MD}
\end{figure}

\paragraph{Model discovery} A particular advantage of the white-box CANN is that its discovered strain-energy density is directly interpretable, since each output term is associated with a single phenomenological building block (cf.\ Section~\ref{sec:theory}). To complement the training--testing matrix discussed in the previous paragraph, Figure~\ref{fig:FH_training_testing_matrix_MD} reproduces the same comparison for the white-box CANN only, while additionally decomposing each predicted stress--stretch curve into the contributions of its individual energy terms. This decomposition makes it possible to assess, at a glance, both the quality of the model predictions and the compactness (i.e.\ sparsity) of the underlying constitutive ansatz.
The results confirm the broader observation that the interpretability of the identified constitutive model is governed by the kinematic richness of the training data. Moreover, they reveal that the regularized white-box CANN trained on TT and the unregularized white-box CANN trained on UT~$\&$~PS essentially discover the same constitutive model. This agreement is also evident from the explicit strain-energy density functions listed in Table~\ref{tab:discovered models}, where all terms with $w_{2,j} < 10^{-10}$ have been omitted. In contrast, the model identified using only UT exhibits a different structure, reflecting the insufficient kinematic information available during training.
A comparison with the invariant-based constitutive model identified in \cite{meunier2008mechanical} further highlights the similarity in both the functional form and the identified parameter values. This agreement is particularly noteworthy because the model proposed in \cite{meunier2008mechanical} was obtained by fitting the strain-energy density directly and is therefore not expected to satisfy polyconvexity, whereas the discovered white-box CANN models are constructed to preserve this property.

\begin{table}[H]
    \centering
    \footnotesize
    \caption{\textbf{Case \hyperref[sec:results:case2]{2},  strain energy density functions, identified with the white-box CANN.} The identified strain energy functions for the models trained on TT and UT\&PS show great similarities in structure and identified parameters. The identified model from \cite{meunier2008mechanical}, based on energy fitting, shows similarities as well, for example the coefficient of the $(I_1-3)$ term.  }
    \label{tab:LS_1D_quantities}
    \begin{tabularx}{\textwidth}{lXl}
    \toprule
      Training data &  Training regime &  Strain energy density $\Psi$ [Pa]  \\

    \midrule
         Tensile test (TT) &  \mbox{ADE-FEMU} &  $\Psi = 0.142\cdot(I_1-3) + 0.006\cdot(e^{0.937\cdot(I_1-3)}-1)+0.011\cdot(I_2-3)$ \\
          Uniaxial tension (UT) \& Pure shear (PS)  &  homogeneous &  $\Psi = 0.141\cdot(I_1-3) + 0.012\cdot(e^{0.726\cdot(I_1-3)}-1)+0.010\cdot(I_2-3)$ \\
          Uniaxial tension (UT)  &  homogeneous &  $\Psi = 0.015\cdot(I_1-3)^2 +0.174\cdot(I_2-3)+0.033\cdot(I_2-3)^2$ \\
     \midrule
     \midrule

         Uniaxial deformation,  Plane strain defo. &  energy &  $\Psi = 0.14\cdot(I_1-3) -0.0026\cdot(I_1-3)^2 +0.0038\cdot(I_1-3)^3$ \\  
        \& Biaxial deformation& fitting \cite{meunier2008mechanical}&  $ +0.033\cdot(I_2-3) +0.00095\cdot(I_2-3)^2 -0.0049\cdot(I_1-3)(I_2-3)$ \\
    \bottomrule
    \end{tabularx}
    \label{tab:discovered models}
\end{table}

\paragraph{Stretch along the measurement path} Figure~\ref{fig:FH_stretch_along_path} compares the experimentally measured maximum principal stretch along the path through holes C1--C2 with the corresponding FE predictions. Models trained on TT reproduce the left and central portions of the profile well for both architectures, with the largest residuals appearing on the right segment of the path. Models trained on UT~$\&$~PS recover the central peak reasonably but slightly overestimate the stretch in the outer regions of the path, especially the grey-box model. For the models trained on UT~$\&$~PS the spread between models is the largest. The models trained on UT tend to overestimate the stretch in the outer regions and to underestimate it in the central region. The inter-architecture variability on the stretch profile is markedly larger than on the displacement field of Case \hyperref[sec:results:case1]{1}, in line with the higher derivative order of the stretch measure, and reinforces the conjecture that the stretch is a more informative training signal whenever it is experimentally accessible.

\begin{figure}[htbp]
        \centering
        \includegraphics[width=\linewidth]{ 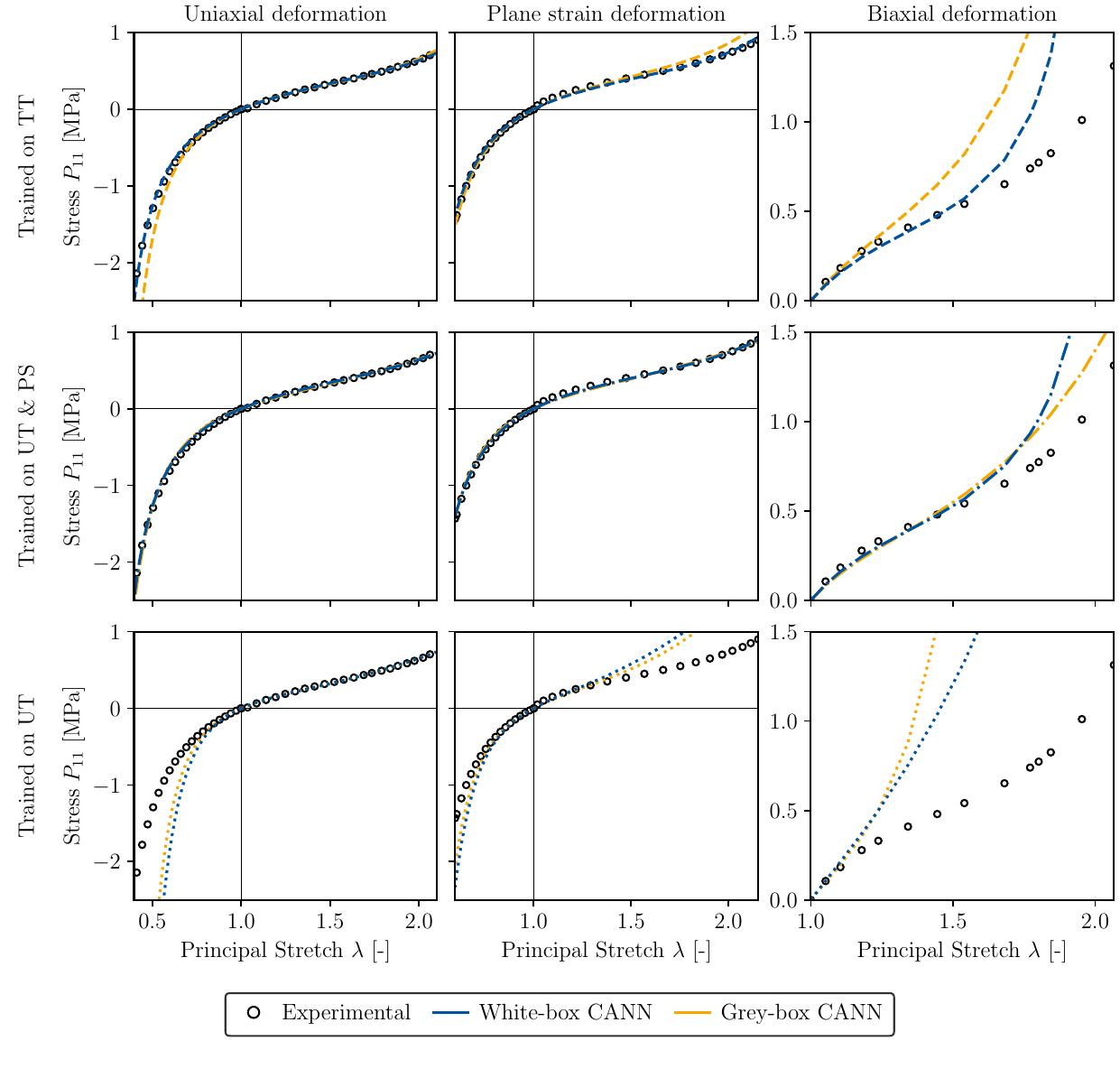}
        \caption{\textbf{Case \hyperref[sec:results:case2]{2}, full material response of the identified models.} Stress--stretch responses of all six trained models on a broad battery of homogeneous deformations: uniaxial tension and compression (UT, UC), plane-strain tension and compression (PS, PSC), and equibiaxial bulge test (BL). The models trained on UT~$\&$~PS, as well as the white-box model trained on TT, reproduce UD and PSD most accurately and remain closest to the experimental bulge curve; the grey-box model trained on TT remains reasonable for UD and PSD but shows pronounced deviations on the bulge test; the models trained on UT alone are reliable only in the immediate vicinity of the training regime.
        \label{fig:FH_full_material_behavior}}
\end{figure}

\paragraph{Full material response} To probe the constitutive content of the identified models beyond the training regime, Figure~\ref{fig:FH_full_material_behavior} evaluates each model on a broader range of homogeneous deformations: uniaxial tension and compression (UT and UC, with $\boldsymbol{F}_{\mathrm{UD}} = \mathrm{diag(\lambda,1/\sqrt\lambda,1/\sqrt\lambda)}$), plane-strain tension and compression (PS and PSC, $\boldsymbol{F}_{\mathrm{PSD}} = \mathrm{diag(\lambda,1,1/\lambda)}$), and an equibiaxial bulge test (BL, with $\boldsymbol{F}_{\mathrm{BL}} = \mathrm{diag(\lambda,\lambda,1/\lambda^2)}$). The configurations trained on UT~$\&$~PS, as well as the regularized white-box model trained on TT, reproduce uniaxial deformation (UD) and plane strain deformation (PSD) most accurately and remain closest to the experimental bulge curve, with the grey-box CANN slightly outperforming its white-box counterpart. The grey-box CANN trained on TT still tracks UD and PSD reasonably but shows pronounced deviations on the bulge test. Models trained on UT alone are reliable only in the immediate vicinity of the training regime ($0.25 \lesssim \lambda \lesssim 1.25$) and exhibit the largest deviation on the bulge test. We note that the comparison between Cases~1 and~2 is not entirely symmetric, since the UT training data of Case \hyperref[sec:results:case1]{1} cover a substantially wider stretch range ($\lambda \in [1, 3.5]$) than the UT training data available here ($\lambda \lesssim 2$).

\paragraph{Quantitative summary} The consolidated error metrics of Case \hyperref[sec:results:case2]{2} are reported in Table~\ref{tab:FH_result_table} of Appendix~\ref{app:case2}. As anticipated, the models trained on UT alone explode on compressive and bulge deformations, with the white-box CANN reaching errors well above $10^{2}\,\%$ on BL. The regularized white-box model trained on TT achieves a satisfactory reproduction of the local stretch ($\sim 2\,\%$) and the reaction force ($2.16\,\%$) of TT itself and can reasonably reproduce  compressive deformations, yet explodes on the bulge test.  Whereas the grey-box CANN remains slightly more robust on the bulge test, it performs significantly worse for uniaxial compression and pure shear. Models trained on UT~$\&$~PS, as well as the regularized white-box CANN trained on TT, provide the most balanced behavior: their extrapolation to compression and to the bulge test is dramatically better, especially in contrast to the grey-box model, trained on TT.

Taken together, Case \hyperref[sec:results:case2]{2} demonstrates that the proposed framework can be successfully trained from sparse, derivative-type kinematic measurements consisting of a single one-dimensional stretch profile and a global force--displacement curve, and that the resulting models retain a usable level of generalization. In contrast to Case \hyperref[sec:results:case1]{1}, we can show that by making use of regularization, the white-box CANN can outperform the grey-box CANN. It also shows that we can essentially find the same model from two very different data scenarios with the white-box CANN. Although the data availability ultimately matters, especially for the grey-box CANN, which benefits most from richer training data while suffering most from data starvation.

\subsection{Case 3: Heterogeneous material systems with full-field displacement data}\label{sec:results:case3}

\paragraph{Data and experimental setting} The previous two cases addressed homogeneous materials. Case \hyperref[sec:results:case3]{3} demonstrates that the proposed framework extends naturally to heterogeneous specimens whose phases possess distinct constitutive behavior. We consider thin plates consisting of a soft matrix and a stiffer inclusion of variable shape, derived from the MNIST digit dataset and reported as an open-access experimental benchmark \cite{726791,LEJEUNE2020100659, MechanicalMNISTChallenge}, based on groundwork from \cite{tepole2026inclusion}. The dataset comprises twenty-three specimens of identical outer dimensions $40\,\mathrm{mm} \times 40\,\mathrm{mm} \times 2\,\mathrm{mm}$, each loaded in uniaxial tension and measured by DIC; the material parameters of both phases are identified simultaneously from a single training specimen and the resulting model is subsequently evaluated on the remaining twenty-two specimens. 

\begin{figure}[htbp]
    \centering
    \includegraphics{ 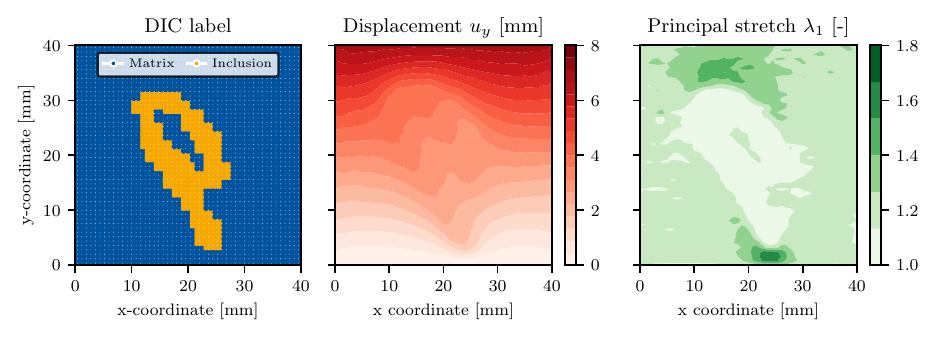}
    \put(-370,0){\small (a)}
    \put(-240,0){\small (b)}
    \put(-90,0){\small (c)}
    \caption{\textbf{Case \hyperref[sec:results:case3]{3}, ground-truth fields of the heterogeneous material system.} (a) DIC labeling of a representative thin plate specimen, with blue indicating the matrix and orange the inclusion \cite{MechanicalMNISTChallenge}. (b) Corresponding measured vertical displacement field $u_y$ at the final load step. (c) Corresponding FE-derived maximum principal stretch field $\lambda_1$ at the final load step, used as the structural ground truth against which the ADE-FEMU predictions are compared.}
    \label{fig:ground_truth}
\end{figure}

A representative DIC field is shown in Figure~\ref{fig:ground_truth}(a), in which the color coding distinguishes the two material phases (blue: matrix, orange: inclusion). The corresponding measured vertical displacement field at the final load step is shown in panel~(b). A three-dimensional finite element model is constructed with dimensions matching the experimental specimen, and the FE discretization is chosen to be consistent with the DIC spatial resolution, so that the DIC data points coincide with the nodal coordinates on the specimen's planar surface. In panel~(c), the maximum principal stretch field obtained from the FE simulation at the final load step is shown.  The material-phase labels attached to the DIC field are mapped to the FE quadrature points by nearest-neighbor sampling, which assigns each integration point unambiguously to the matrix or to the inclusion. 

A discretization with $2,500$ elements and $5,202$ nodes is used for the boundary value problem considered during training. The resulting FE system comprises $15,606$ displacement DOFs and $2,500$ pressure DOFs, yielding a total of $18,106$ DOFs. The optimization is performed using only the nodes located on the planar surface and their in-plane displacement components, resulting in $5{,}202$ displacement DOFs entering the displacement-based loss function.

\paragraph{Constitutive setup and loss} Two independent CANNs are assigned to the matrix and to the inclusion, with trainable parameters $\btheta_{M}$ and $\btheta_{I}$, respectively. Both networks are pre-trained on synthetic Mooney--Rivlin data to provide a physically meaningful initial state and to ensure that the forward problem remains solvable from the first epoch, in line with the safeguarding strategy discussed in Section~\ref{sec:framework:impl}. The forward FE solve yields the displacement field $\bU = \mathcal{S}(\Gamma_D; \btheta_{M}, \btheta_{I})$ for the applied Dirichlet boundary condition $\Gamma_D$ and material parameters $\btheta$, and the reaction forces $\boldsymbol{R}$. The loss specializes the generic form \eqref{eq:loss_total} to the present setting by combining a normalized mean-squared displacement mismatch over the DIC points with a normalized reaction-force mismatch,
\begin{equation}\label{eq:case3_loss}
    \mathcal{L}(\btheta_{M}, \btheta_{I})
    = \frac{\omega_{\kappa}}{2} \cdot
    \frac{\frac{1}{N_x N_y}\sum_{x,y}\bigl\|\bU^{(j)}_{xy}-\bu^{\mathrm{obs},(j)}_{xy}\bigr\|^{2}}{\frac{1}{N_x N_y}\sum_{x,y}\bigl\|\bu^{\mathrm{obs},(j)}_{xy}\bigr\|^{2}+\varepsilon}
    + \frac{\omega_F}{2} \cdot \frac{\bigl(\boldsymbol{Q}_F\boldsymbol{R}^{(j)}-{F}^{\mathrm{obs},(j)}\bigr)^{2}}{\bigl(F^{\mathrm{obs},(j)}\bigr)^{2}+\varepsilon},
\end{equation}
where $\omega_{\kappa} = 2$ and $\omega_F = 1$ balances the two contributions and $\varepsilon > 0$ regularizes the normalization. Here, $N_x$ and $N_y$ denote the numbers of nodes along the $x$- and $y$-directions of the domain, respectively.

\begin{figure}[htbp]

 \includegraphics{ 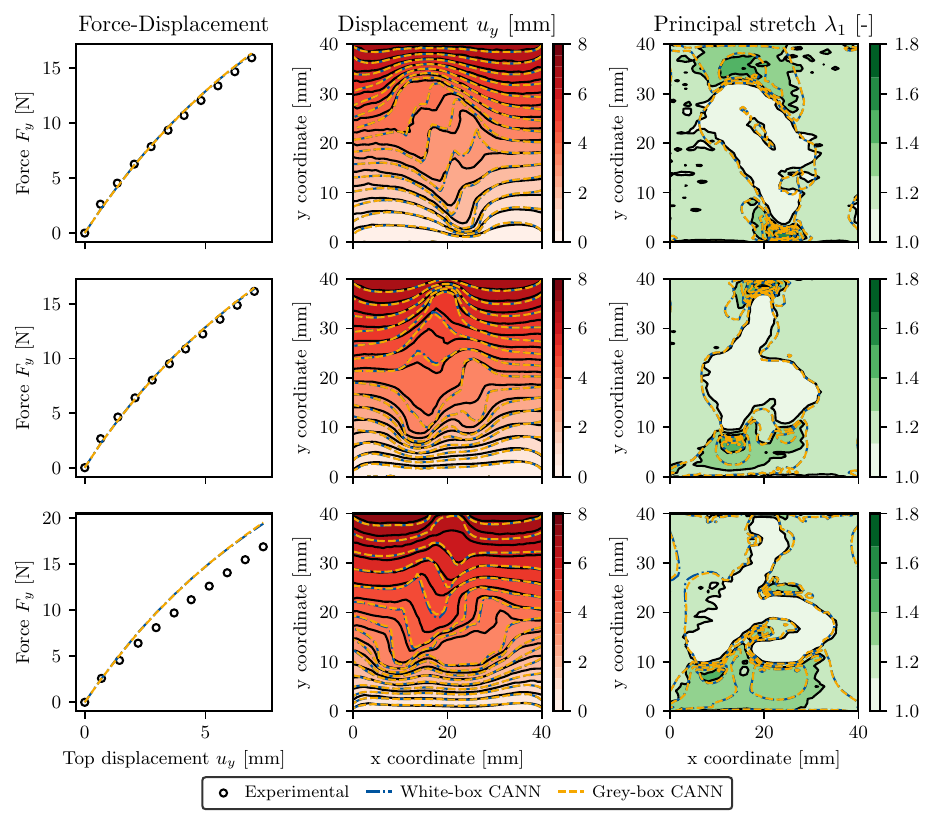}
    \small
    \put(-413,357){
    \tcbox[
        colback=white,
        colframe=Petrol,
        boxrule=.8pt,
        arc=.5mm,
        boxsep=1mm,
        left=.1mm,
        right=.1mm,
        top=0.1mm,
        bottom=0.1mm
    ]{
        \small
        \shortstack{\textcolor{Petrol}{Training}}
    }}
    \put(-413,232){
    \tcbox[
        colback=white,
        colframe=Bordeaux,
        boxrule=.8pt,
        arc=.5mm,
        boxsep=1mm,
        left=.1mm,
        right=.1mm,
        top=0.1mm,
        bottom=0.1mm
    ]{
        \small
        \shortstack{\textcolor{Bordeaux}{Testing} \\ \textcolor{Bordeaux}{(best case)}}
    }
}
    \put(-410,120){\tcbox[
        colback=white,
        colframe=Bordeaux,
        boxrule=.8pt,
        arc=.5mm,
        boxsep=1mm,
        left=.1mm,
        right=.1mm,
        top=0.1mm,
        bottom=0.1mm
    ]{
        \small
        \shortstack{\textcolor{Bordeaux}{Testing} \\ \textcolor{Bordeaux}{(worst case)}}
    }}
    
    \caption{\textbf{Case \hyperref[sec:results:case3]{3}, identified heterogeneous material system applied to training and unseen specimens.} Rows: training specimen (top), best-case unseen specimen (middle), worst-case unseen specimen (bottom). Columns: force--displacement curve (left), vertical displacement field $u_y$ (center), maximum principal stretch field $\lambda_1$ (right). Black lines and contours indicate the experimental ground truth, orange lines and contours indicate the ADE-FEMU prediction. The constitutive parameters identified on the single training specimen transfer to the unseen specimens, with errors increasing only for the worst-case phase morphology.}
    \label{fig:results_MNIST}
\end{figure}

\paragraph{Training and testing} The networks are trained using the Adam optimizer with a cyclic learning-rate schedule oscillating between a base value of $0.002$ and a peak value of $0.02$ over a step size of $25$ epochs. For both architectures, the models are trained for $1000$ epochs on a single representative specimen and subsequently evaluated on the remaining twenty-two unseen samples. The predictions for the training specimen as well as for the best- and worst-case unseen specimens are summarized in Figure~\ref{fig:results_MNIST}, and the corresponding error metrics are reported in Table~\ref{tab:error_metrics_MNIST} of Appendix~\ref{app:case3}.

\paragraph{Training and generalization to unseen specimens} Figure~\ref{fig:results_MNIST}, together with the consolidated error metrics reported in Table~\ref{tab:error_metrics_MNIST} of Appendix~\ref{app:case3}, shows that both architectures reproduce the training specimen with displacement and stretch errors of about $2.8\,\%$ and $1.5\,\%$, respectively, and a reaction-force error of approximately $3.5\,\%$. More importantly, the same identified material parameters generalize to the unseen specimens: on the best-case unseen sample, the displacement-, stretch-, and force-errors remain essentially indistinguishable from those of the training case, while on the worst-case unseen sample they rise to about $10\,\%$ for the displacement, $3\,\%$ for the stretch, and $15\,\%$ for the reaction force, with virtually no difference between the two CANN architectures. The fact that even the worst-case specimen is predicted within errors that are well below the geometric tolerance of the DIC field suggests that the residual discrepancy is dominated by inevitable variations in the as-manufactured specimen geometry rather than by a deficiency of the identified constitutive model.

\begin{figure}[htbp]
    \centering
    \includegraphics[width=0.5\linewidth]{ 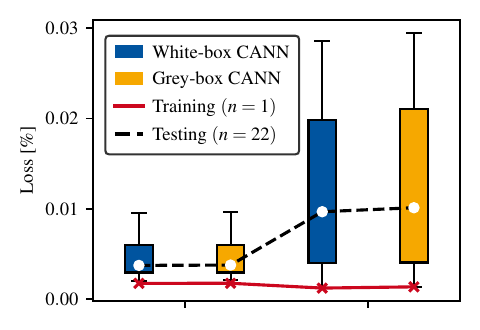}
    \small
    \put(-153,0){$\mathcal{L}_u$}
    \put(-60,0){$\mathcal{L}_F$}
    \caption{\textbf{Case \hyperref[sec:results:case3]{3}, distribution of testing losses across unseen specimens.} Empirical distributions of the kinematic loss $\mathcal{L}_u$ (left) and the reaction-force loss $\mathcal{L}_F$ (right) across the twenty-two unseen specimens, for the grey-box and white-box CANN trained simultaneously on the matrix and on the inclusion. Red markers denote the training losses on the single training specimen; white dots indicate the median of the testing losses; boxes span the interquartile range ($25\,\%$--$75\,\%$); whiskers extend from the $5\,\%$ to the $95\,\%$ percentile. The median testing losses are smaller than the training loss, demonstrating that the identification does not overfit.}
    \label{fig:loss_comparison_MNIST}
\end{figure}

\paragraph{Loss statistics across the test population} Figure~\ref{fig:loss_comparison_MNIST} aggregates the testing performance into empirical distributions of the kinematic loss $\mathcal{L}_u
    =\frac{1}{N_j N_x N_y}\sum_{j,x,y}\bigl\|\bU^{(j)}_{xy}-{\bu}^{\mathrm{obs},(j)}_{xy}\bigr\|^{2}/\left( \frac{1}{N_j N_x N_y}\sum_{j,x,y}\bigl\|{\bu}^{\mathrm{obs},(j)}_{xy}\bigr\|^{2}+\varepsilon\right)$
and the reaction-force loss $\mathcal{L}_F
    = \frac{1}{N_j}\sum_j \bigl(\boldsymbol{Q}_F\boldsymbol{R}^{(j)}-{F}^{\mathrm{obs},(j)}\bigr)^{2}/ \left(\frac{1}{N_j}\sum_j \bigl({F}^{\mathrm{obs},(j)}\bigr)^{2}+\varepsilon\right)$ evaluated over $N_j$ load steps across the twenty-two unseen specimens; the corresponding per-sample values are reported in Table~\ref{tab:loss_metrics_MNIST} of Appendix~\ref{app:case3}. The white-box CANN performs marginally better than the grey-box CANN, but the difference is well within the spread of the test population. Crucially, the median testing losses remains the same order of magnitude as the training loss on the single specimen used for identification, which is a strong indicator that the procedure does not overfit to the training specimen and that the simultaneously identified constitutive laws for the matrix and the inclusion transfer to the entire population of phase morphologies considered here. The remaining variability between unseen specimens correlates with discrepancies between the nominal and the as-manufactured DIC field dimensions, again pointing to geometric idealization as the dominant remaining source of error.

In summary, Case \hyperref[sec:results:case3]{3} establishes that the ADE-FEMU framework can simultaneously identify the constitutive laws of multiple phases of a heterogeneous material system from full-field DIC measurements on a single specimen, and that the resulting model generalizes to a substantial population of unseen phase morphologies of the same material class.

	\section{Discussion}\label{sec:discussion}

\paragraph{Summary of findings} The three cases of Section~\ref{sec:results} jointly establish that ADE-FEMU provides a robust, general, and easy-to-implement route from heterogeneous experimental observations to physically consistent hyperelastic constitutive neural networks. At its core, the method embeds a constitutive neural network directly into a nonlinear, differentiable JAX-FEM forward solver and treats constitutive identification as a PDE-constrained optimization problem in which the trainable parameters of the material network are updated by gradient descent on a measurement-mismatch loss. The gradient is propagated automatically through the entire forward pipeline---FE assembly, Newton solve, observation operators, and loss evaluation---by reverse-mode automatic differentiation, so that no analytic adjoint or tangent operators need to be derived and no surrogate model has to be trained offline. By design, the same training loop accommodates arbitrary observation operators and therefore adapts seamlessly to the experimental measurement modality at hand. To probe the role of the constitutive neural network model within this framework, we compared two architectural that span between model flexibility vs. interpretability. The grey-box CANN is a general feed-forward network with arbitrary depth and width that satisfies all constitutive constraints by construction and is therefore highly flexible and expressive, but offers only limited interpretability of its parameters. The white-box CANN, in contrast, is built from a small, predefined library of phenomenologically motivated terms (e.g., Neo-Hookean-, Mooney-Rivlin-type) and yields a strain-energy density whose active terms can be read off directly, at the price of a more restricted state space. This contrast is itself an experimental design choice: it allows us to disentangle the influence of constitutive flexibility from that of training-data richness across the three cases. Case \hyperref[sec:results:case1]{1} demonstrates identification from a complete DIC field of a perforated specimen and shows that the identified models generalize to a geometrically different unseen tensile test. Case \hyperref[sec:results:case2]{2} demonstrates that the framework remains operational when only a one-dimensional stretch profile and a global force--displacement curve are available, by replacing the displacement-based observation operator in the loss \eqref{eq:loss_total} by a stretch-based one. Case \hyperref[sec:results:case3]{3} demonstrates that the same workflow extends without modification to heterogeneous geometries, identifying the constitutive parameters of two phases simultaneously and generalizing across a population of twenty-two test specimens of varying inclusion topology, while training only on a single test specimen.

\paragraph{Training data matter more than architecture} A consistent observation across all three cases is that the predictive performance of the identified models is governed by the diversity of deformation modes contained in the training data more than by the choice between the grey-box and the white-box CANN. Within a given training configuration, both architectures yield essentially indistinguishable displacement and stretch fields. The grey-box CANN, with its higher capacity, benefits the most from data-rich training (heterogeneous DIC or carefully selected combinations of homogeneous tests) but is also the most exposed when the training data are restricted to a single homogeneous deformation mode. This data-hungry character of higher-capacity neural constitutive ansatzes echoes the recommendations of \cite{Tan2026,Fuhg2024Review} on multi-modal data acquisition, and is in turn consistent with the practical experience of the FEMU and EUCLID communities \cite{RomerEtAl2025,FlaschelKumarDeLorenzis2023,Abbasi2026}.

\paragraph{Global versus local data} Cases~\hyperref[sec:results:case1]{1} and~\hyperref[sec:results:case2]{2} also show a complementary effect of homogeneous and heterogeneous training data: homogeneous data primarily constrain the global stress--stretch behavior, which is reflected in their excellent reproduction of stress on UT and PS but a slight underestimation of the heterogeneous reaction force. Heterogeneous full-field data primarily constrain the local kinematic response, which is reflected in their excellent reproduction of full-field displacements and stretches but in their overestimation of the reaction force on an unseen geometry. The combination of homogeneous and heterogeneous data therefore appears as the most informative training set, in line with the recent literature on multi-modal identification \cite{Tan2026}. Among the kinematic observables, the stretch is consistently more sensitive to the constitutive parameters than the displacement; whenever stretch data are accessible, they should be preferred as a training target.

\paragraph{Regularization and scaling matters for model discovery}
As demonstrated in Case \hyperref[sec:results:case2]{2}, sparse and interpretable expressions for the strain energy density function can be identified from heterogeneous mechanical tests by applying L1 regularization to the weights of the last layer of the white-box CANN. Despite relying on heterogeneous rather than homogeneous loading conditions, the discovered constitutive model is essentially identical to the expression obtained from homogeneous data. This result highlights the robustness of the proposed framework and demonstrates that physically meaningful constitutive laws can be recovered even from more complex and experimentally realistic datasets, without sacrificing interpretability or predictive capability.
Regarding the scaling parameters of the loss terms, we recommend selecting them in a problem-dependent manner, taking into account the relative data richness of the individual fields. A suitable approach is to assign a higher scaling factor to the physical field that contains the greatest amount of information about the underlying material behavior. For example in Case \hyperref[sec:results:case1]{1}, this corresponds to the force–displacement curve, whereas in Case \hyperref[sec:results:case2]{2}, the stretch plot provides more material-specific information and should therefore receive greater emphasis.

\paragraph{Relation to existing approaches} The proposed framework occupies a complementary position in the existing landscape of full-field identification methods. Building on classical FEMU \cite{CollinsEtAl1974,WiesheierEtAl2024,WiesheierEtAl2026,Roux2020}, ADE-FEMU retains the generality of simulation-based calibration while removing the bottleneck of manually derived sensitivities through end-to-end automatic differentiation, in line with the physics-augmented FEMU paradigm of \cite{Tan2026} and the broader differentiable-implicit-layer trend \cite{Bleyer2025}. Compared with weak-form discovery methods such as VFM \cite{Grediac1989,GrediacEtAl2006} or EUCLID and its recent extensions \cite{FlaschelKumarDeLorenzis2021,FlaschelKumarDeLorenzis2022,FlaschelKumarDeLorenzis2023,Thakolkaran2022,Joshi2022,Marino2023,Abbasi2026}, it avoids the direct differentiation of noisy measured displacement fields and instead differentiates the computational model itself, in the spirit of the Neural-DFEM framework of \cite{Regazzoni2026}. In contrast to two-stage workflows that train a constitutive surrogate offline and re-import it into a FE code \cite{XueEtAl2023}, ADE-FEMU performs forward simulation, loss evaluation, and parameter update within a single differentiable program, which we found to be one of the main practical enablers of the present results. Conceptually, it also differs from optimization-free discovery strategies such as Material Fingerprinting \cite{Flaschel2026MF}, which trade flexibility against the cost of building an offline database, while ADE-FEMU remains anchored to a continuous, gradient-based identification.

\paragraph{Limitations} Several limitations of the present work deserve explicit acknowledgement. First, the white-box CANN passes its weights through a nonlinear activation to keep gradients informative during the inverse optimization (i.e., counteracting the vanishing-gradient effect \cite{hu2021handling}); this stabilizes training but complicates direct sparsification, since a classical $L_{1}$ penalty acts on the non-activated weights and does not necessarily promote sparsity in the activated terms that enter the strain-energy density. Achieving a compact, interpretable model therefore requires penalties tailored to the activated contributions or alternative parametrizations preserving stable gradients. We advice to generally use an alternative weight activation, as shown in Case \hyperref[sec:results:case2]{2}, where a quadratic activation is used. Second, for Cases~\hyperref[sec:results:case1]{1} and~\hyperref[sec:results:case2]{2} the FE mesh is built from specimen drawings reported in the original experimental publications, so the identification depends on the agreement between nominal and as-manufactured geometries; several of the systematic, model-independent error concentrations along the hole boundaries are consistent with this idealization and could be mitigated by reconstructing the actual geometry from the DIC reference image.  Third, the relative weighting between displacement, stretch, and reaction-force terms in the loss \eqref{eq:loss_total} has been calibrated heuristically; a more systematic, Bayesian or adaptive strategy would likely improve the identification, especially when observables differ by orders of magnitude in magnitude. Fourth, all examples concern incompressible isotropic hyperelasticity, for which homogeneous tests can cover the entire invariant plane; extending ADE-FEMU to anisotropic, compressible, or inelastic materials, where such coverage is harder to obtain, is an important next step \cite{Holthusen2026a,Holthusen2026,Abdolazizi2024,Rosenkranz2024,Kalina2025}. A further transversal observation is that displacement-driven boundary-value problems leave only a small optimization margin to the constitutive model, because the displacement field is largely fixed by the boundary conditions and the structural geometry; future experimental designs should therefore prefer load-controlled or mixed-control conditions whenever possible.

\paragraph{Outlook} The natural extensions of the present work follow directly from these limitations. A first direction concerns more expressive yet sparsifiable constitutive ansatzes that combine the discovery capability of EUCLID \cite{FlaschelKumarDeLorenzis2023,Flaschel2025NS} with the structural inductive biases of HNNs \cite{Regazzoni2026}, the convex-by-construction representations recently proposed for generalized standard materials \cite{Flaschel2025JMPS}, and the multi-modal training strategy of paFEMU \cite{Tan2026}. A second direction concerns the systematic embedding of uncertainty quantification, for instance through Bayesian constitutive neural networks \cite{Linka2025} or Bayesian-EUCLID-type approaches \cite{Joshi2022}, into the differentiable training loop. A third direction concerns the extension of ADE-FEMU to inelastic, viscoelastic, and anisotropic materials, building on existing complement-based and dual-potential architectures \cite{Holthusen2026a,Holthusen2026,Abdolazizi2024,Rosenkranz2024,Benady2024,Boes2026,Friedrichs2026,Jadoon2025,Jones2026,Ji2026} as well as on multiscale neural-network frameworks for finite-strain magneto-elasticity and lattice structures \cite{Kalina2023FEANN,Kalina2024MS,Stollberg2025}. A fourth direction concerns the application of the proposed simulation-native discovery pipeline to active biological tissues such as in growth-and-remodeling settings \cite{Holthusen2025TH} and to inverse design of anisotropic microstructures \cite{Jadoon2025ID}. Finally, the open-source nature of the underlying JAX-FEM stack makes it straightforward to integrate the proposed framework with optimal experimental design \cite{PierronGrediac2020,Ghouli2025} and with full three-dimensional DIC, paving the way toward routine simulation-native constitutive discovery from realistic experiments.

\section{Conclusion}\label{sec:conclusion}

We have presented ADE-FEMU, an end-to-end differentiable finite element framework for learning hyperelastic constitutive neural network models from heterogeneous full-field deformation data. By embedding the constitutive network directly into a JAX-FEM forward solver and propagating gradients through the entire pipeline by automatic differentiation, the proposed approach unifies the generality of simulation-based identification, the architectural rigor of physics-augmented neural networks, and the implementational simplicity of differentiable programming. We have deliberately instantiated the framework with two complementary architectures that span the flexibility--interpretability trade-off: a high-capacity grey-box CANN and an interpretable, expert-system-type white-box CANN. The three demonstration cases reveal a consistent and actionable pattern: the flexible grey-box CANN consistently performs best when the training data are rich, that is, when full-field DIC measurements or heterogeneous specimens cover a sufficiently informative portion of the deformation space, whereas the more constrained white-box CANN is the more reliable choice for genuine model discovery from sparser data such as a few homogeneous stress--stretch tests, where its built-in inductive biases compensate for the lack of data. Across all three settings, ADE-FEMU reliably delivered mechanically consistent constitutive descriptions that generalized beyond the training experiments. We anticipate that the open-source nature of the underlying stack and the modular design of the loss and observation operators will make ADE-FEMU a useful starting point for further extensions toward inelastic, anisotropic, and uncertainty-aware constitutive discovery from structural-scale experiments.

\appendix

\section{Supplementary error metrics and full-field results}\label{app:metrics}

For compactness, several supporting tables and figures referenced in Section~\ref{sec:results} have been collected in this appendix. Specifically, Appendix~\ref{app:case1} reports the quantitative error metrics of Case \hyperref[sec:results:case1]{1} in tabular form and includes the stretch field and the corresponding error-distribution plots on the unseen specimen TT2. Appendix~\ref{app:case2} reports the supplementary training--testing matrix and the consolidated error metrics of Case \hyperref[sec:results:case2]{2}, as well as results for varying scaling parameter ratios and regularization penalties. Appendix~\ref{app:case3} reports the error metrics of Case \hyperref[sec:results:case3]{3} for the training and best- and worst-case unseen specimens as well as the per-sample testing losses over the full set of unseen specimens.

\subsection{Case 1: Quantitative error metrics and supplementary TT2 fields}\label{app:case1}

\begin{table}[H]
    \centering
    \footnotesize
    \caption{\textbf{Case \hyperref[sec:results:case1]{1}, one-dimensional error metrics.} Relative $\mathcal{L}^{2}$ error of the first Piola--Kirchhoff stress on the homogeneous tests UT and PS, and relative $\mathcal{L}^{2}$ error of the global reaction force on TT1 (training fit) and TT2 (unseen) for each combination of constitutive architecture (white-box or grey-box CANN) and training set. The training set used for each row is indicated in the second column.}
    \label{tab:LS_1D_quantities}
    \begin{tabularx}{\textwidth}{lXXXXXX}
    \toprule
       Model & Training data & UT & PS & TT1-$F_y$ & TT2-$F_y$  \\
        & &  $\mathcal{L}^2(P_{11})$ & $\mathcal{L}^2(P_{11})$ & $\mathcal{L}^2(F_y)$ & $\mathcal{L}^2(F_y)$    \\
    \midrule
        White-box CANN & TT1     & 1.7E+3\,\% & 21.99\,\% & 0.54\,\% & 9.51\,\% \\
        Grey-box CANN  & TT1     & 38.95\,\%  & 11.63\,\% & 0.54\,\% & 8.38\,\% \\
        White-box CANN & UT~\&~PS & 1.99\,\%  & 6.09\,\%  & 6.86\,\% & 3.23\,\% \\
        Grey-box CANN  & UT~\&~PS & 1.48\,\%  & 5.76\,\%  & 6.34\,\% & 3.45\,\% \\
        White-box CANN & UT      & 1.37\,\%  & 36.25\,\% & 4.46\,\% & 5.28\,\% \\
        Grey-box CANN  & UT      & 0.63\,\%  & 41.47\,\% & 4.76\,\% & 6.54\,\% \\
    \bottomrule
    \end{tabularx}
\end{table}

\begin{table}[H]
    \footnotesize
    \centering
    \caption{\textbf{Case \hyperref[sec:results:case1]{1}, two-dimensional error metrics.} Median pointwise relative error of the displacement field $u_y$ and of the maximum principal stretch $\lambda_1$, evaluated at the DIC points of the heterogeneous specimens TT1 and TT2 for each combination of constitutive architecture and training set. Notation: $\hat{u}_y$ and $\hat{\lambda}_1$ denote experimentally measured values.}
    \label{tab:LS_2D_quantities}
    \begin{tabularx}{\textwidth}{lXXXXXX}
    \toprule
    Model & Training data & TT1-$u_y$ & TT1-$\lambda_1$ & TT2-$u_y$ & TT2-$\lambda_1$ \\
       &  & median$|1-\frac{u_y}{\hat u_y}|$ & median$|1-\frac{\lambda_1}{\hat \lambda_1}|$ & median$|1-\frac{u_y}{\hat u_y}|$ & median$|1-\frac{\lambda_1}{\hat \lambda_1}|$  \\
    \midrule
        White-box CANN & TT1     & 1.42\,\% & 4.09\,\% & 1.90\,\% & 3.53\,\% \\
        Grey-box CANN  & TT1     & 1.29\,\% & 3.91\,\% & 1.75\,\% & 3.38\,\% \\
        White-box CANN & UT~\&~PS & 1.63\,\% & 4.43\,\% & 1.74\,\% & 3.39\,\% \\
        Grey-box CANN  & UT~\&~PS & 1.70\,\% & 4.65\,\% & 1.81\,\% & 3.49\,\% \\
        White-box CANN & UT      & 1.96\,\% & 4.81\,\% & 2.20\,\% & 3.76\,\% \\
        Grey-box CANN  & UT      & 2.05\,\% & 5.16\,\% & 2.41\,\% & 3.98\,\% \\
    \bottomrule
    \end{tabularx}
\end{table}

\begin{figure}[H]
    \centering
    \includegraphics[width=\linewidth]{ 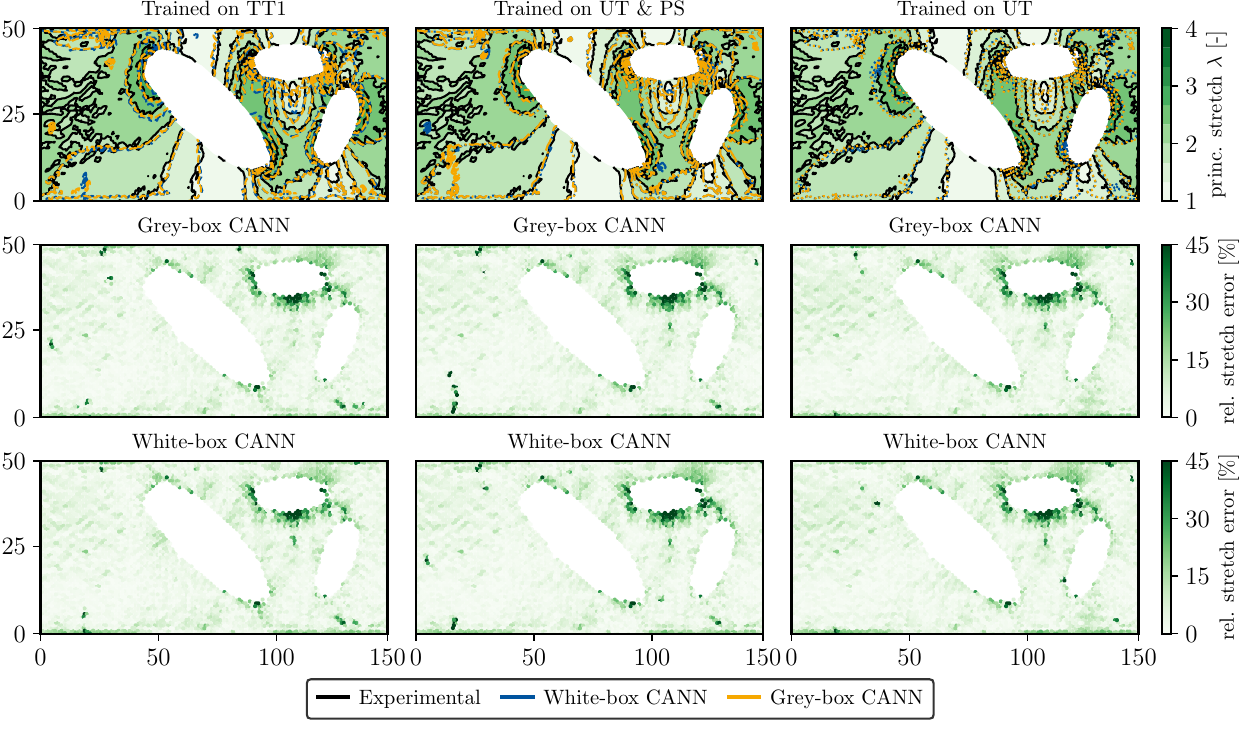}
    \caption{\textbf{Case \hyperref[sec:results:case1]{1}, supplementary stretch field on the unseen specimen TT2.} For each training configuration (columns: UT, UT~$\&$~PS, TT1) and each architecture (grey-box CANN and white-box CANN), the figure compares the DIC-measured maximum principal stretch $\lambda_1$ (top row, black contour lines on filled colormap) with the FE prediction (overlaid blue contours for the white-box CANN, yellow contours for the grey-box CANN), and reports the pointwise relative error of the grey-box prediction (middle row) and of the white-box prediction (bottom row) with respect to the DIC reference. Companion of Figure~\ref{fig:all_models_on_TT1_stretch}.}
    \label{fig:all_models_on_TT2_stretch}
\end{figure}

\begin{figure}[H]
    \centering
    \includegraphics[width=\linewidth]{ 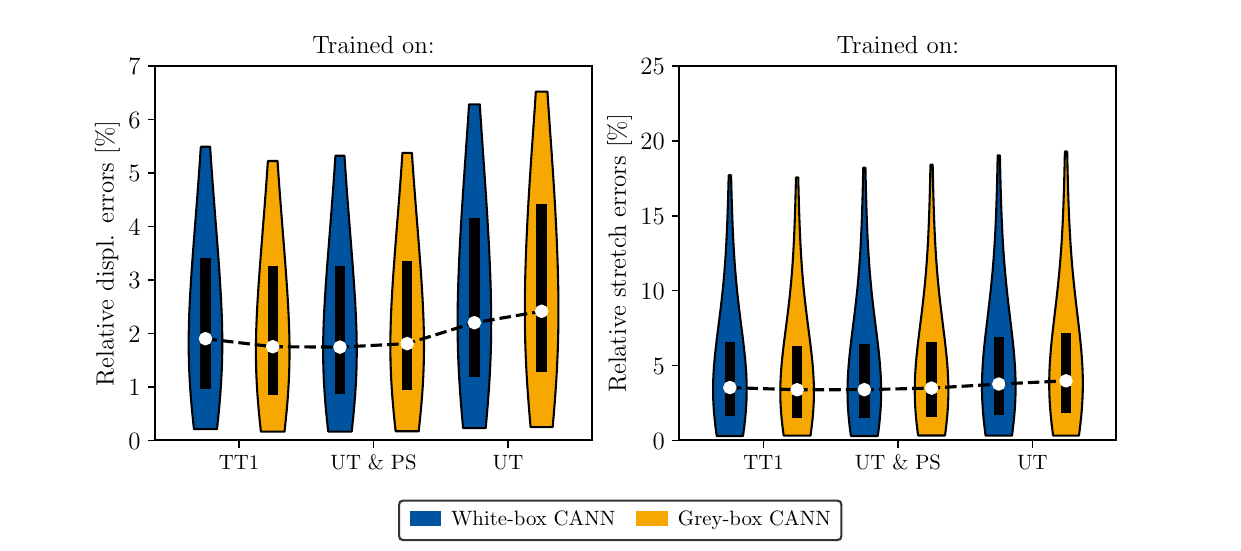}
    \caption{\textbf{Case \hyperref[sec:results:case1]{1}, supplementary error distributions on the unseen specimen TT2.} Empirical distributions of the pointwise relative errors of the predicted displacement field (left) and of the maximum principal stretch (right) on TT2, evaluated for each combination of architecture and training set. Each violin shows the central $90\,\%$ of the data (clipped between the $5\,\%$ and $95\,\%$ percentile); the black bar indicates the central $50\,\%$ (clipped between the $25\,\%$ and $75\,\%$ percentile); the white dot marks the median. Companion of Figure~\ref{fig:TT1_violins}.}
    \label{fig:TT2_violins}
\end{figure}

\subsection{Case 2: Quantitative error metrics,  supplementary training--testing matrix, loss-scaling \& regularization study and geometric information}\label{app:case2}

\begin{table}[H]
    \centering
    \footnotesize
    \caption{\textbf{Case \hyperref[sec:results:case2]{2}, relevant coordinates of the heterogeneous tensile test experiment} }
    \label{tab:FH_result_table}
    \begin{tabularx}{\linewidth}{lXXXXXXXXX}
    \toprule
      Coordinate / Point & A & B & C & D  & C1 & C2 & C3 & C4 & C5\\
      
    \midrule
        x [mm] & 0.0     & -62.0 & -61.5 &  0.0 & -47.5 & -14 & -31.5 &  -47.5 & -14.0 \\
        y [mm]  & 0.0      & 0.0 & -82.5  & -82.5 & -21.2 & -23.0 & -41.0 &-59.0 & -58.0 \\

    \bottomrule
    \end{tabularx}
        \label{tab:case2_coors}
\end{table}

\begin{table}[H]
    \centering
    \footnotesize
    \caption{\textbf{Case \hyperref[sec:results:case2]{2}, error metrics for the reduced-data setting.} Relative $\mathcal{L}^{2}$ error of the first Piola--Kirchhoff stress on uniaxial tension and compression (UT, UC), pure shear (PS), plane-strain compression (PSC), and bulge test (BL); relative reaction-force error and relative $\mathcal{L}^{2}$ error of the maximum principal stretch along the measurement path of TT. Errors above $100\,\%$ reflect extrapolation beyond the training range.}
    \label{tab:FH_result_table}
    \begin{tabularx}{\linewidth}{lXXXXXXXX}
    \toprule
      Model & Training & UT & UC & PS  & PSC & BL & TT-$F_y$ & TT-$\lambda_1$\\
       & data & $\mathcal{L}^2(P_{11})$ & $\mathcal{L}^2(P_{11})$  & $\mathcal{L}^2(P_{11})$ & $\mathcal{L}^2(P_{11})$ & $\mathcal{L}^2(P_{11})$ & $|1-F/\hat F|$  & $\mathcal{L}^2(\lambda_1)$ \\
    \midrule
        White-box CANN & TT      & 2.08 \,\% & 4.86 \,\% &  5.19 \,\% & 4.41 \,\% & 159.27 \,\% & 2.78 \,\% &  2.16 \,\% \\
        Grey-box CANN  & TT      & 2.33\,\% & 39.94\,\%  & 17.65\,\% & 4.31\,\%  & 142.79\,\% & 1.72\,\% & 2.05\,\% \\
        White-box CANN & UT~\&~PS &  2.70\,\% &  4.17\,\%   &   4.22 \,\%  &  3.31 \,\%  & 69.93 \,\%  &  2.61 \,\% &  3.12 \,\% \\
        Grey-box CANN  & UT~\&~PS & 2.73\,\% & 6.92\,\%   & 5.57\,\%  & 4.92\,\%  & 21.32\,\%  & 5.01\,\% & 5.17\,\% \\
        White-box CANN & UT      &  2.52\,\% &  348.81\,\% &  90.3\,\% &  49.71\,\% & 183.05\,\% & 5.19\,\% &  4.56\,\% \\
        Grey-box CANN  & UT      & 2.02\,\% & 306.62\,\% & 83.72\,\% & 33.12\,\% & 1.5E+3\,\% & 4.58\,\% & 3.71\,\% \\
    \bottomrule
    \end{tabularx}
\end{table}

\begin{figure}[H]
    \centering
    \includegraphics[width=\linewidth]{ 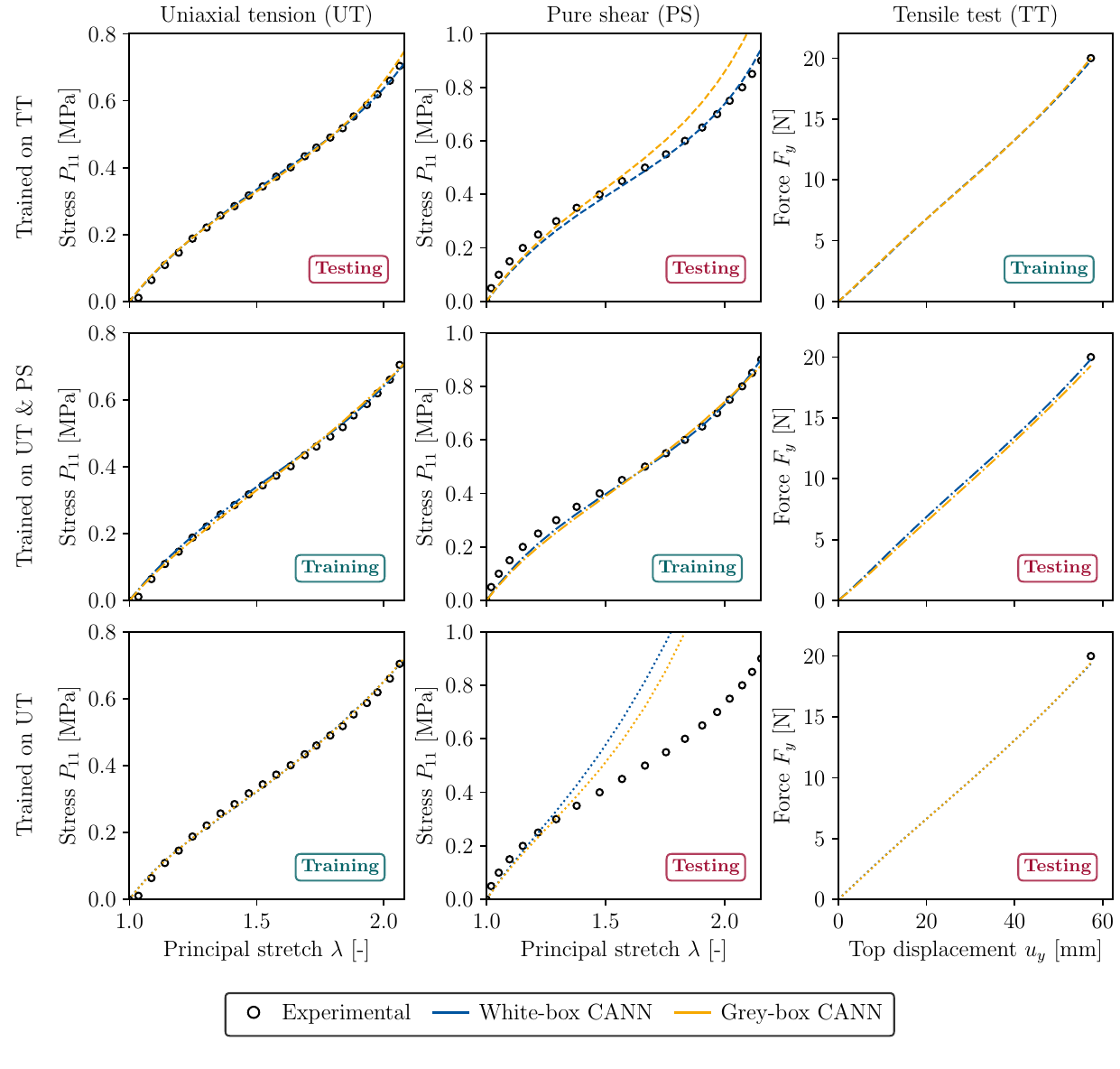}
    \caption{\textbf{Case \hyperref[sec:results:case2]{2}, supplementary training--testing matrix.} Stress--stretch responses of all six trained models (grey-box and white-box CANN; three training configurations TT, UT~$\&$~PS, UT) on the homogeneous tests UT and PS, and reaction-force response on the heterogeneous specimen TT. Each row corresponds to a training configuration; the training data for each row is explicitly labeled. The grey-box and the white-box CANN trained on the same data deliver very similar curves, with the largest inter-architecture differences appearing for models trained on UT alone. Companion of Figure~\ref{fig:FH_training_testing_matrix_MD}, which decomposes the white-box responses into their individual term contributions.}
    \label{fig:FH_training_testing_matrix}
\end{figure}

\begin{figure}[htbp]
        \centering
        \includegraphics[width=\linewidth]{ 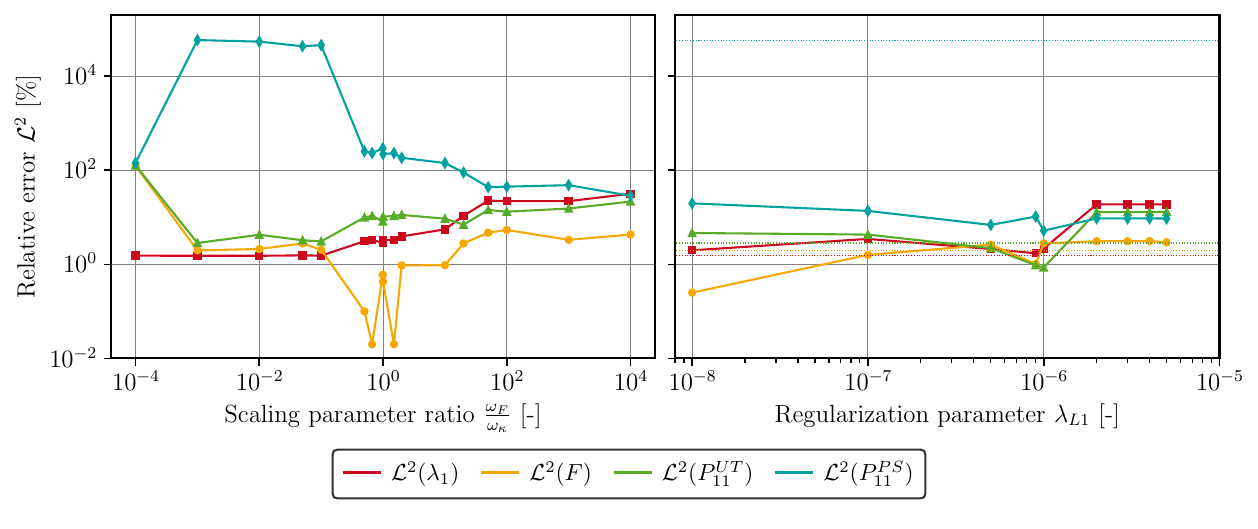}
        \caption{\textbf{Case \hyperref[sec:results:case2]{2}, influence of scaling and regularization parameters for the white-box CANN.} On the left: investigation of the unregularized optimization problem for scaling parameters $\omega_F$ and $\omega_\kappa$. The study shows choosing a far larger scaling parameter for the kinematic term than for the force term seems favorable in reducing all error metrics: $10^{-3} < \frac{\omega_{F}}{\omega_{\kappa}} <10^{-1}$. On the right: investigation of the regularization parameter for a scaling parameter ratio $\frac{\omega_{F}}{\omega_{\kappa}} = 10^{-3}$. The error metrics for $\lambda_{L1} = 0$ (unregularized) is marked with dotted lines (..). The best performance can be seen for $\lambda_{L1} = 10^{-6}$. The error plateau for $\lambda_{L1} > 10^{-6}$ is due to the elimination of all but one white-box CANN term.}
        \label{fig:case_2_norm_and_reg_evolution}
\end{figure}

\begin{figure}[htbp]
        \centering
        \includegraphics[width=\linewidth]{ 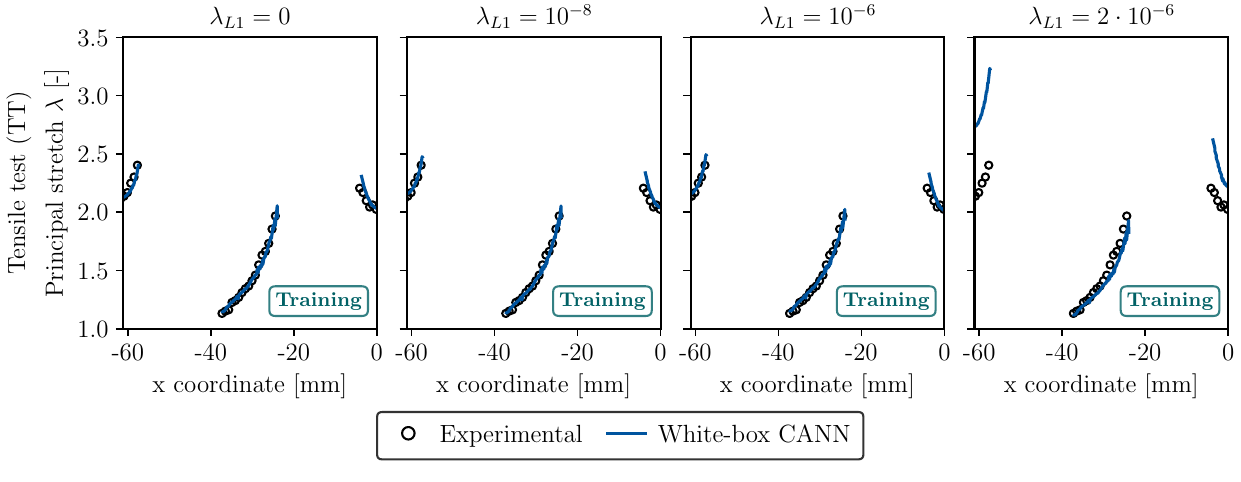}
        \caption{\textbf{Case \hyperref[sec:results:case2]{2}, fits of the stretch path, calculated with regularized white-box CANNs.} }
        \label{fig:stretch_reg}
\end{figure}

\begin{figure}[htbp]
        \centering
        \includegraphics[width=\linewidth]{ 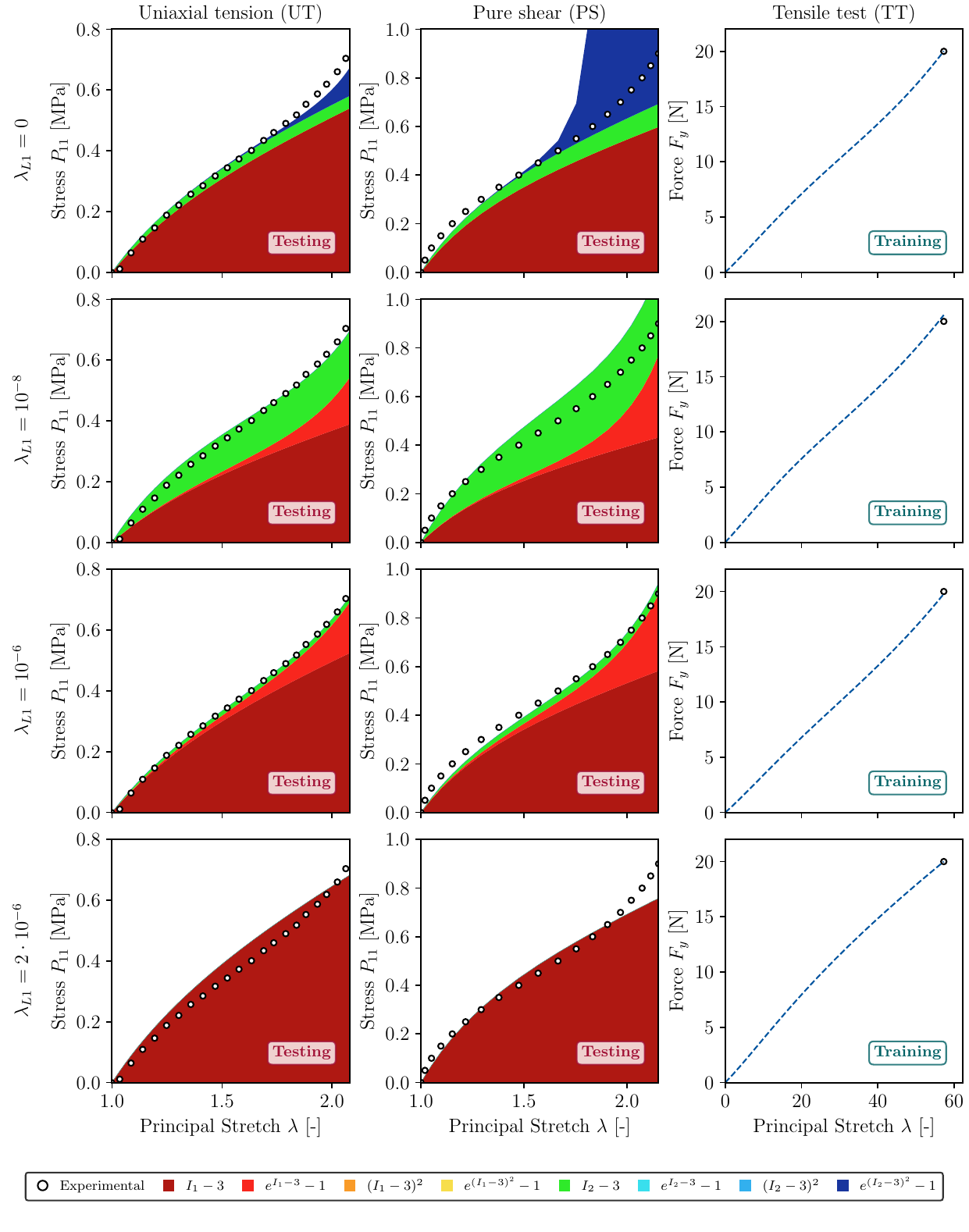}
        \caption{\textbf{Case \hyperref[sec:results:case2]{2}, reproduction of the homogeneous tests and fits of the force---displacement data, calculated with regularized white-box CANNs.} }
        \label{fig:UT_PF_F_reg}
\end{figure}

\subsection{Case 3: Quantitative error metrics across training and unseen specimens}\label{app:case3}

\begin{table}[H]
    \centering
    \footnotesize
    \caption{\textbf{Case \hyperref[sec:results:case3]{3}, error metrics for the heterogeneous material system.} Relative reaction-force error, relative displacement error, and relative maximum principal stretch error on the training specimen and on the best- and worst-case unseen specimens, for the white-box and grey-box CANN trained simultaneously on the matrix and on the inclusion.}
    \label{tab:error_metrics_MNIST}
    \begin{tabularx}{\textwidth}{lXXXXXX}
    \toprule
    \multirow{2}{*}{Case} & \multicolumn{3}{c}{White-box CANN} & \multicolumn{3}{c}{Grey-box CANN} \\
    \cmidrule(lr){2-4} \cmidrule(lr){5-7}
    & Force error [\%] & Disp. error [\%] & Stretch error [\%] & Force error [\%] & Disp. error [\%] & Stretch error [\%] \\
    \midrule
    Training             & 3.42  & 2.78  & 1.54 & 3.60  & 2.75  & 1.54 \\
    Testing (best case)  & 2.87  & 2.76  & 1.62 & 3.03  & 2.77  & 1.60 \\
    Testing (worst case) & 15.10 & 10.64 & 2.95 & 15.26 & 10.34 & 2.88 \\
    \bottomrule
    \end{tabularx}
\end{table}

\begin{table}[H]
    \centering
    \footnotesize
    \caption{\textbf{Case \hyperref[sec:results:case3]{3}, per-sample testing losses on the unseen specimens.} Relative kinematic loss $\mathcal{L}_u$, relative reaction-force loss $\mathcal{L}_F$, and total loss $\mathcal{L} \frac{\omega_{\kappa}{2}\mathcal{L}_u} + \frac{\omega_F}{2} \mathcal{L}_F$ (with $\omega_{\kappa} = 2$} and $\omega_F = 1$) for each of the twenty-two unseen specimens, evaluated with the white-box and the grey-box CANN trained on sample~22. Sample~22 is the training specimen.
    \label{tab:loss_metrics_MNIST}
    \begin{tabularx}{\textwidth}{lXXXXXX}
    \toprule
    \multirow{2}{*}{Sample} & \multicolumn{3}{c}{White-box CANN} & \multicolumn{3}{c}{Grey-box CANN} \\
    \cmidrule(lr){2-4} \cmidrule(lr){5-7}
    & $\mathcal{L}_u$ [\%] & $\mathcal{L}_F$ [\%] & $\mathcal{L}$ [\%] & $\mathcal{L}_u$ [\%] & $\mathcal{L}_F$ [\%] & $\mathcal{L}$ [\%] \\
    \midrule
    Sample 2             & 0.95 & 0.40 & 1.15 & 0.97 & 0.39 & 1.16 \\
    Sample 6             & 0.59 & 0.42 & 0.80 & 0.59 & 0.45 & 0.82 \\
    Sample 10            & 0.26 & 0.50 & 0.51 & 0.26 & 0.49 & 0.50 \\
    Sample 18            & 1.13 & 0.11 & 1.18 & 1.13 & 0.12 & 1.19 \\
    Sample 22 (training) & 0.18 & 0.12 & 0.24 & 0.18 & 0.14 & 0.24 \\
    Sample 26            & 0.33 & 0.09 & 0.37 & 0.33 & 0.10 & 0.38 \\
    Sample 30            & 0.36 & 0.18 & 0.45 & 0.36 & 0.18 & 0.45 \\
    Sample 34            & 0.60 & 0.51 & 0.86 & 0.61 & 0.47 & 0.84 \\
    Sample 38            & 0.14 & 0.96 & 0.62 & 0.14 & 0.95 & 0.62 \\
    Sample 42            & 0.52 & 0.16 & 0.60 & 0.52 & 0.16 & 0.59 \\
    Sample 46            & 0.82 & 2.47 & 2.05 & 0.82 & 2.61 & 2.13 \\
    Sample 50            & 0.35 & 1.52 & 1.11 & 0.35 & 1.54 & 1.12 \\
    Sample 54            & 1.30 & 3.15 & 2.88 & 1.27 & 3.23 & 2.89 \\
    Sample 58            & 0.28 & 0.98 & 0.77 & 0.28 & 1.07 & 0.81 \\
    Sample 62            & 0.79 & 1.91 & 1.74 & 0.79 & 2.01 & 1.79 \\
    Sample 66            & 0.29 & 3.24 & 1.91 & 0.29 & 3.27 & 1.93 \\
    Sample 70            & 0.39 & 2.86 & 1.81 & 0.39 & 2.94 & 1.86 \\
    Sample 73            & 0.34 & 1.99 & 1.34 & 0.35 & 2.07 & 1.38 \\
    Sample 77            & 0.21 & 2.34 & 1.37 & 0.21 & 2.37 & 1.39 \\
    Sample 81            & 0.40 & 1.96 & 1.38 & 0.40 & 2.11 & 1.45 \\
    Sample 85            & 0.31 & 0.79 & 0.71 & 0.31 & 0.86 & 0.74 \\
    Sample 89            & 0.51 & 1.42 & 1.22 & 0.51 & 1.47 & 1.25 \\
    \bottomrule
    \end{tabularx}
\end{table}


	\section*{Acknowledgements}
	K. Linka gratefully acknowledges financial support from Deutsche Forschungsgemeinschaft (DFG, German Research Foundation) – 517243167.

    \section*{Declaration of generative AI and AI-assisted technologies in the writing process}
    During the preparation of this work the authors used OpenAI's ChatGPT-5.2 in order to improve language and readability. After using this tool, the authors reviewed and edited the content as needed and take full responsibility for the content of the published article.

    \section*{Code availability}
    The source code that reproduces all results of this manuscript, together with the example datasets used in Cases~1--3, will be made publicly available upon publication.


	\bibliography{library.bib}
	\bibliographystyle{unsrt}
	
\end{document}